\documentclass[a4paper,11pt]{article}
\pdfoutput=1 

\usepackage{jheppub} 

\usepackage[T1]{fontenc} 

\title{ Mass distributions of dijet resonances from excited quarks at proton-proton colliders}

\author[a]{Emine Gurpinar Guler,}
\author[a]{Yalcin Guler,}
\author[b]{and Robert M. Harris}

\affiliation[a]{Konya Technical University,\\Road 42250, Selcuklu/Konya, Turkey}
\affiliation[b]{Fermilab,\\P.O. Box 500, Batavia, IL, 60510, U.S.A.}

\emailAdd{emine.gurpinar@cern.ch}
\emailAdd{yalcin.guler@cern.ch}
\emailAdd{rharris@fnal.gov}

\abstract{We study the expected experimental mass distributions of dijet resonances from excited quarks in proton-proton collisions at energies $\sqrt{s}=$ 13, 14, 27, 100, 300, and 500 TeV. 
We explore in detail the expected shapes at both the generator and experimental levels, and identify within the distributions the effects of the excited quark natural width, parton momentum distributions of the proton, radiation, and experimental resolution. We present both differential and cumulative probability distributions as a function of dijet mass, and the signal acceptance of a window in dijet mass centered on each resonance. We find that for a range of resonance masses, between 10\% and 50\% of $\sqrt{s}$, the dijet mass distributions and window acceptance are practically universal, approximately invariant under changes in resonance mass and $\sqrt{s}$. This work supports our Snowmass 2021 study on the sensitivity to dijet resonances at proton-proton colliders. }

\begin{document} 
\begin{flushright}
FERMILAB-PUB-22-301-PPD-SCD \\
May 5, 2022
\end{flushright}
\maketitle
\flushbottom

\section{Introduction}
\label{sec:intro}

A significant benchmark for discovery at a proton-proton ($pp$) collider is the sensitivity to a dijet resonance~\cite{Harris:2022kls}, $X$, the intermediate state of the s-channel process $pp \rightarrow X \rightarrow 2\mbox{ jets}$. To search for new particles with the highest values of resonance mass ($M$), hadron collider experiments have used the classic technique of searching for bumps in the invariant mass spectrum of dijets~\cite{Harris:2011bh}, the two individually resolved jets with the largest transverse momentum. The most common benchmark model for dijet resonances is the excited quark, $q^*$~\cite{ref_qstar,Baur:1989kv}, explicitly sought and constrained by every LHC search for resolved dijet resonances~\cite{Aad:2011aj,Khachatryan:2010jd,ATLAS2010,Chatrchyan2011123,ATLAS:2012pu,CMS:2012yf,CMS:2013qha,ATLAS:2014aqa,CMS:2015sja,CMS:2015dcf,ATLAS:2015nsi,CMS:2016ecr,CMS:2016iap,ATLAS:2017yvp,ATLAS:2018fzt,CMS:2018xlo,ATLAS:2019hjw,CMS:2019vgj}. We expect the $q^*$ model will also be a benchmark of dijet resonance searches at future $pp$ colliders. 

In this paper we explore the expected shape at future $pp$ colliders of the distribution in dijet mass ($m$) of the $q^*$ signal. This shape determines the fraction of signal events that form a characteristic resonant peak, a bump, that experiments search for on top of the large continuum background of dijet events from quantum chromodynamics (QCD). The signal shape as a function of dijet mass is used by the experiments to determine statistical significance of potential signals, determine upper limits on the production cross section, and set mass limits on the $q^*$ model. To provide a foundation to better understand the experimental sensitivity of $pp$ colliders to excited quarks, this paper will explore the generator and experiment level contributions to the shapes of expected dijet mass distributions from excited quarks. We will establish the approximate invariance of the resonance shapes as a function of both $M$ and $\sqrt{s}$, and discuss the cases where that invariance is violated. We expect this exploration and discussion will be of interest to searches for dijet resonances at both current and future $pp$ colliders. 
In a previous paper~\cite{Harris:2022kls} we estimated the sensitivity of future $pp$ colliders to dijet resonances, including $q^*$.  This study will use the shapes of excited quarks at future $pp$ colliders to support that paper, demonstrating the range of validity of the dijet mass window acceptance we used.

We study existing, planned, proposed and envisioned $pp$ colliders at six values of $\sqrt{s}$. The energy $\sqrt{s}=13$ TeV, used to collect data from 2015 to 2018 at the Large Hadron Collider (LHC) at CERN, connects our studies with published searches to provide deeper understanding of the signal shapes that were used. There is also a strong interest in $\sqrt{s}=14$ TeV, the planned energy of the LHC in its final high luminosity operation (HL-LHC) anticipated to begin in 
2029 and continue for about a decade\footnote{See the LHC schedule at \href{https://lhc-commissioning.web.cern.ch/schedule/LHC-long-term.htm}{https://lhc-commissioning.web.cern.ch/schedule/LHC-long-term.htm}}. We present just a few results for $\sqrt{s}=27$ TeV, a previously studied high energy option for the LHC (HE-LHC)~\cite{FCC:2018bvk}, an energy which also has been considered for a $pp$ collider to fill the Fermilab site~\cite{Bhat:2022hdi}. Our studies are more complete for $\sqrt{s}=100$ TeV, because it is a large next step in energy to a proposed and widely studied $pp$ collider stage of a Future Circular Collider (FCC-hh) at CERN ~\cite{FCC:2018vvp}. One result is presented for $\sqrt{s}=300$ TeV, because it was a high energy option discussed at Snowmass 2021 and included in our previous paper~\cite{Harris:2022kls}. Finally, to understand how the excited quark shapes evolve to the highest possible collision energies, we include complete studies for $\sqrt{s}=500$ TeV, the envisioned energy of a Collider in the Sea~\cite{McIntyre:2017ibd}.

\section{Generator samples}
\label{sec:samples}

We have produced generator samples for the sub-process $qg \rightarrow q^* \rightarrow qg$ in $pp$ collisions using PYTHIA8 Version 8.243~\cite{Sjostrand:2014zea}. Each sample contains $20$ thousand events with $q^*$ mass $M$ equal to 10\%, 20\%, 30\%, 40\%, 50\% and 60\% of the considered $pp$ collision energy, $\sqrt{s}=13$, $14$, $27$, $100$, $300$ and $500$ TeV. The PYTHIA8 settings for all the samples were identical to those used by previously published CMS simulations~\cite{CMS:2018xlo} for $\sqrt{s}=13$ TeV, and to directly compare with those simulations, we also produced samples at $M=2$, $4$, $6$ and $8$ TeV for $\sqrt{s}=13$ TeV. As in Ref.~\cite{CMS:2018xlo}, we used the CTEQ6L1 parton distribution function (PDF) set~\cite{refCTEQ}, the CUETP8M1 event tune~\cite{Khachatryan:2015pea,Skands:2014pea}, and the standard assumptions about excited quarks: mass degenerate $u^*$ and $d^*$ only, compositeness scale $\Lambda=M$, and the same couplings of excited quarks to quarks and gauge bosons as in the standard model ($f=f^{\prime}=f_s=1$).

\section{Dijet types}
\label{sec:jet}

In this paper we consider dijet mass distributions for four different types of dijets, representing different steps in the production, reconstruction and potential observation of the excited quark.  

First, we construct "di-parton" dijet mass distributions, the invariant mass of the quark-gluon state within the sub-process $qg \rightarrow q^* \rightarrow qg$. Di-parton mass is equivalent to the sub-process total energy in the center-of momentum frame, often denoted as $\sqrt{\hat{s}}$. 

Second, we construct "AK4 genjet" dijet mass distributions, the invariant mass of the two AK4 genjets with highest transverse momentum ($p_T$) in the event, also called the two leading AK4 genjets. Here, AK4 genjets are found by clustering all the particles within the event using the anti-$k_T$ algorithm~\cite{Cacciari:2008gp,Cacciari:2011ma} with a distance parameter $R=0.4$, the effective radius of a narrow jet cone. This AK4 algorithm is a jet reconstruction algorithm, chosen as the LHC standard because it is useful for the widest range of analyses that make use of jets, but the narrow jet cone size makes it suboptimal for dijet resonance event reconstruction. 

Third, we construct "wide genjet" dijet mass distributions, where we implement the wide jet algorithm discussed in Ref.~\cite{CMS:2018xlo} using AK4 genjets as input. In this algorithm the two leading AK4 genjets are used as seeds for each of the two wide genjets, and every event is required to have exactly two wide genjets. Each wide genjet contains exactly one leading AK4 genjet, as well as all other AK4 genjets with pseudorapidity $|\eta|<2.5$ and $p_T>(\sqrt{s}/13\ \mbox{TeV}) \times 30$ GeV found within a distance $\Delta R=\sqrt{(\Delta\eta)^2 + (\Delta\phi)^2}<1.1$ from the leading AK4 genjet.  The genjet $p_T$ requirement above is generalized for all pp collision energies from the value in Ref.~\cite{CMS:2018xlo}, scaled with $\sqrt{s}$ from the requirement $p_T>30$ GeV at $\sqrt{s}=13$ TeV. The wide jet algorithm effectively widens the cone size to $R=1.1$, recovering the energy lost to jets radiated outside that distance, without including low energy jets from pileup.  It is an event algorithm, forcing the dijet interpretation on each event, and it has been optimized for dijet resonance  event reconstruction~\cite{CMS:2018xlo}.

Fourth, we construct "smeared wide genjet" dijet mass distributions, by convoluting the dijet mass distribution from wide genjets with the anticipated experimental dijet mass resolution of wide jets estimated in section~\ref{sec:resolution}. This final set of dijet mass distributions are the ones with all estimated experimental effects.

\section{Generator-level resonances}
\label{sec:genjets}

\begin{figure}[tbp]
\centering 
\includegraphics[width=.32\textwidth]{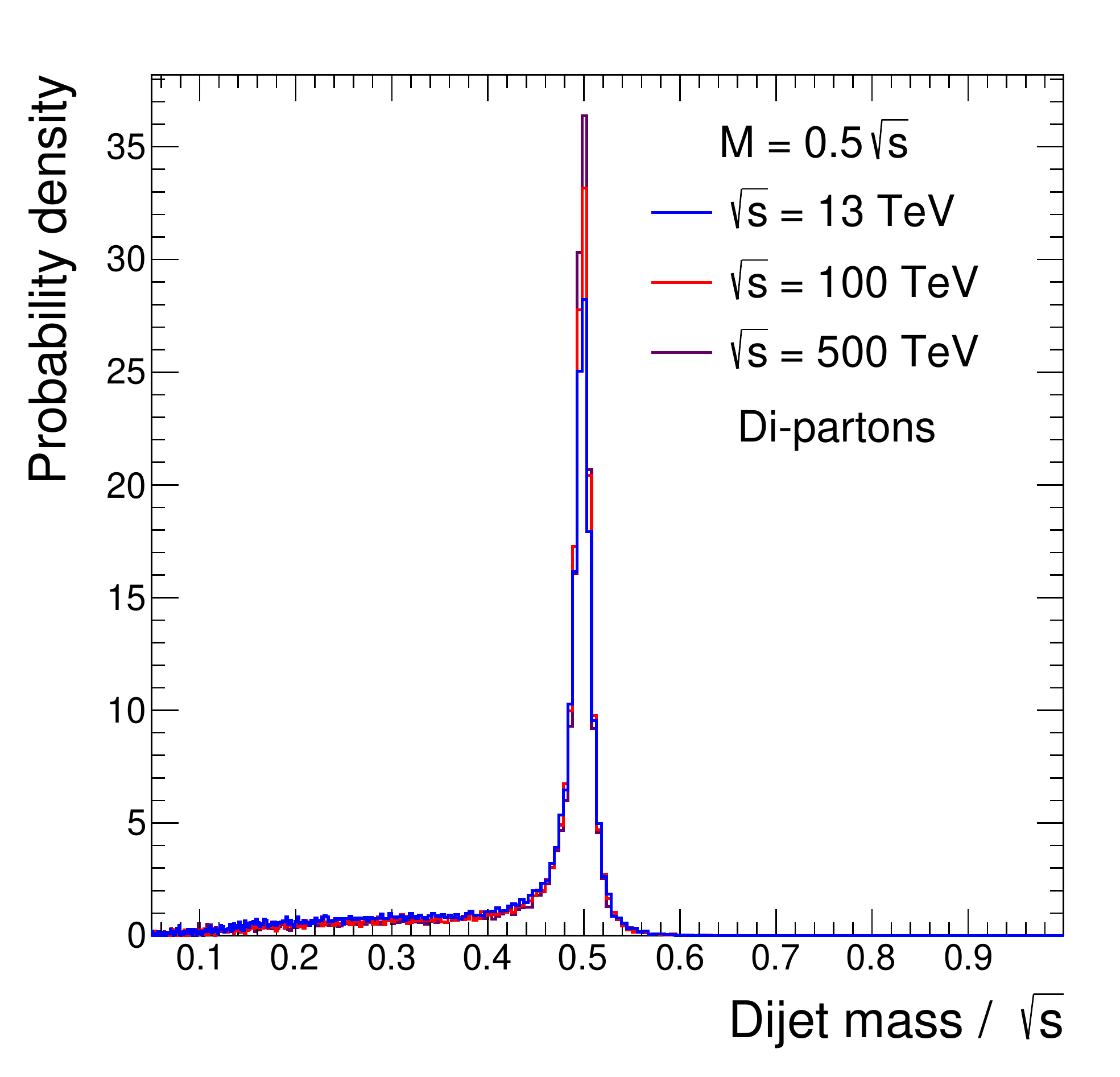}
\includegraphics[width=.32\textwidth]{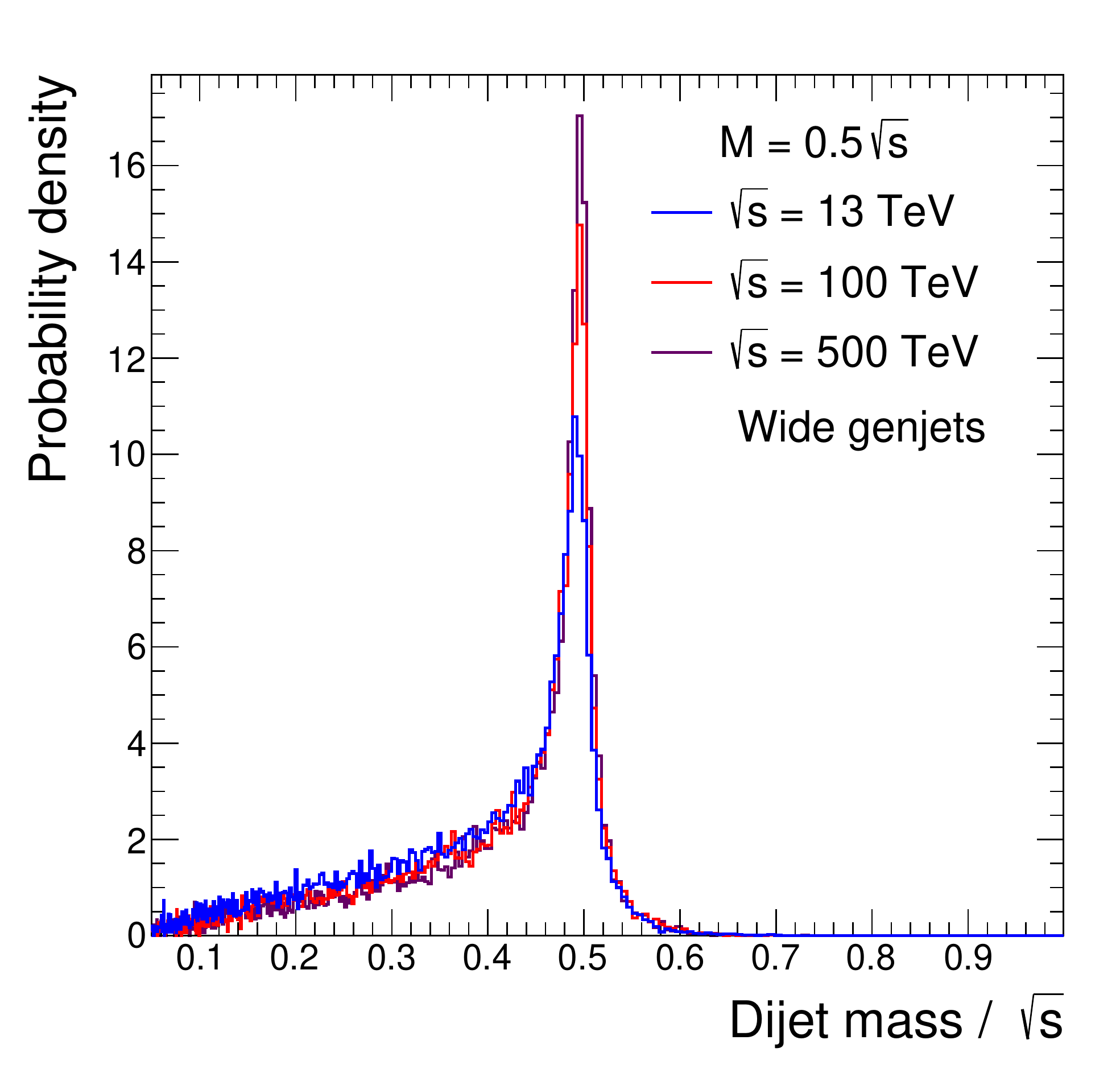}
\includegraphics[width=.32\textwidth]{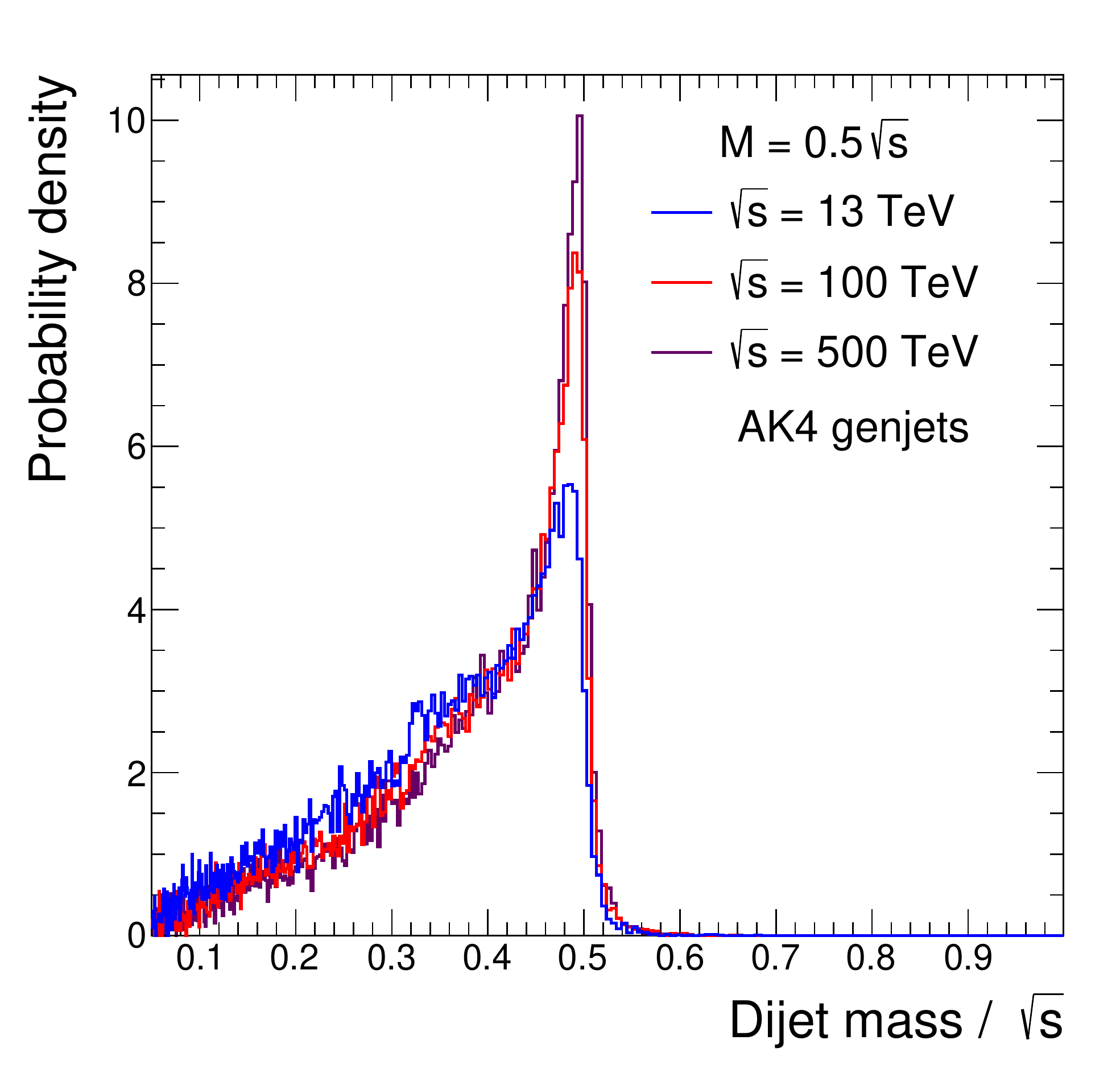}
\includegraphics[width=.32\textwidth]{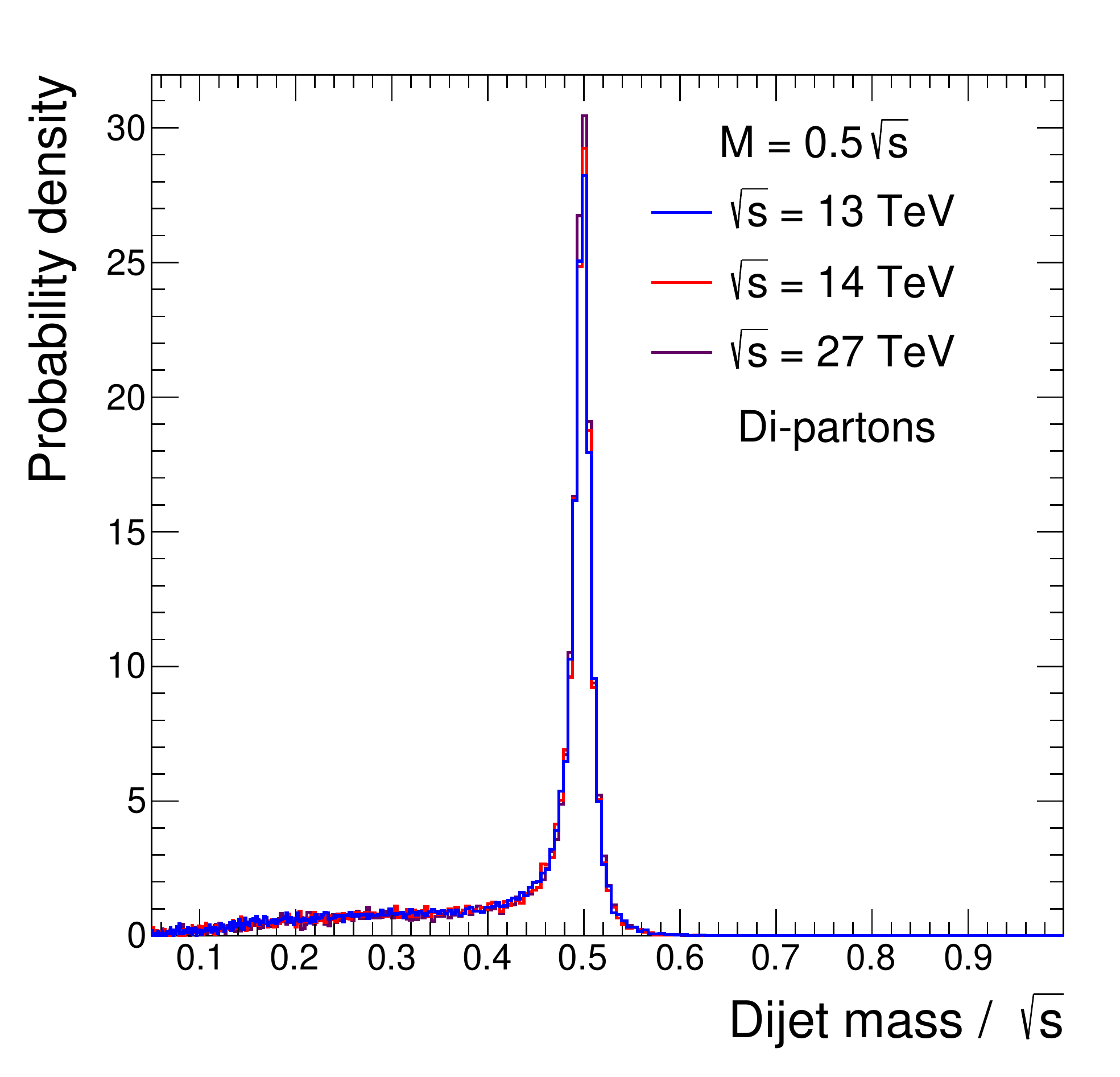}
\includegraphics[width=.32\textwidth]{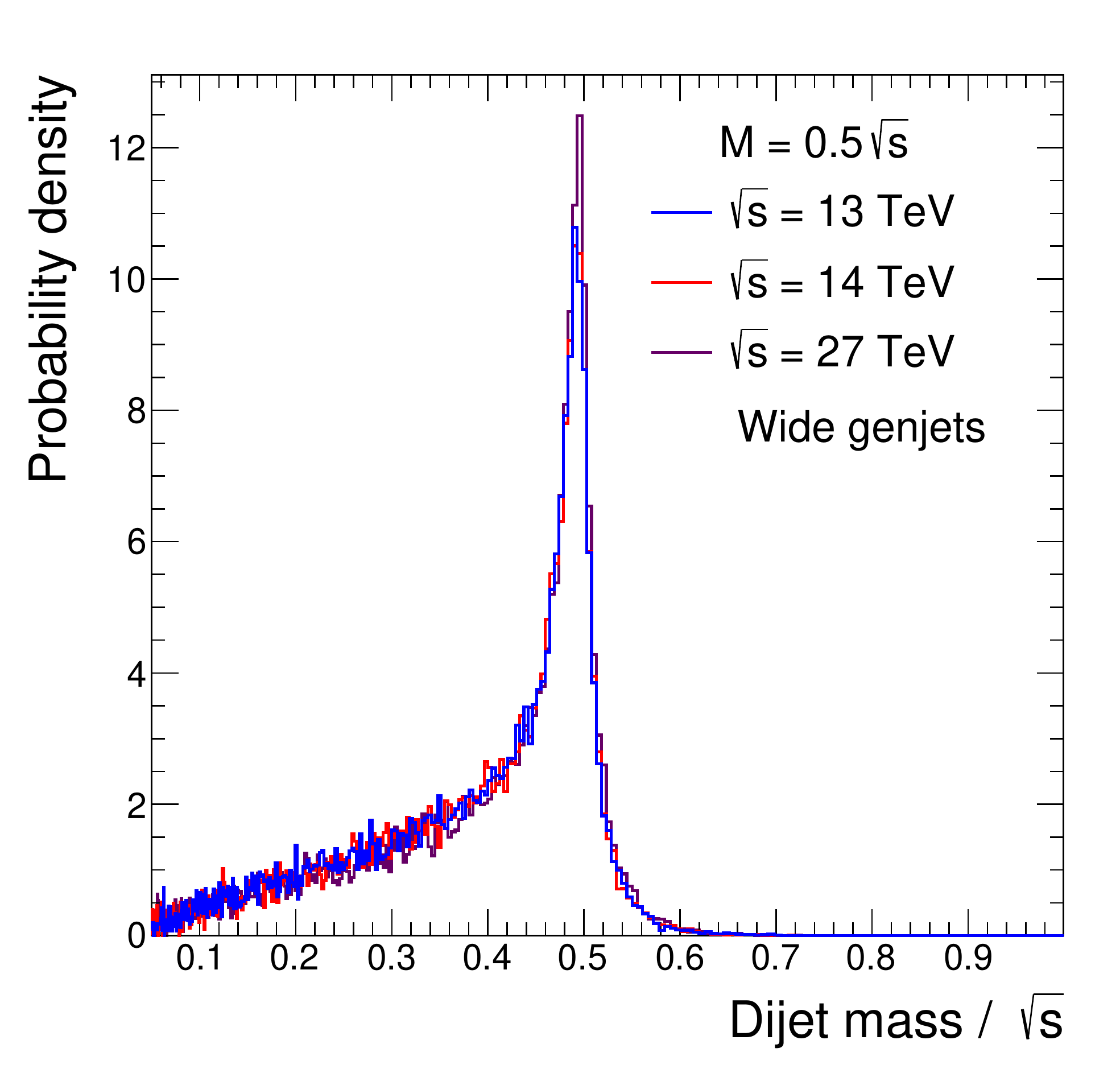}
\includegraphics[width=.32\textwidth]{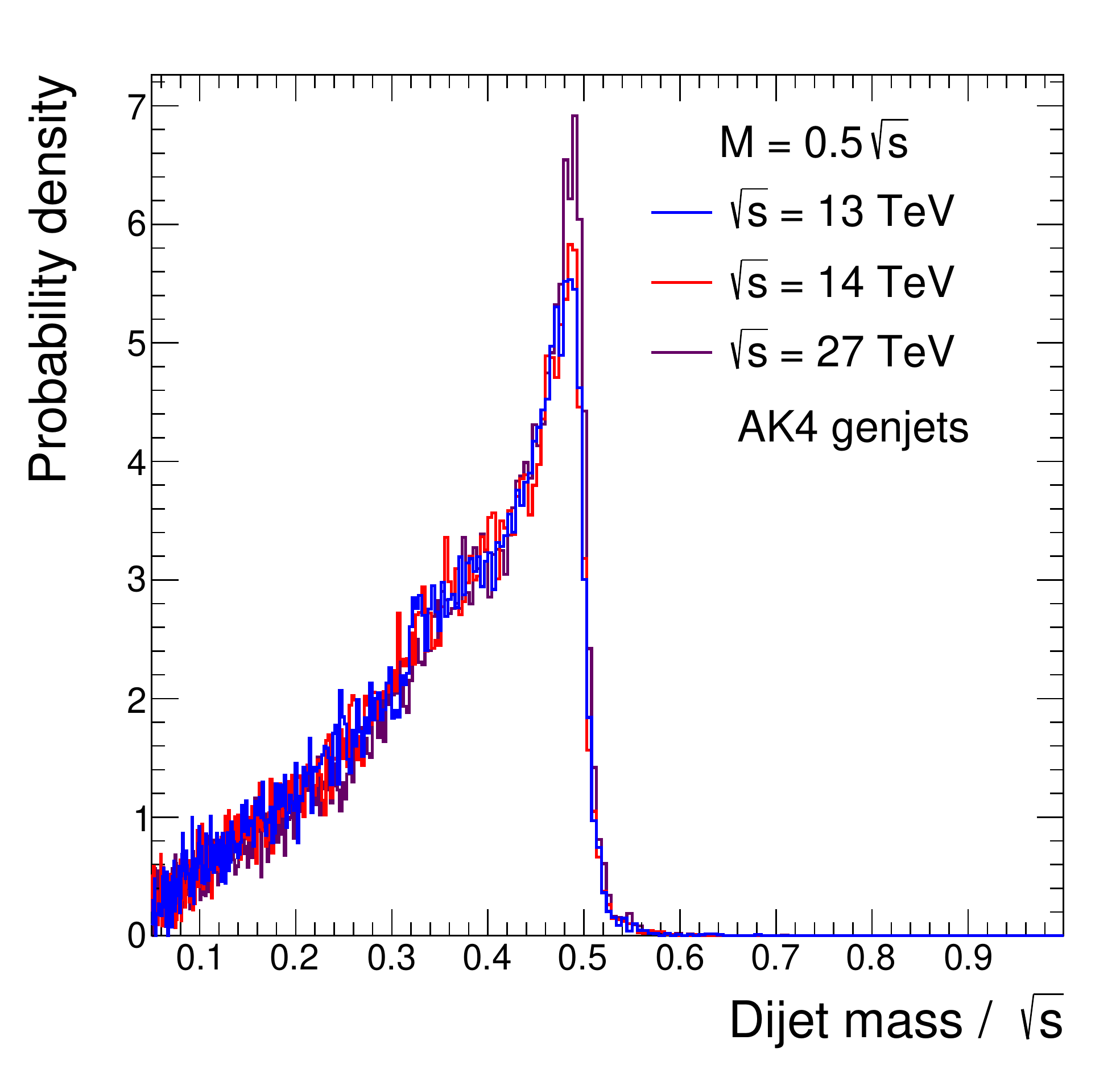}
\caption{\label{fig:GenjetSoloPlots} Dijet mass distributions for excited quarks with a resonance mass equal to 50\% of $\sqrt{s}$, from $pp$ collisions at $\sqrt{s}$ equal to 13, 100, and 500 TeV (top row) and 13, 14, and 27 TeV (bottom row), for the cases where the dijet mass is calculated from the Lorentz vectors of the two final-state partons (left column), the two wide genjets (middle column), and the two leading AK4 genjets in the event (right column).}
\end{figure}

Generator-level dijet mass distributions predict the shape of dijet resonances before the effects of experimental resolution. These distributions are valuable for understanding the physics of the shape of dijet resonances at future $pp$ colliders. They are determined by the properties of the underlying model and the parton shower within the generator, which are relatively well known compared to the significant uncertainties in the experimental resolution of future detectors.  Generator level dijet mass distributions for excited quarks are presented in Fig.~\ref{fig:GenjetSoloPlots}, \ref{fig:GenjetThreePlots} and \ref{fig:GenjetAllPlots}, and we discuss in this section what we learn from them about the physics determining the shape of dijet resonances.

\subsection{Collision energy invariance}
\label{sec:energyInvariance}

Figure~\ref{fig:GenjetSoloPlots} demonstrates the approximate invariance of the dijet mass distribution to the choice of $\sqrt{s}$. For a fixed value of the $q^*$ mass ratio, $M/\sqrt{s}=0.5$, Fig.~\ref{fig:GenjetSoloPlots} shows the dijet mass distributions of di-partons, wide genjets and AK4 genjets.  This choice of $q^*$ mass ratio was shown to be a critical value in Ref.~\cite{Harris:2022kls}, the approximate mass limit for excited quarks at $pp$ colliders at the baseline integrated luminosity. All of the distributions in Fig.~\ref{fig:GenjetSoloPlots} are for the dimensionless ratio, $m/\sqrt{s}$, which allows a fair comparison of the shapes for various choices of collider $\sqrt{s}$. Variable width bins are used corresponding to 1\% of the dijet mass and starting at a value $m/\sqrt{s}=0.05$: the bin edges are $0.05$, $0.0505$, $0.051005$, etc.

Figure~\ref{fig:GenjetSoloPlots} (left) shows the di-parton mass distributions from the generator, which are almost completely invariant with changes in $\sqrt{s}$. They originate from the differential cross section which has the general form:
\begin{equation}
  \frac{d\sigma}{dm} \sim \int_0^1\int_0^1{dx_1 dx_2 (q(x_1)g(x_2) + q(x_2)g(x_1))\left[ \frac{\Gamma^2}{\left(m^2-M^2\right)^2+M^2\Gamma^2} \right] \delta(x_1 x_2-m^2/s)}
\label{eq:BW}
\end{equation}
where $q(x)$ and $g(x)$ are the PDFs of the initial state quark and gluon, respectively, and $x_1$ and $x_2$ are their fractional momenta within each of the two colliding protons. The term in square brackets in Eq.~\ref{eq:BW} is the relativistic Breit-Wigner distribution of the resonant sub-process with width $\Gamma$. Further details of how it is treated within PYTHIA are given in Eq.(7.47) of Ref.~\cite{Sjostrand:2006za}.
For the $q^*$ model, $\Gamma$ can be written as
\begin{equation}
\Gamma/M=\frac{\alpha_s}{3} + k\alpha\approx.03
\label{eq:width}
\end{equation}
where $\alpha_s$ is the strong gauge coupling of the $q^*\rightarrow qg$ decay that produces dijets. In Eq.~\ref{eq:width}, $\alpha$ is the fine structure constant, the coupling of electroweak interactions. It is multiplied by the factor $k\approx0.81$ to obtain the electroweak partial width $k\alpha$, for the sum of the decays to an electroweak gauge boson ($q^*\rightarrow qW,qZ, \mbox{and } q\gamma$).
The $q^*$ width evolves only very slightly as a function of $\sqrt{s}$, due to the logarithmic running of the strong coupling $\alpha_s$ as a function of the excited quark mass $M$, as shown in table~\ref{tab:width}. Here $\alpha \approx 1/125$ at $M=6.5$ TeV, and the running of $\alpha$ is very small and has negligible affect on the $q^*$ width. As a result, the di-parton resonance width narrows only slightly with increasing $\sqrt{s}$ in Fig.~\ref{fig:GenjetSoloPlots} (left). Finally, notice the long tail at low dijet mass, which arises at di-parton level from the larger values of the PDFs at lower values of fractional momenta $x$, making these distributions slightly asymmetric. This effect is also discussed in the paragraphs following Eq.(7.49) in Ref.~\cite{Sjostrand:2006za}. We note it is fairly independent of $\sqrt{s}$. Throughout this paper we will discuss similar PDF effects on the distributions, and we will elaborate more on their cause and size.

\begin{table}[htbp]
\centering
\begin{tabular}{|c|c|c|c|}
\hline
$\sqrt{s}$ & $M$ & $\Gamma/M$ & $\frac{q^*\rightarrow qg}{q^*\rightarrow \mbox{all}}$ \\
(TeV) & (TeV) & & \\
\hline
13 & 6.5 & 0.032 & 0.80 \\
100 & 50 & 0.028 & 0.77 \\
500 & 250 & 0.026 & 0.75 \\ \hline
\end{tabular}
\caption{\label{tab:width} As a function of $\sqrt{s}$, we list a corresponding choice of resonance mass $M/\sqrt{s}=0.5$, the full $q^*$ width $\Gamma/M$, and the branching fraction for the decay to dijets $q^*\rightarrow qg$.}
\end{table}

In Fig.~\ref{fig:GenjetSoloPlots} (middle) the wide genjet dijet mass distributions again show almost no variation with collider energy in the narrow range $13<\sqrt{s}<27$ TeV (bottom plot), and a small variation with collider energy in a very wide range $13<\sqrt{s}<500$ TeV (top plot). Comparing with di-partons (left), we see that for wide genjets the width has grown and the tail to low mass has increased noticeably.  At this value of M, the tail is no longer predominantly due to PDFs, it is now dominated by the energy loss from very wide angle QCD radiation emitted by the final state quark and gluon. Here the radiation is also decreasing as $\sqrt{s}$ increases, again this is due to the logarithmic decrease of $\alpha_s$ with $M$.

In Fig.~\ref{fig:GenjetSoloPlots} (right) the AK4 genjet dijet mass distributions show a little variation with collider energy.  Comparing to wide genjets (middle), we see that for AK4 genjets both the width and the tail to low mass has increased significantly.  The narrow effective cone size $R=0.4$ of AK4 genjets results in a significant energy loss from moderate angle QCD radiation emitted by the final state quark and gluon, greatly increasing the tail to low mass and the width of the AK4 genjet distribution.  The wide genjet algorithm, with its larger effective cone size $R=1.1$, collects more final state radiation in the genjet than the AK4 genjet algorithm, and therefore results in dijet mass distributions that are significantly closer to the di-parton mass distribution and with less variation as a function of $\sqrt{s}$ 

\begin{figure}[tbp]
\centering 
\includegraphics[width=.32\textwidth]{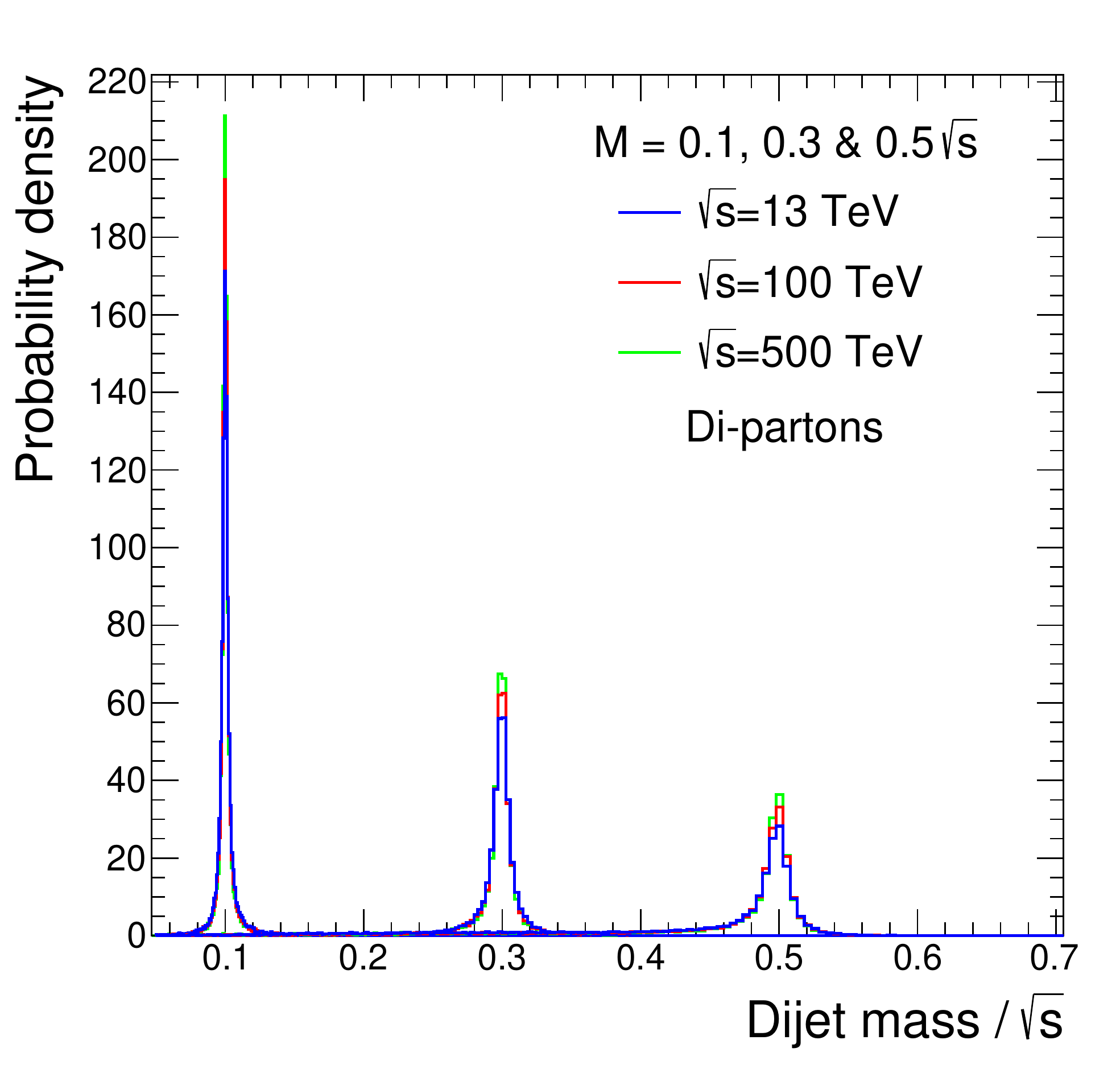}
\includegraphics[width=.32\textwidth]{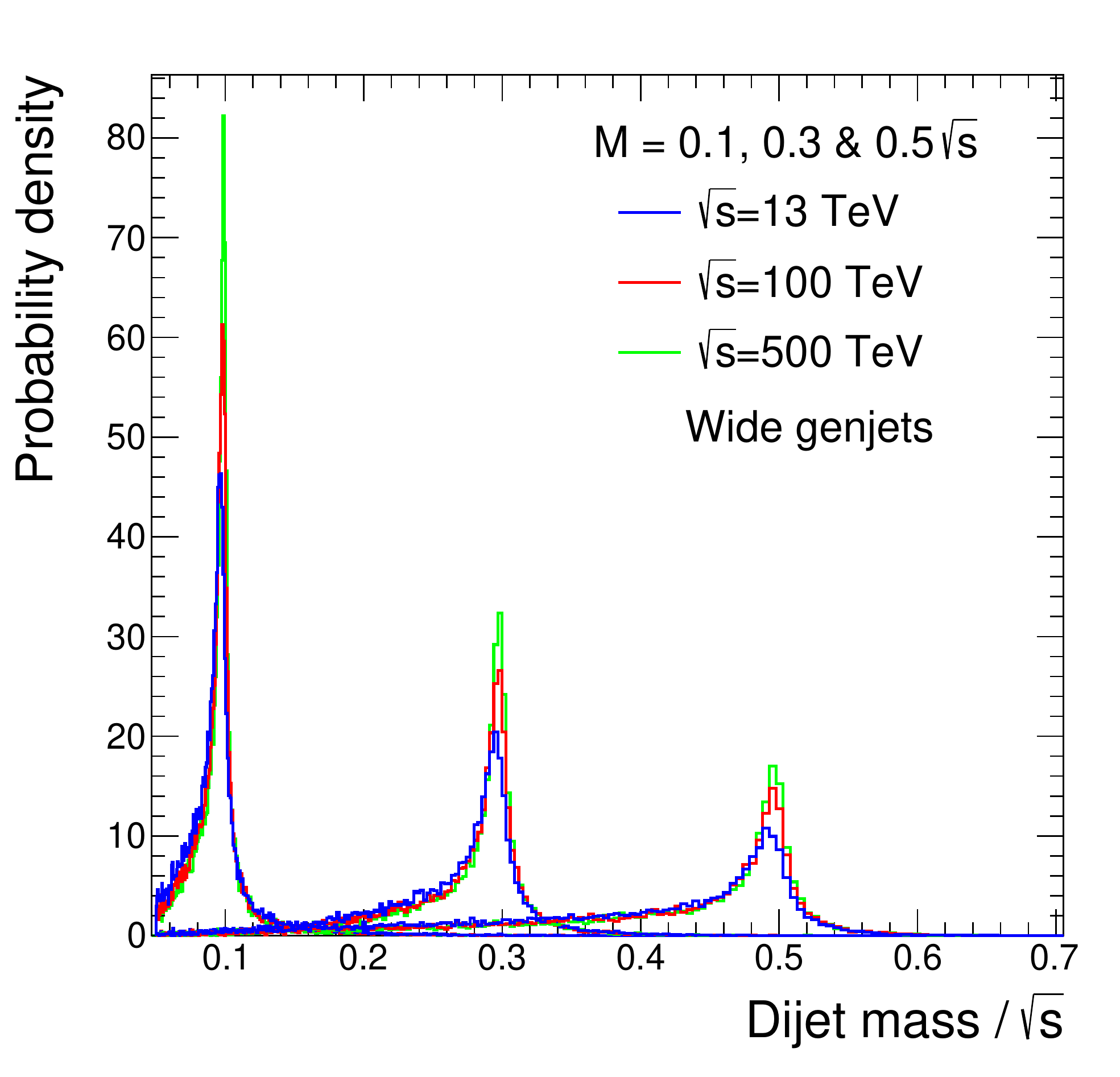}
\includegraphics[width=.32\textwidth]{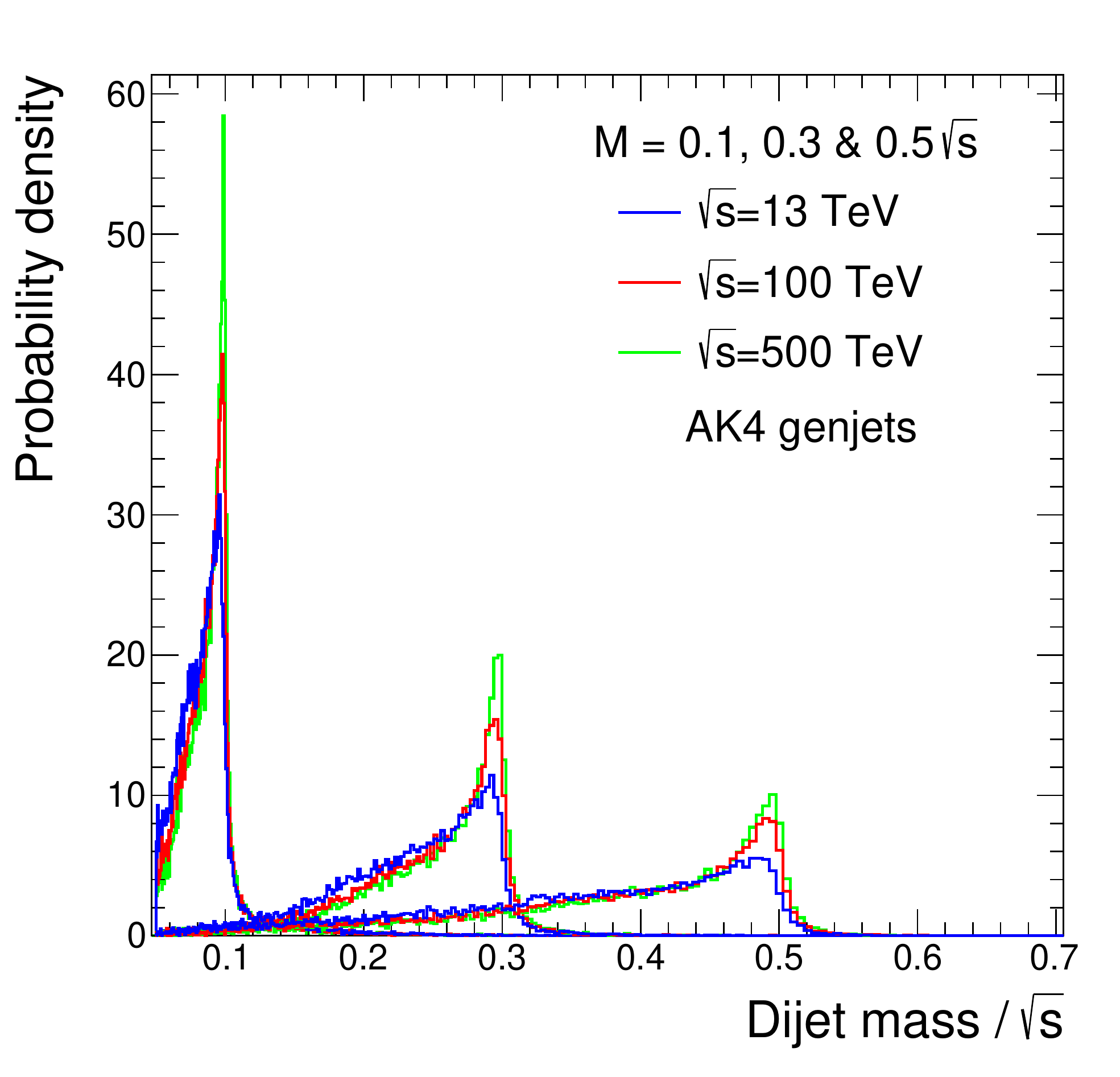}
\includegraphics[width=.32\textwidth]{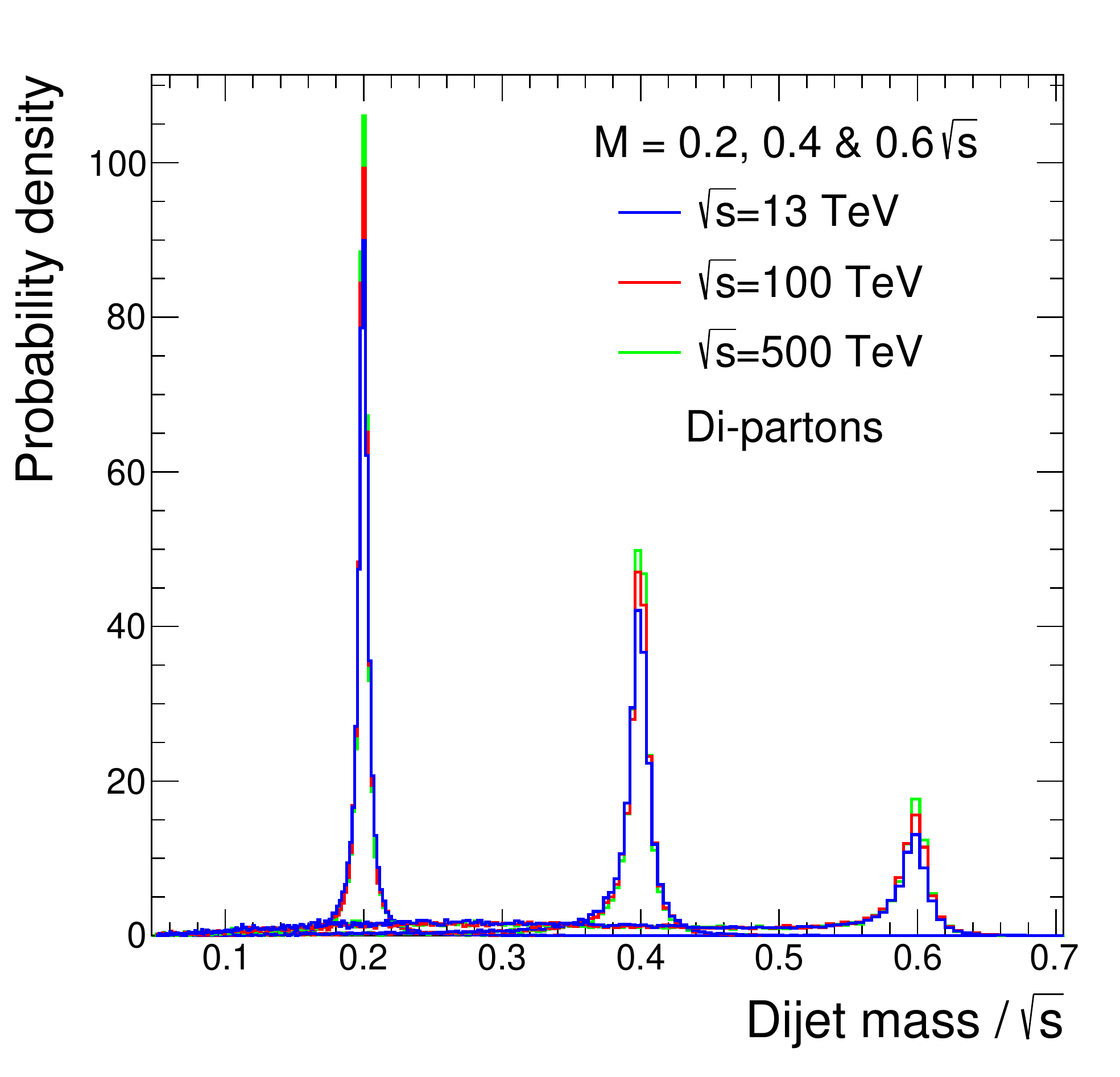}
\includegraphics[width=.32\textwidth]{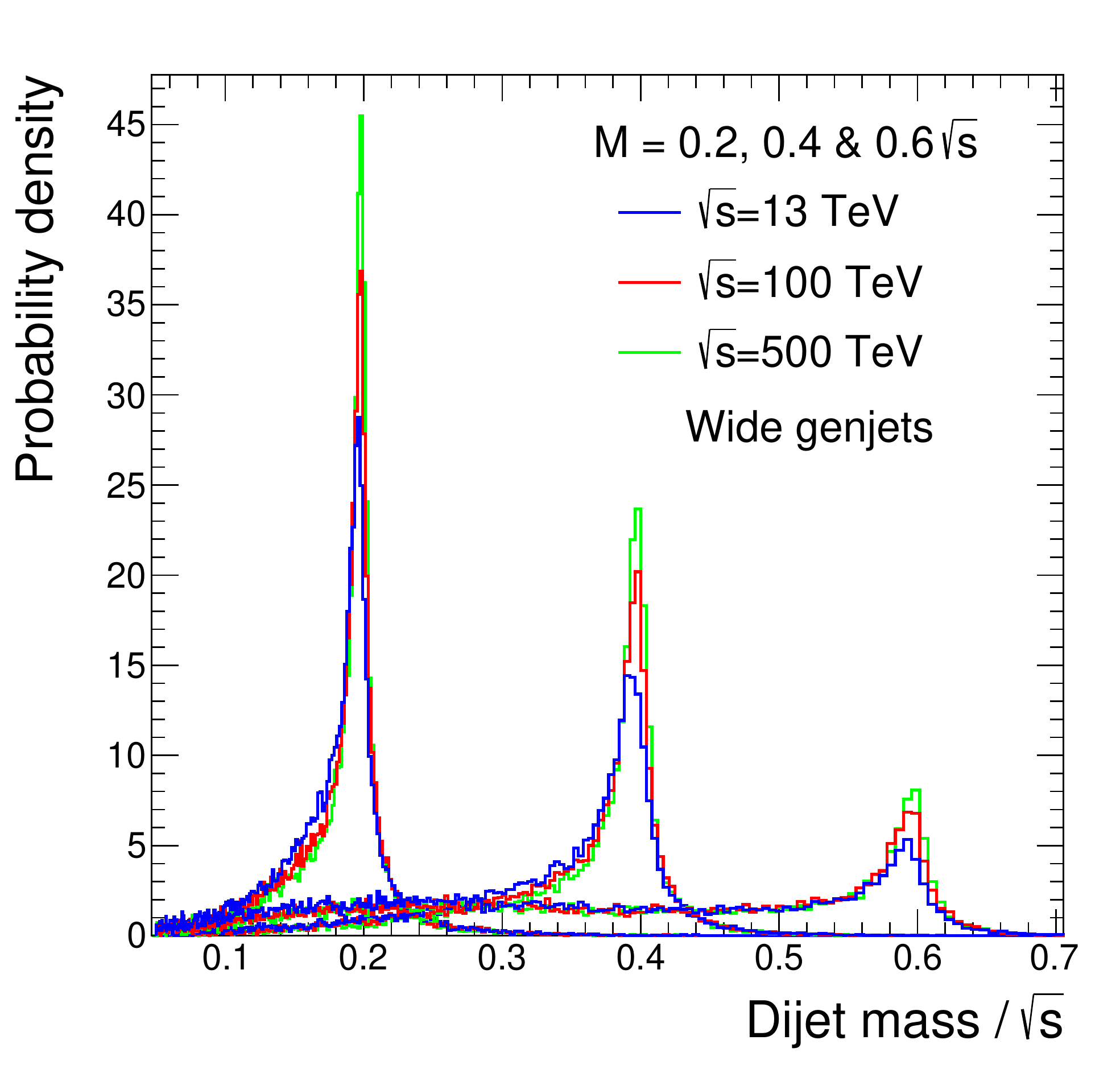}
\includegraphics[width=.32\textwidth]{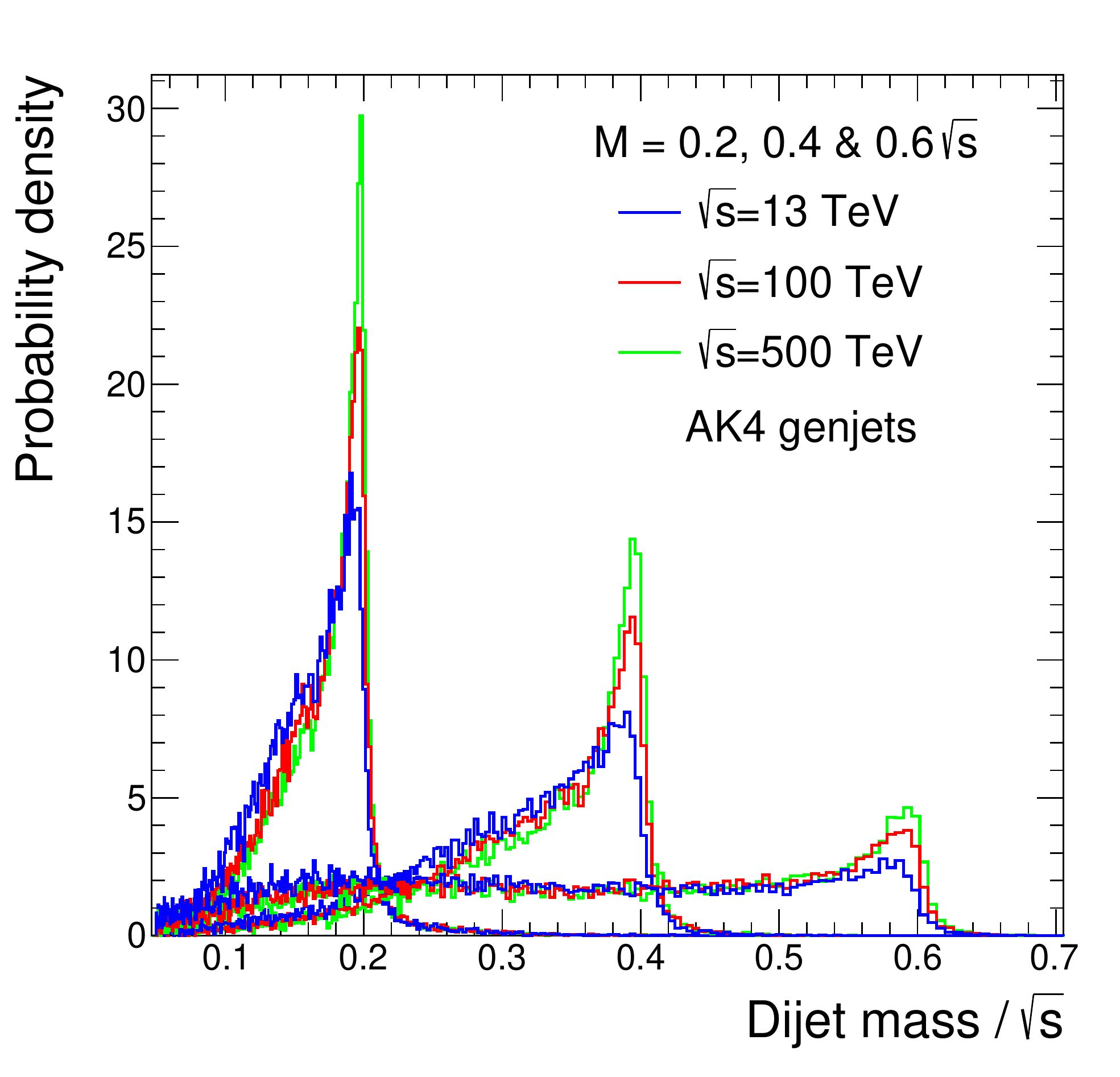}
\caption{\label{fig:GenjetThreePlots} Dijet mass distributions for excited quarks 
from $pp$ collisions at $\sqrt{s}$ equal to 13, 100, and 500 TeV, for resonance mass equal to 10\%, 30\%, and 50\% of $\sqrt{s}$ (top row) and 20\%, 40\%, and 60\% of $\sqrt{s}$ (bottom row), for the cases where the dijet mass is calculated from the Lorentz vectors of the two final-state partons (left column),  the two wide genjets (middle column), and the two leading AK4 genjets in the event (right column).}
\end{figure}

Figure~\ref{fig:GenjetThreePlots} also demonstrates the approximate insensitivity of the dijet mass distributions to the choice of $\sqrt{s}$, but does so for all six values of excited quark mass we consider, in the range $0.1\leq M/\sqrt{s} \leq 0.6$.  Once again the distributions are of the dimensionless ratio $m/\sqrt{s}$, which spaces the dijet mass distributions across the plot, because the dijet mass, $m$, peaks at the excited quark mass, $M$, for each distribution.  We can already begin to see that there is something significantly different about the distribution for $M/\sqrt{s}=0.6$, which we will explore in more detail in the next section using a more appropriate choice of dimensionless variable to study this effect.  For di-partons (left), the long tail to low mass we had noted previously for $M/\sqrt{s}=0.5$, has evolved into a longer tail with significantly larger probability for $M/\sqrt{s}=0.6$. The same is true for wide genjets (middle) and AK4 genjets (right). For the case of $\sqrt{s}=13$ TeV the $M/\sqrt{s}=0.6$ distribution for AK4 genjets is practically flat as a function of dijet mass without a significant peak all the way to the kinematic edge at $m/\sqrt{s}=0.6$. 

\subsection{Resonance mass invariance}
\label{sec:massInvariance}

\begin{figure}[tbp]
\centering 
\includegraphics[width=.32\textwidth]{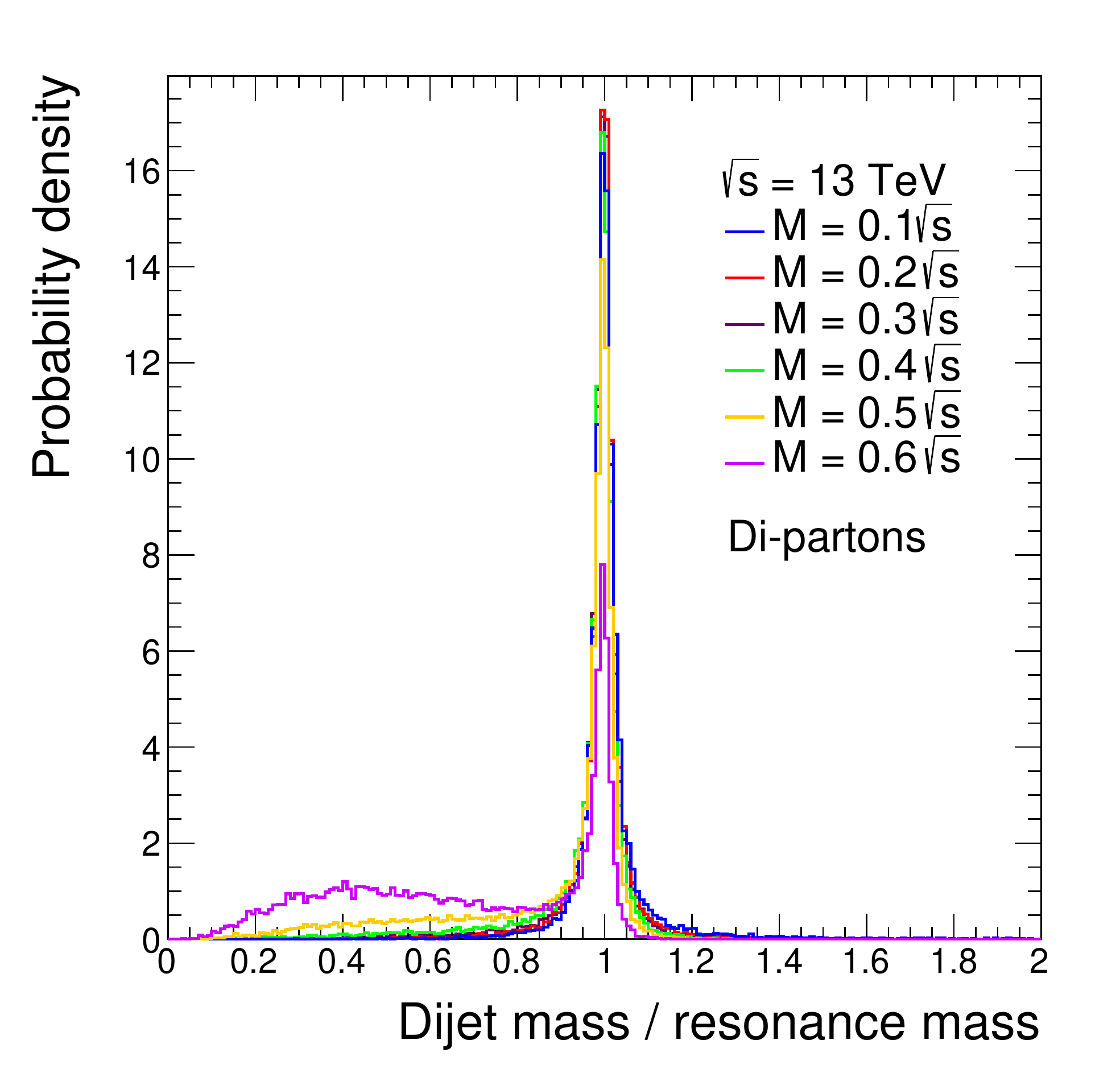}
\includegraphics[width=.32\textwidth]{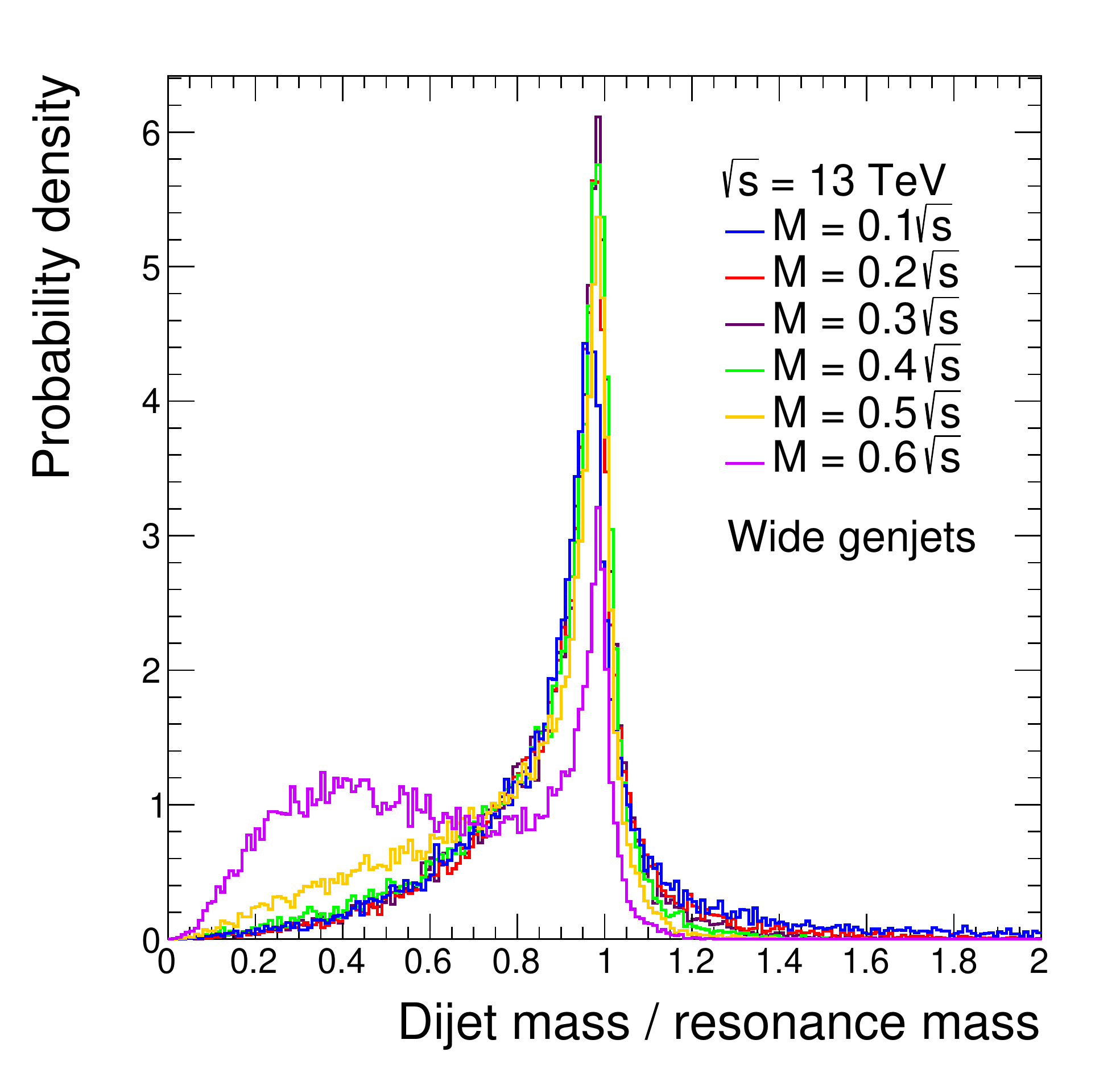}
\includegraphics[width=.32\textwidth]{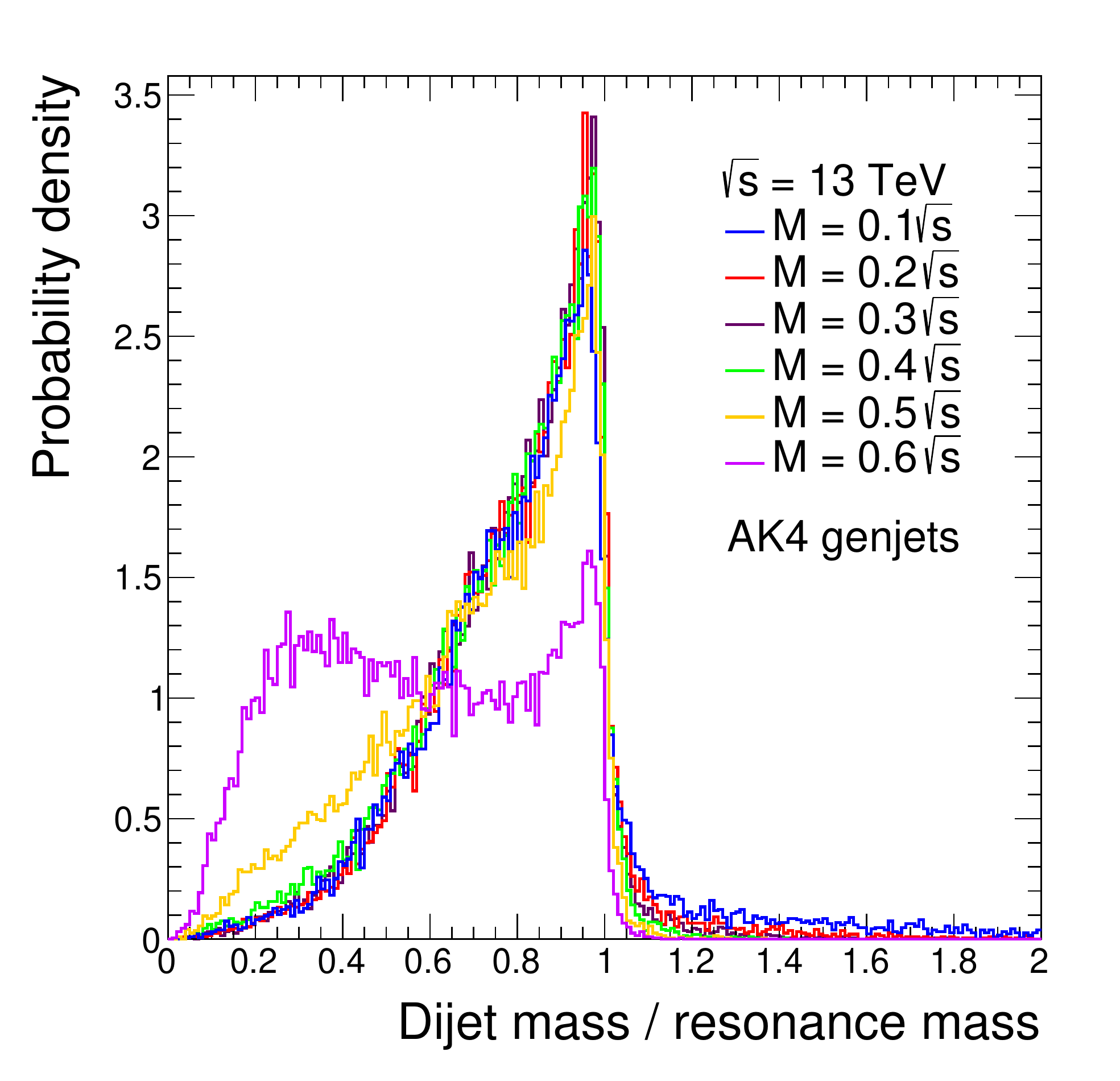}
\includegraphics[width=.32\textwidth]{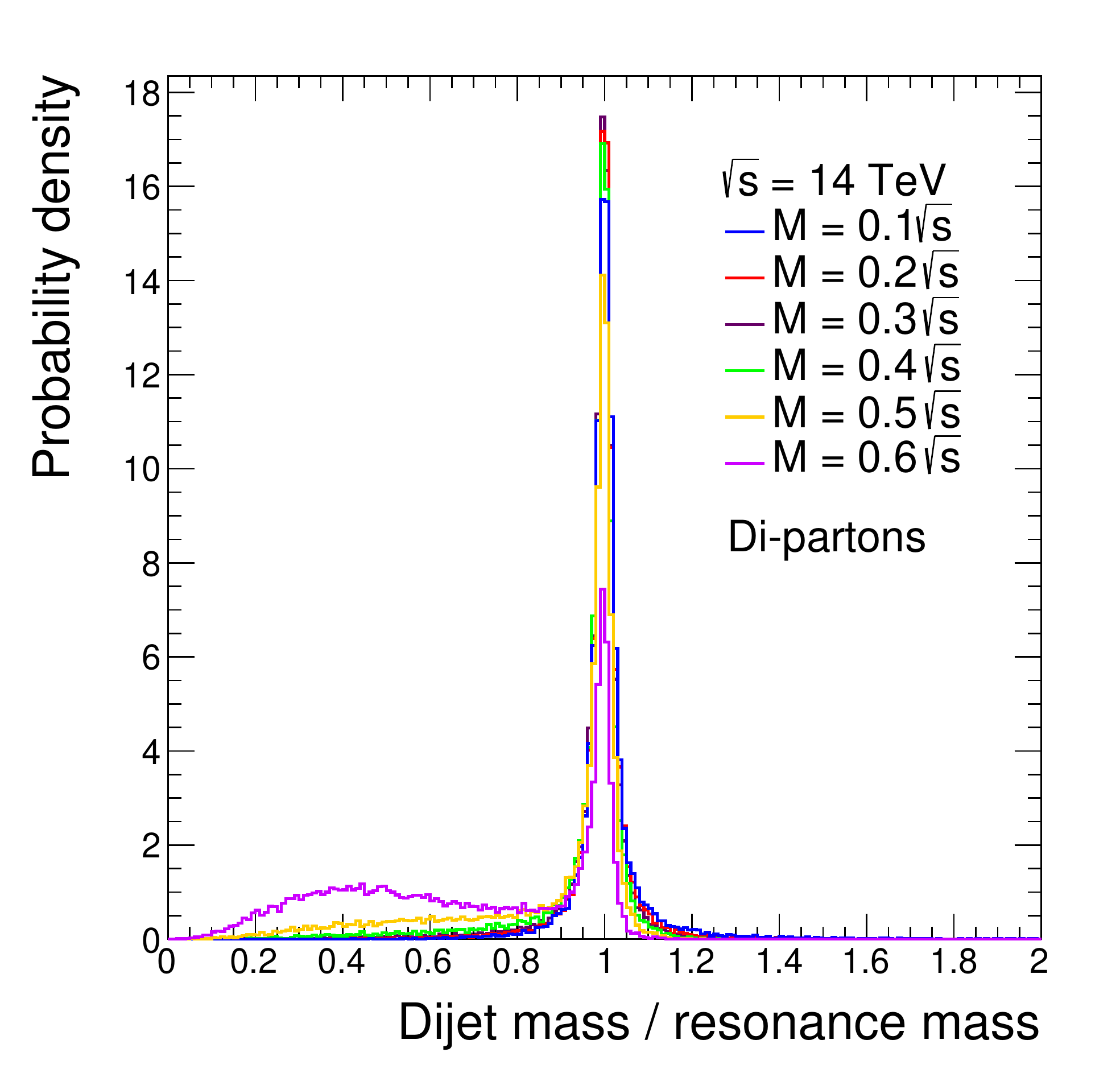}
\includegraphics[width=.32\textwidth]{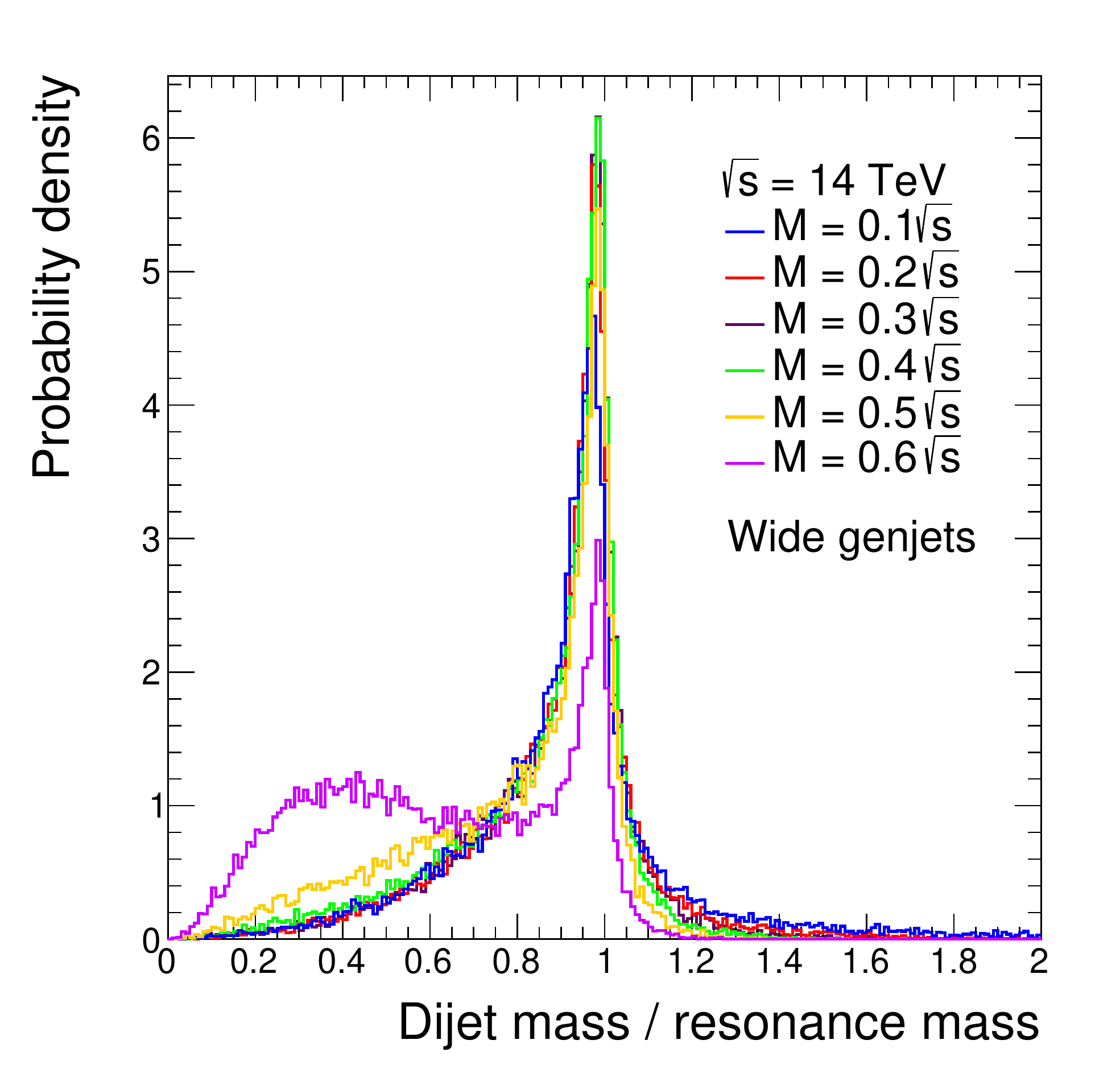}
\includegraphics[width=.32\textwidth]{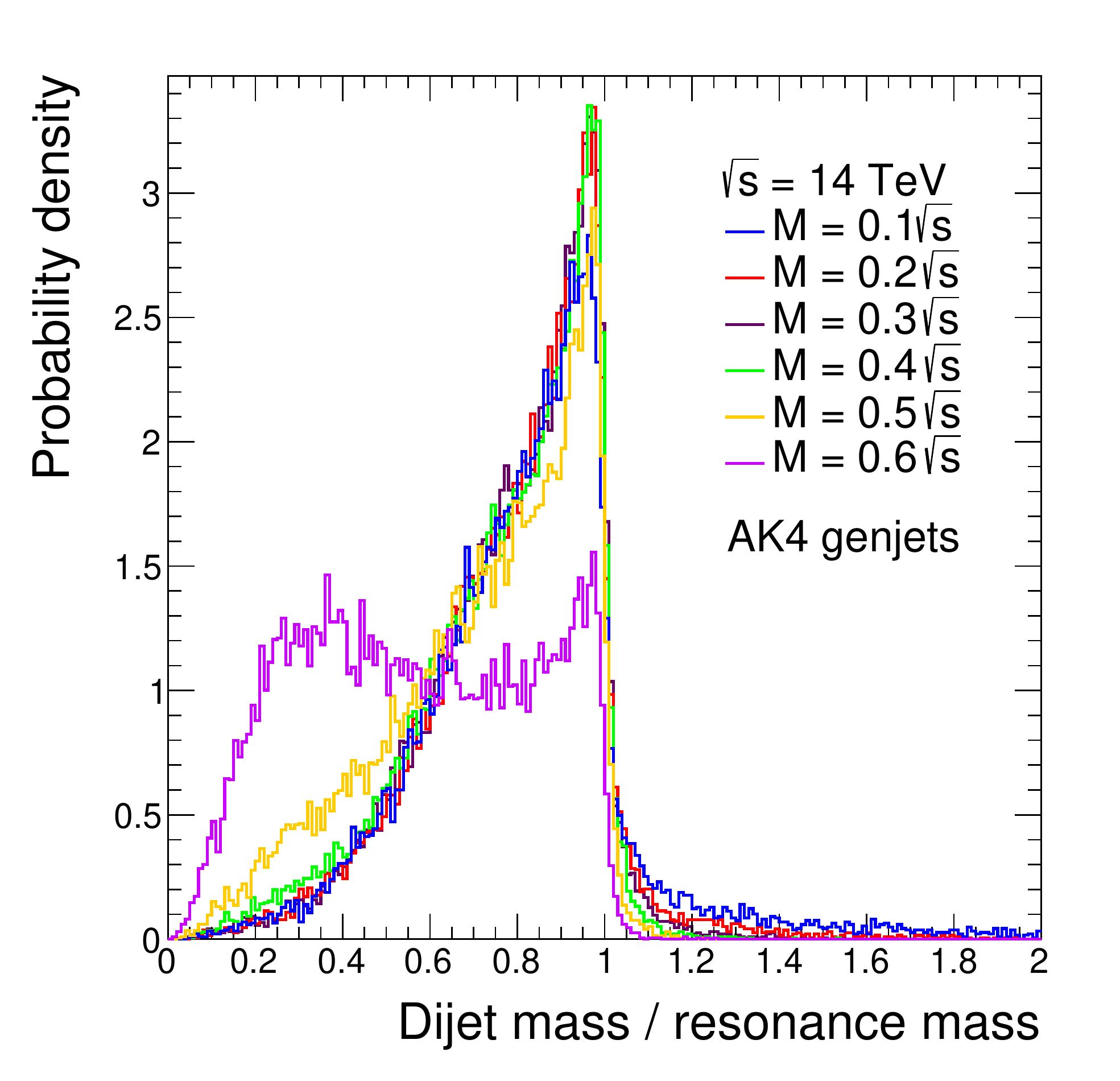}
\includegraphics[width=.32\textwidth]{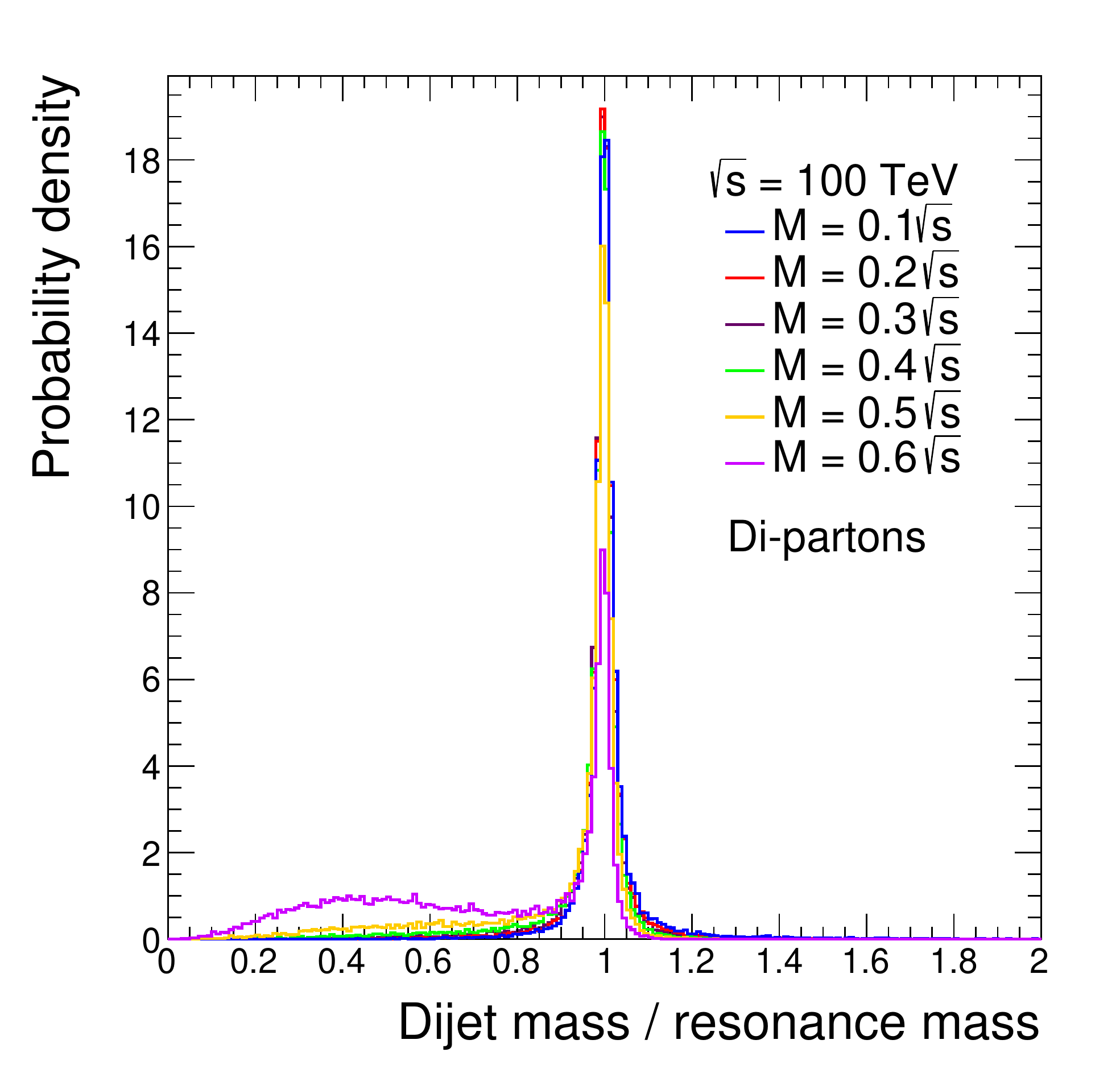}
\includegraphics[width=.32\textwidth]{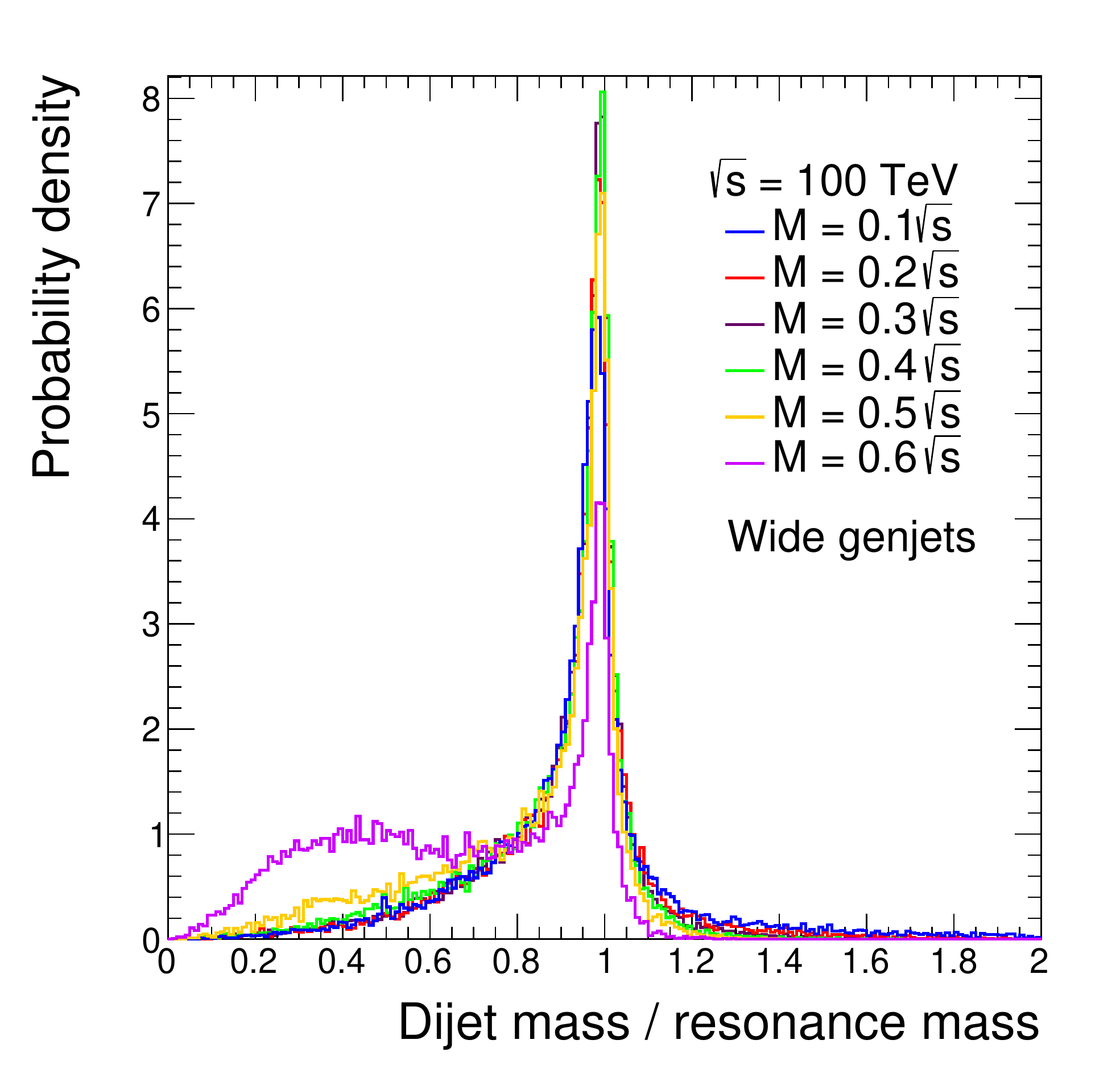}
\includegraphics[width=.32\textwidth]{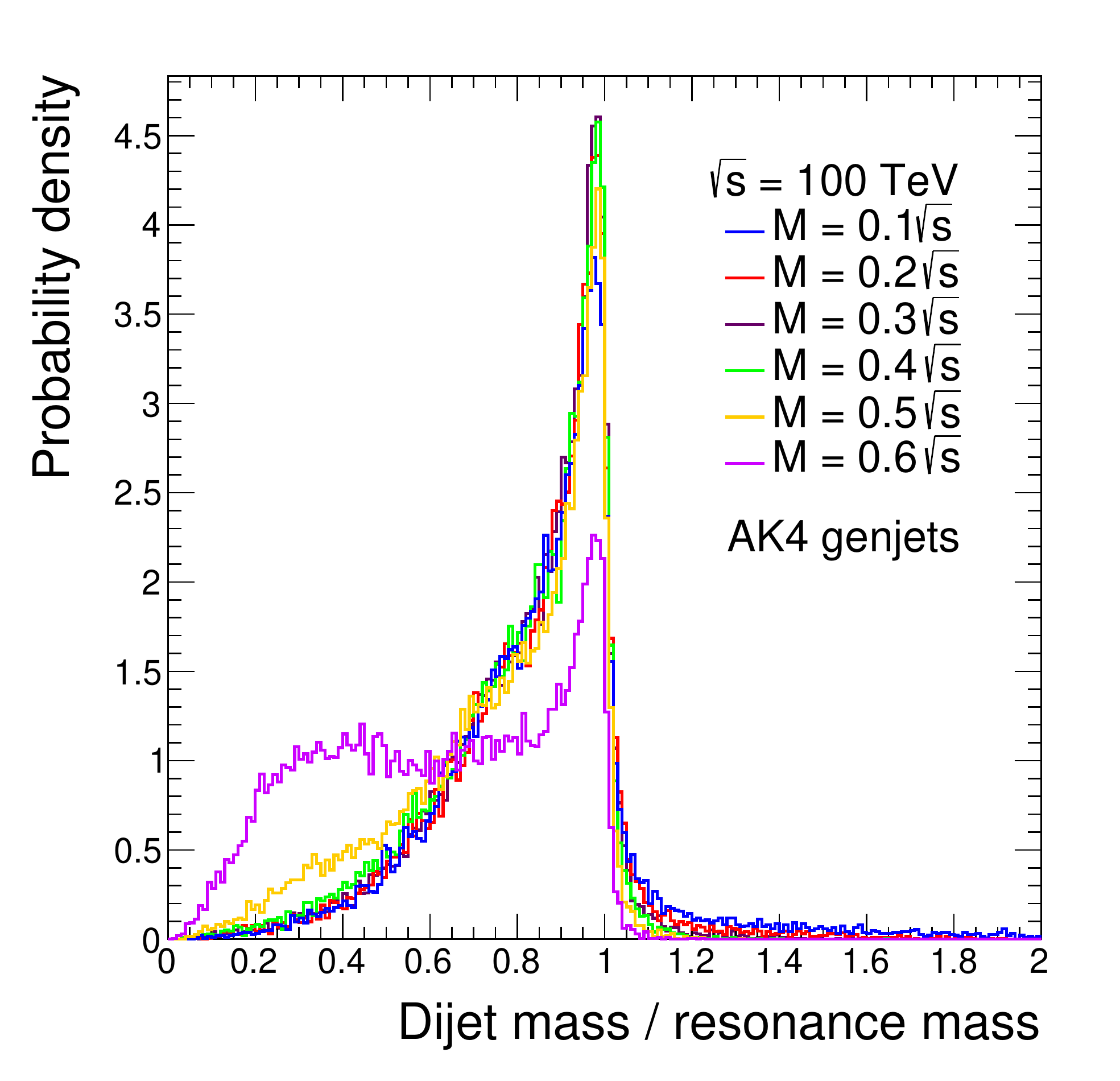}
\includegraphics[width=.32\textwidth]{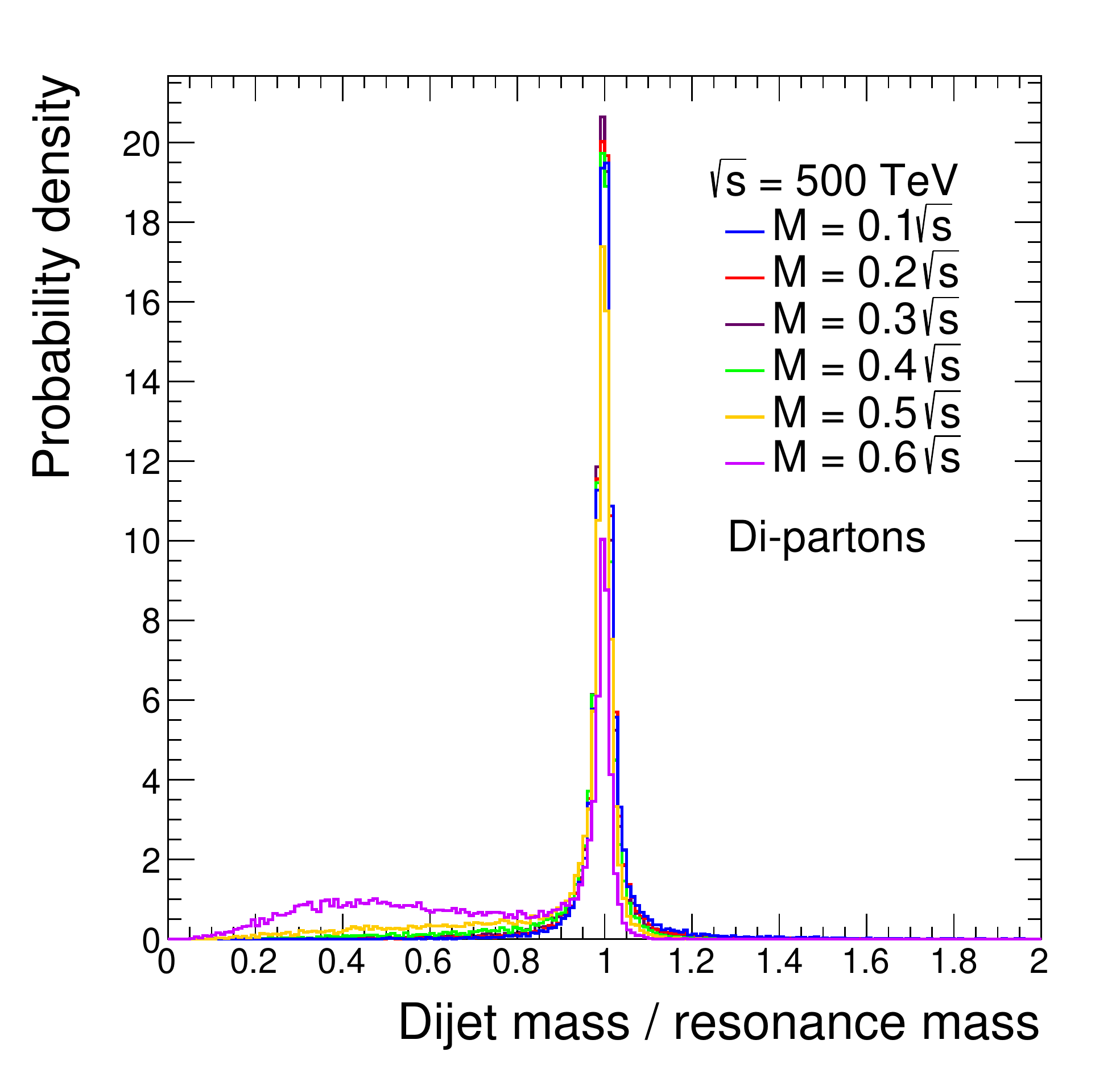}
\includegraphics[width=.32\textwidth]{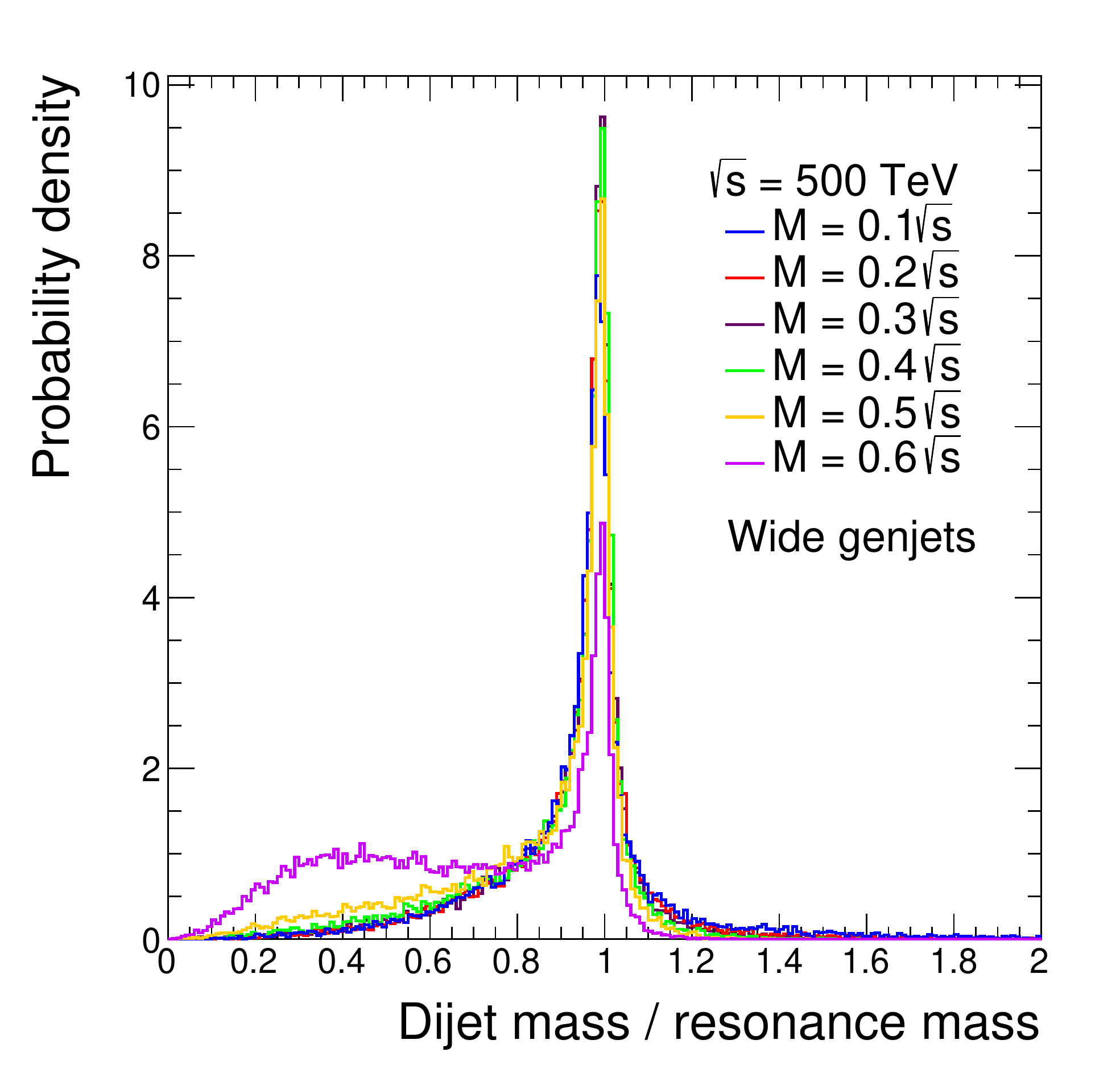}
\includegraphics[width=.32\textwidth]{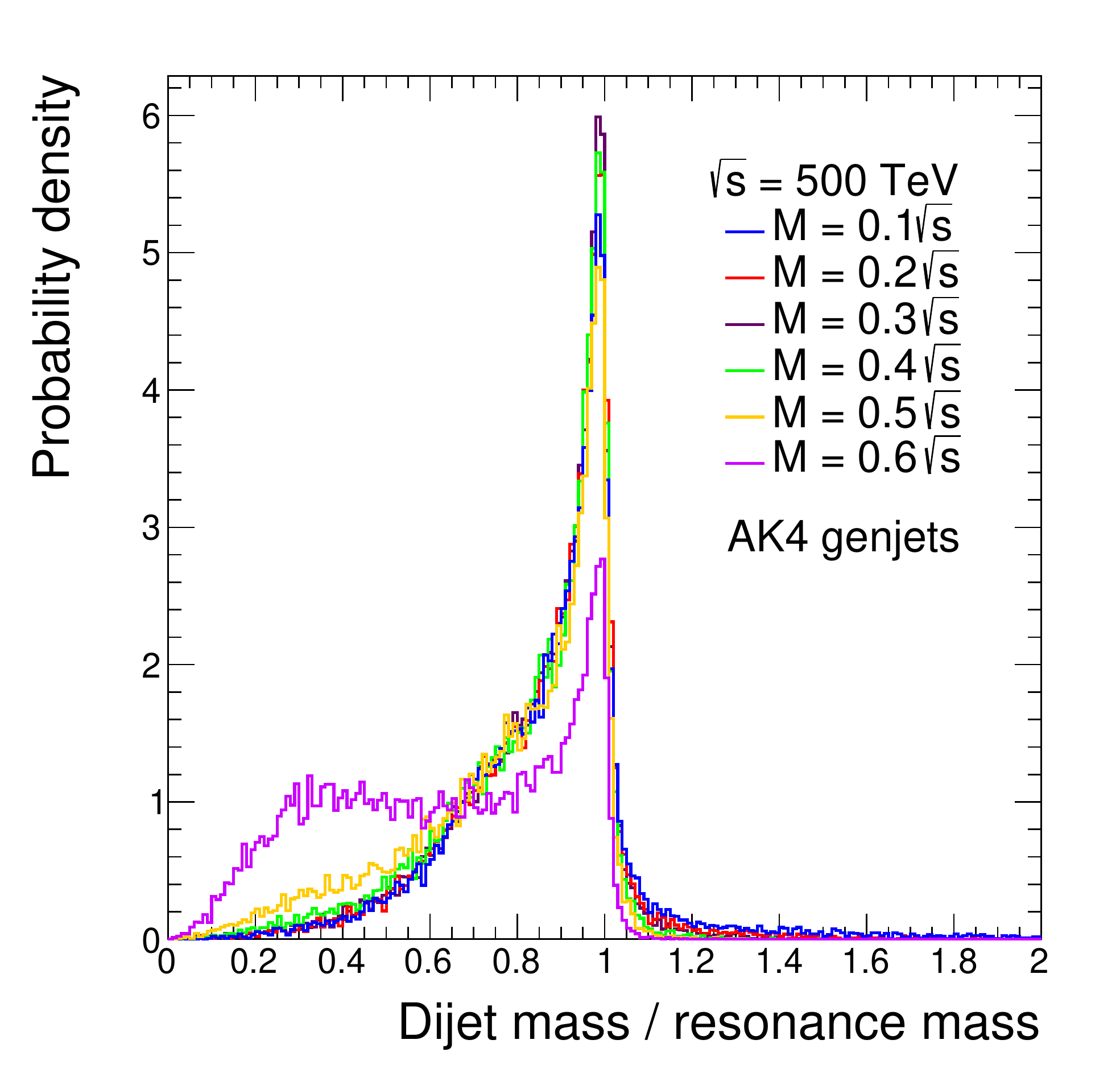}
\caption{\label{fig:GenjetAllPlots} Dijet mass distributions for excited quarks with resonance mass equal to 10\%, 20\%, 30\%, 40\%, 50\% and 60\% of $\sqrt{s}$, 
from $pp$ collisions at $\sqrt{s}$ equal to 13 TeV (top row), 14 TeV (2nd row), 100 TeV (3rd row), and 500 TeV (bottom row), for the cases where the dijet mass is calculated from the Lorentz vectors of the two final-state partons (left column), the two wide genjets (middle column), and the two leading AK4 genjets in the event (right column).}
\end{figure}

Figure~\ref{fig:GenjetAllPlots} demonstrates the approximate invariance of the dijet mass distribution to changes in resonance mass. Here we show distributions of another dimensionless ratio, dijet mass divided by resonance mass ($m/M$), which allows a more direct comparison of the resonance shapes at different values of resonance mass. For each of the three types of dijets reconstructed at generator-level (di-partons, wide genjets, and AK4 genjets), regardless of $\sqrt{s}$, the four distributions within the resonance mass range $0.1 < M/\sqrt{s} < 0.4$ are almost indistinguishable, and for practical purposes the five distributions within the slightly larger range $0.1 < M/\sqrt{s} < 0.5$ are approximately invariant. 

Here we discuss why the shape of the dijet mass distribution is different at high values of resonance mass.
The low mass tail begins to become noticeable for $M/\sqrt{s}=0.5$, and is significant for $M/\sqrt{s}=0.6$. This is consistent with it's source, dominated by the product of the resonance Breit-Wigner and the steeply falling PDFs of the proton, which decrease rapidly as a function of fractional momentum, $x$, and are much steeper at high values of $x$. To appreciate this, note that the pole mass of the excited quark resonance occurs at the dijet mass value $m=\sqrt{x_1 x_2s}=M$.  In Eq.~\ref{eq:BW}, the PDFs are steeply falling with increasing $x$, so at the small values of $x$, corresponding to the low dijet mass tail of the resonance, the PDFs are orders of magnitude larger than at the relatively high value $x\approx 0.5$. This multiplies the Breit-Wigner tail at low dijet mass, and begins to compensate the fall in the Breit-Wigner at low mass, creating an observable low mass tail for $M/\sqrt{s}=0.5$. That steepness of the PDFs increases at higher resonance mass, so that the ratio of PDF values between the tail and the peak in dijet mass is roughly an additional order of magnitude larger at $x\approx 0.6$ than they are at $x\approx 0.5$, making the low mass tail practically a second peak at low mass, which is the dominant feature of the excited quark resonance shapes for $M/\sqrt{s}=0.6$. 

The other source of the low mass tail is final-state radiation, which dominates the tail for $0.1 < M/\sqrt{s} < 0.4$, and roughly makes the same amount of low mass tail independent of resonance mass. However, for $M/\sqrt{s}=0.6$, the effect of PDFs discussed above is the dominant source of the low mass tail in Fig.~\ref{fig:GenjetAllPlots}, and final-state radiation does not contribute significantly in comparison. To see this, note that the probability density values on the tail at very low dijet mass for di-partons, wide genjets, and AK4 genjets are approximately the same. This can also be seen even clearer when we compare di-partons and wide genjets directly in Fig.~\ref{fig:AllSimLevels}. It is only the height and width of the resonance peak that are different for the three generator-level types of dijets.  The dijet mass distribution of di-partons is not affected by final state radiation, because the mass is reconstructed from partons before the radiation is emitted. The wide genjet distributions are moderately affected by final state radiation, and the AK4 genjet distributions are significantly affected by final state radiation. The observation that the probability density values on the tail at very low mass for both types of genjets are approximately the same as for di-partons, for excited quarks with $M/\sqrt{s}=0.6$, demonstrates that at this value of resonance mass the tail at very low dijet mass is not significantly affected by radiation, it is dominated by PDF effects.  The excited quark mass value $M/\sqrt{s}=0.5$ is roughly a transition point, where the tail at very low mass has significant contributions from both PDFs and final-state radiation, and neither is dominant.

\section{Experimental resolution}
\label{sec:resolution}

The establishment at the generator-level of the approximate invariance of the dijet mass distribution to changes in $\sqrt{s}$ and $M$, for the range $0.1 < M/\sqrt{s} < 0.5$, accomplishes most of what is needed to support our Snowmass 2021 study~\cite{Harris:2022kls},
which assumed this invariance. Nevertheless, we include estimates of experimental resolution, both to estimate the expected experimental distributions at future $pp$ colliders, and to derive additional understanding of the previously published distributions from CMS~\cite{CMS:2018xlo}. In short, we will extract the effective CMS experimental resolution for dijet resonances from excited quarks, generalize that experimental resolution for all $pp$ colliders, and smear our wide genjet distributions with that resolution to obtain experimental distributions at all $pp$ colliders. This procedure has the advantage that it is simple, quick, and easy to understand. It scales from an existing experiment to obtain a plausible resolution that is sufficient for our study of future pp colliders.  

\subsection{Extraction of Gaussian experimental resolution for CMS}

\begin{figure}[tbp]
\centering 
\includegraphics[width=.32\textwidth]{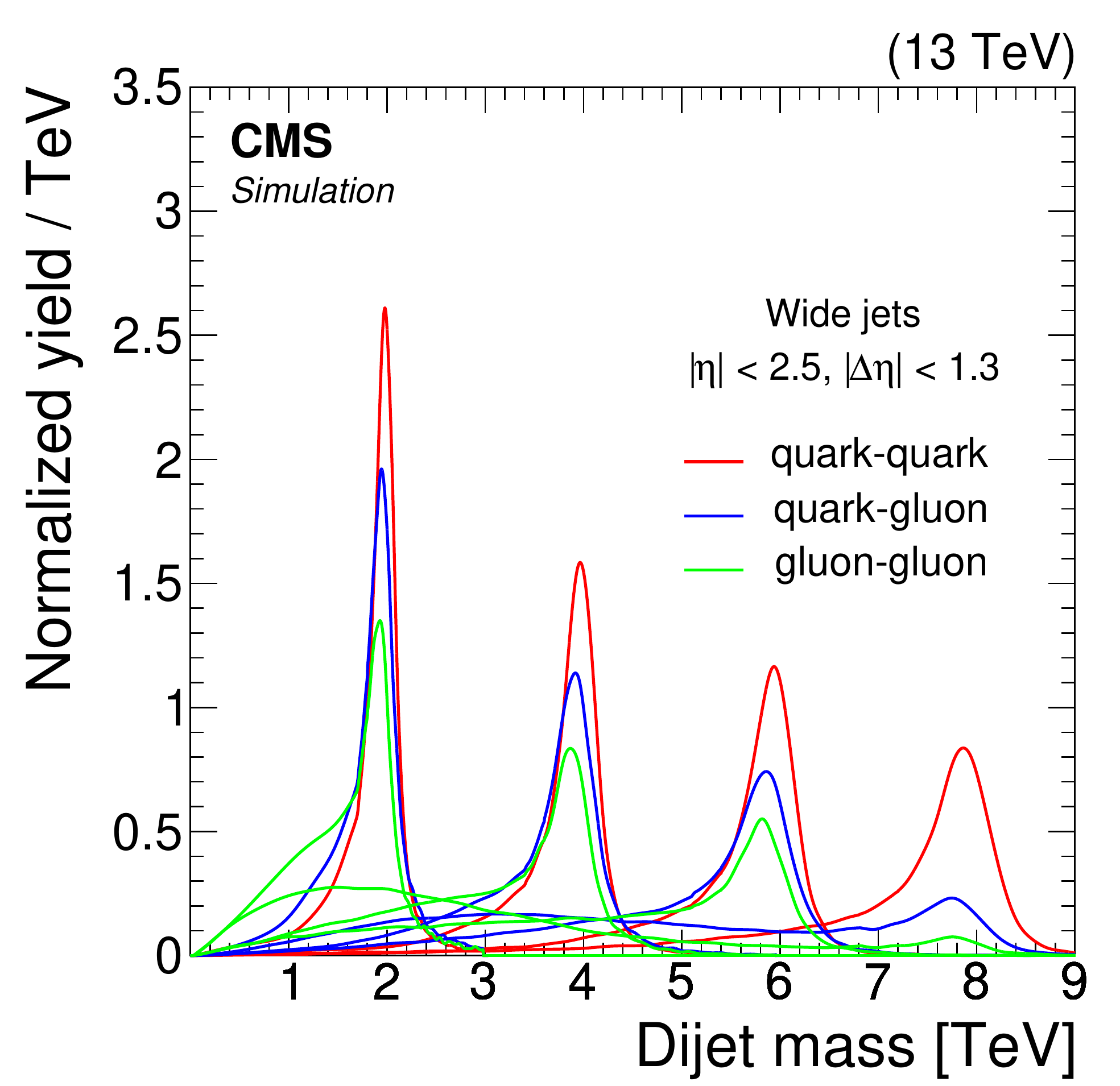}
\includegraphics[width=.32\textwidth]{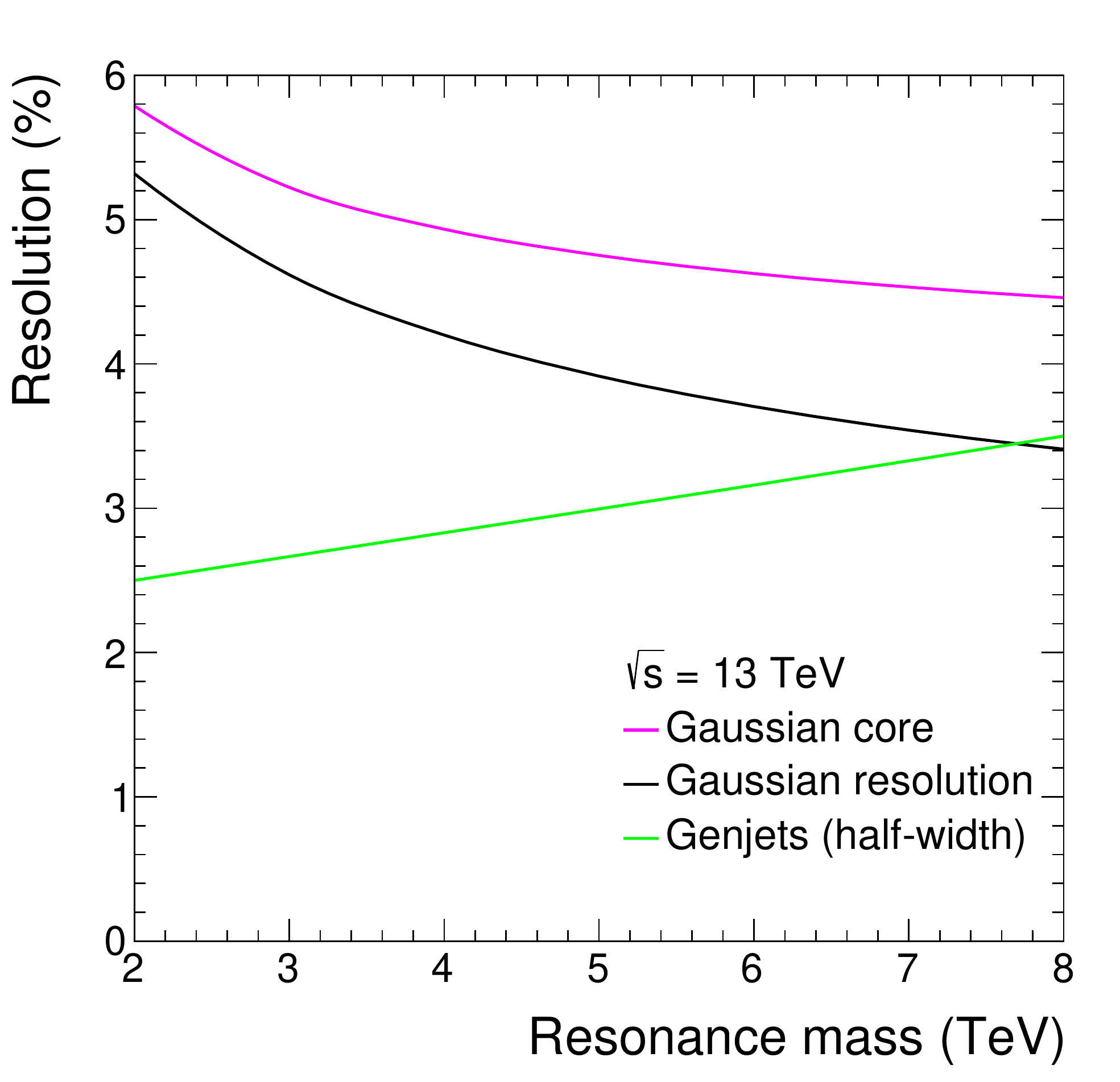}
\includegraphics[width=.32\textwidth]{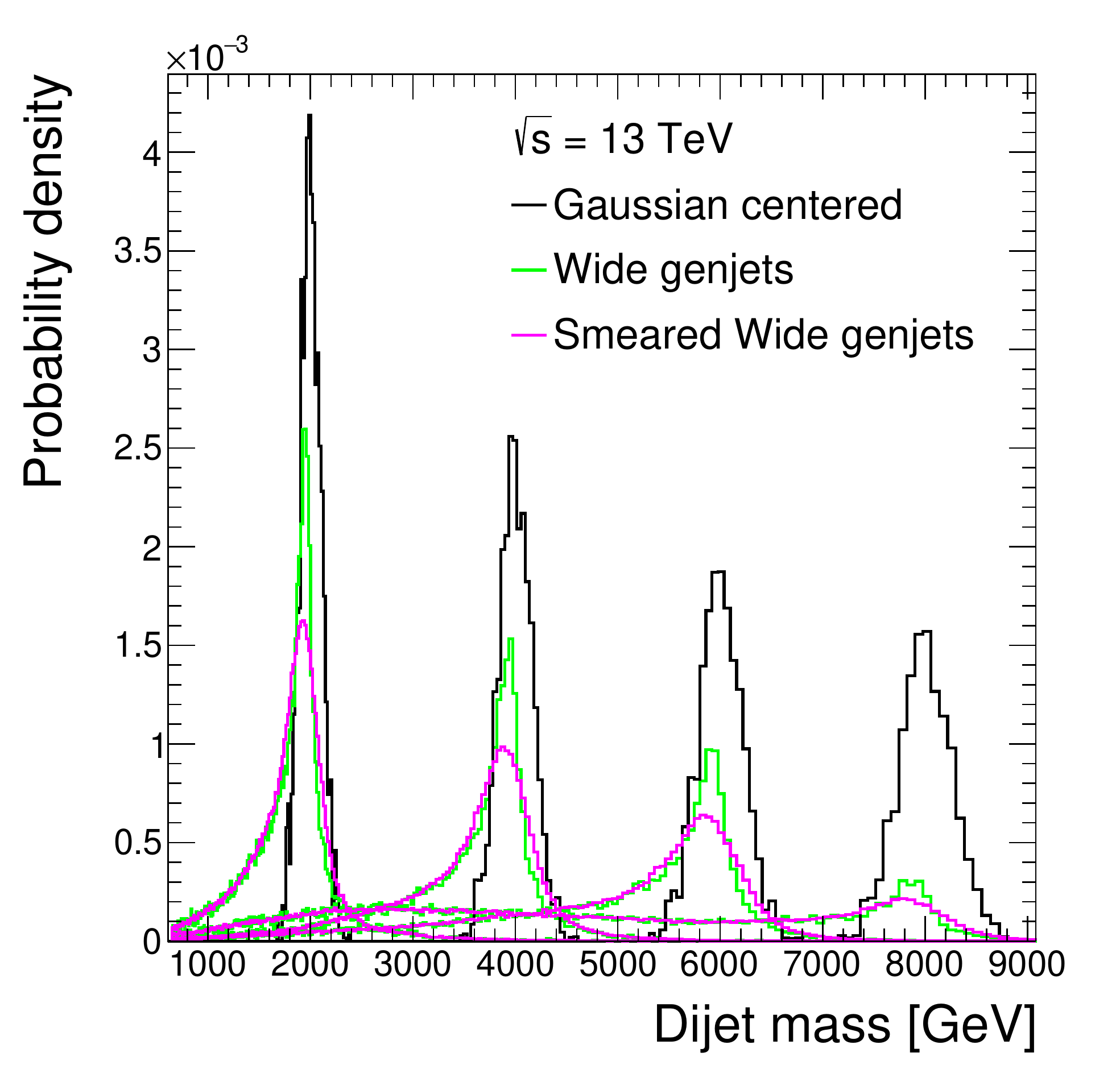}
\caption{\label{fig:resolution} Extraction and re-application of experimental resolution. Left) Reproduced from Ref.~\cite{CMS:2018xlo} are the dijet mass distributions of fully simulated dijet resonances in the CMS experiment, where excited quarks were used to model quark-gluon resonances (blue). Middle) the resolution of the Gaussian core of the CMS fully simulated dijet resonances (violet), the half-width of the wide genjet mass distributions (green), and the approximate Gaussian detector resolution resulting from the deconvolution of these two resolutions (black). Right) the Gaussian distribution as a function of dijet mass from detector resolution alone (black), the excited quark dijet mass distribution for wide genjets (green), and the convolution of this  generator-level distribution with Gaussian detector resolution (violet).  }
\end{figure}

Full simulations of dijet resonances in $pp$ collisions at $\sqrt{s}=13$ TeV are shown in Fig.~\ref{fig:resolution}(left) for excited quarks with resonance mass 2, 4, 6 and 8 TeV. The reconstruction used is the wide jet algorithm, equivalent to our wide genjet algorithm, but using reconstructed experiment-level AK4 jets as input, instead of generator-level AK4 genjets. The CMS publication~\cite{CMS:2018xlo} discussed these dijet mass distributions as having ``\emph{... Gaussian cores from jet energy resolution, and tails towards lower mass values primarily from QCD radiation}'', and presented values for the dijet mass resolution within the Gaussian core for quark-quark and gluon-gluon resonances. These came from fitting the distributions with truncated Gaussians, to estimate the core resolution.  We find that the corresponding values of the RMS deviation for the Gaussian core of quark-gluon resonances in Fig.~\ref{fig:resolution}(left) is well represented by the function
\begin{equation}
\sigma_{\mbox{core}}(\%) = 3.6\% + \frac{2.0\%}{\sqrt{M\mbox{[TeV]}}-0.5}.
\label{eq:GaussCore}
\end{equation}
This function, the pink curve labeled "Gaussian core" in Fig.~\ref{fig:resolution}(middle), gives RMS values for the Gaussian cores of quark-gluon resonances that are the average of the RMS values presented for quark-quark and gluon-gluon resonances in Ref.~\cite{CMS:2018xlo}. We assume this Gaussian core RMS deviation is a convolution of two resolutions: the resolution of the underlying wide genjets convolved with the true Gaussian resolution of the CMS detector for experimentally observing wide jets. The measured half width at half maximum of the wide genjets dijet mass distribution, labeled "Genjets (half width)" in Fig.~\ref{fig:resolution}(middle), increases from 2.5\% to 3.5\% as $M$ increases from 2 to 8 TeV.  Deconvolving the Gaussian core RMS, by subtracting in quadrature the genjet RMS, where genjet $\mbox{RMS}\approx\mbox{(half width)}/1.18$, gives the curve labeled "Gaussian resolution" in Fig.~\ref{fig:resolution}(middle). This estimate of the Gaussian experimental resolution of the CMS experiment for wide jets, in Run 2 at $\sqrt{s}=13$ TeV, is well modeled by the function
\begin{equation}
\sigma_{\mbox{gauss}}(\%) = 1.5\% + \frac{5.4\%}{\sqrt{M\mbox{[TeV]}}}.
\label{eq:GaussRes}
\end{equation}

\subsection{Application of resolution and check of experimental distribution for CMS}
\label{sec:application}

\begin{figure}[tbp]
\centering 
\includegraphics[width=0.5\textwidth]{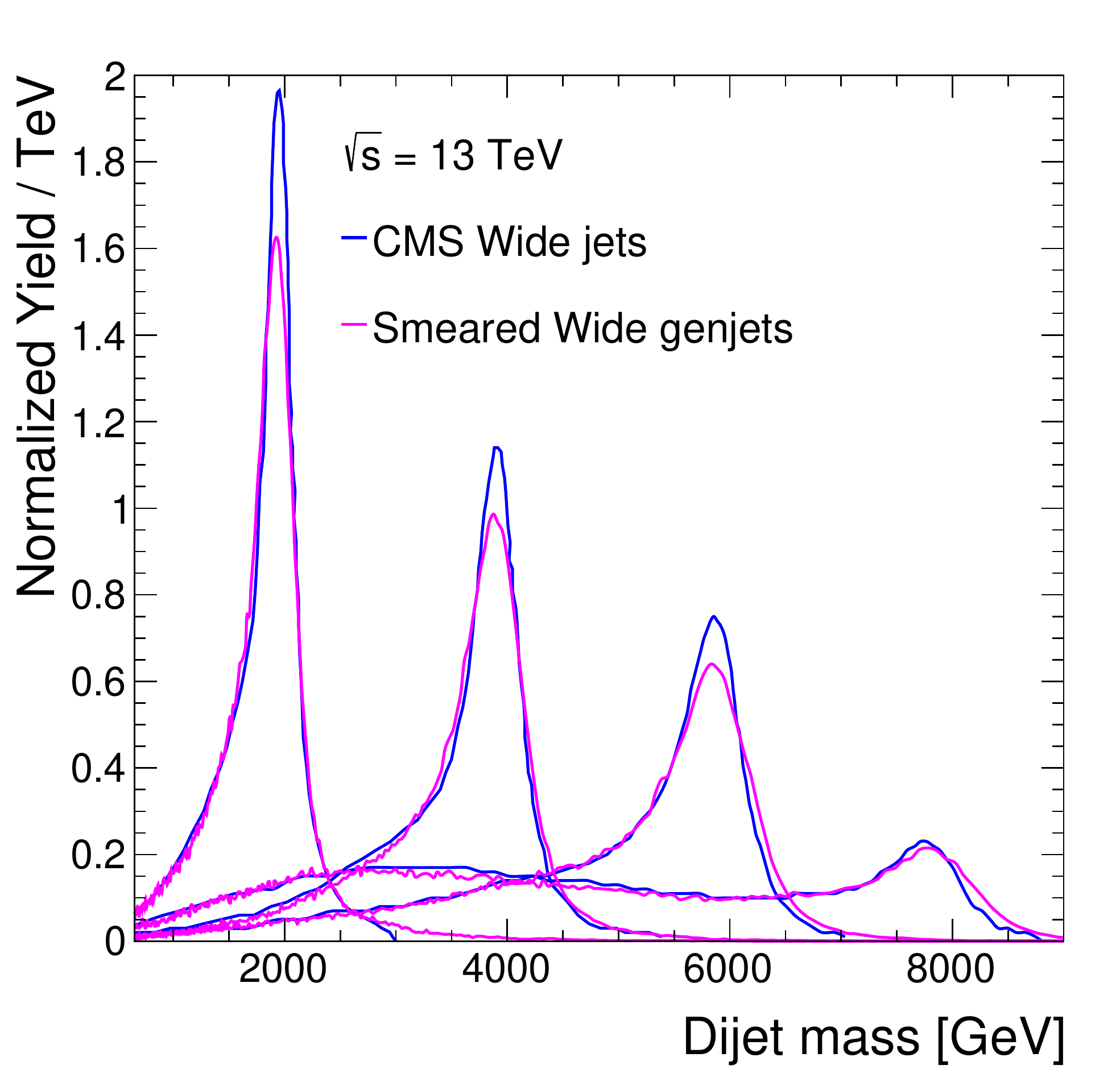}
\caption{\label{fig:check} Comparison of the fully simulated dijet resonance shapes of an excited quark in the CMS experiment from Ref.~\cite{CMS:2018xlo} (blue) and predictions of the dijet resonance shapes using wide genjets smeared with an estimate of the Gaussian resolution of the CMS experiment (violet). }
\end{figure}

In Fig.~\ref{fig:resolution}(right) we show Gaussian distributions (black histograms), with RMS equal to the experimental resolution in Eq.~\ref{eq:GaussRes}, at the four different values of excited quark mass that were simulated in the CMS publication~\cite{CMS:2018xlo}. Also shown are the dijet mass distributions of wide genjets (green histograms) at the same values of excited quark mass, which we convolve with the Gaussian resolution to obtain smeared wide genjets (pink histograms).  This re-application of the Gaussian experimental resolution to our generator-level distributions, produces experiment-level dijet mass distributions, the smeared wide genjets in Fig.~\ref{fig:resolution} (right). We check how well our procedure works in Fig.~\ref{fig:check}, where we compare the smeared wide genjets (pink curves) with the fully-simulated CMS wide jet dijet mass distributions (blue curves).

The level of agreement between our smeared wide genjets and the fully-simulated CMS wide jets in Fig.~\ref{fig:check} is adequate for our purpose in this paper, namely to produce excited quark dijet mass distributions containing approximate experimental resolution. Nevertheless, we can learn a little more from the details of the comparison. The width of the peaks are only slightly overestimated by this procedure. The small difference may be due to biases in directly fitting the fully simulated distribution using truncated Gaussians, to try and obtain the experimental Gaussian core resolution, since final-state radiation will slightly widen the core of the distribution in addition to producing the low mass tail. However, the long tails to low mass of the two distributions are in good agreement. This is likely because the low mass tail is produced almost completely by final-state radiation and PDFs at the generator-level, as discussed in section~\ref{sec:genjets}, and is not affected by the Gaussian experimental resolution. This is especially true for the excited quark mass value 8 TeV, where the overall agreement between the two distributions is very good, because the main feature of this shape is the long tail to low dijet mass. Finally, we note a very small difference in the 2 TeV shapes. The published shape from CMS is truncated at a dijet mass of 3 TeV, while the full tail to high mass is shown for smeared wide genjets.

\subsection{Experimental resolution at pp Colliders}
\label{sec:experimentalRes}

The estimated experimental resolution at $\sqrt{s}=13$ TeV in Eq.~\ref{eq:GaussRes} can be scaled to any pp collision energy by replacing $M\mbox{[TeV]}$ with $M(13/\sqrt{s})$, and rewriting in terms of the dimensionless variable $M/\sqrt{s}$:
\begin{equation}
\sigma_{\mbox{gauss}}(\%) = 1.5\% + \frac{1.5\%}{\sqrt{M/\sqrt{s}}}.
\label{eq:ExpRes}
\end{equation}
Equation~\ref{eq:ExpRes} is a generalized experimental resolution for the Gaussian core of dijet resonances reconstructed with wide jets from excited quarks at pp colliders. This resolution only depends on $M/\sqrt{s}$, so we are assuming that the experimental resolution is constant for a fixed value of $M/\sqrt{s}$. This is the same as assuming that future detectors will be designed to give the same experimental resolution in percent as current detectors, at the values of dijet mass appropriate for high mass search and discovery, which is likely a conservative assumption. We apply this resolution to the generator-level distributions (wide genjets) to obtain experiment-level distributions (smeared wide genjets).

\subsection{From generator-level to experiment-level resonances}
\label{sec:allLevels}

\begin{figure}[tbp]
\centering 
\includegraphics[width=.32\textwidth]{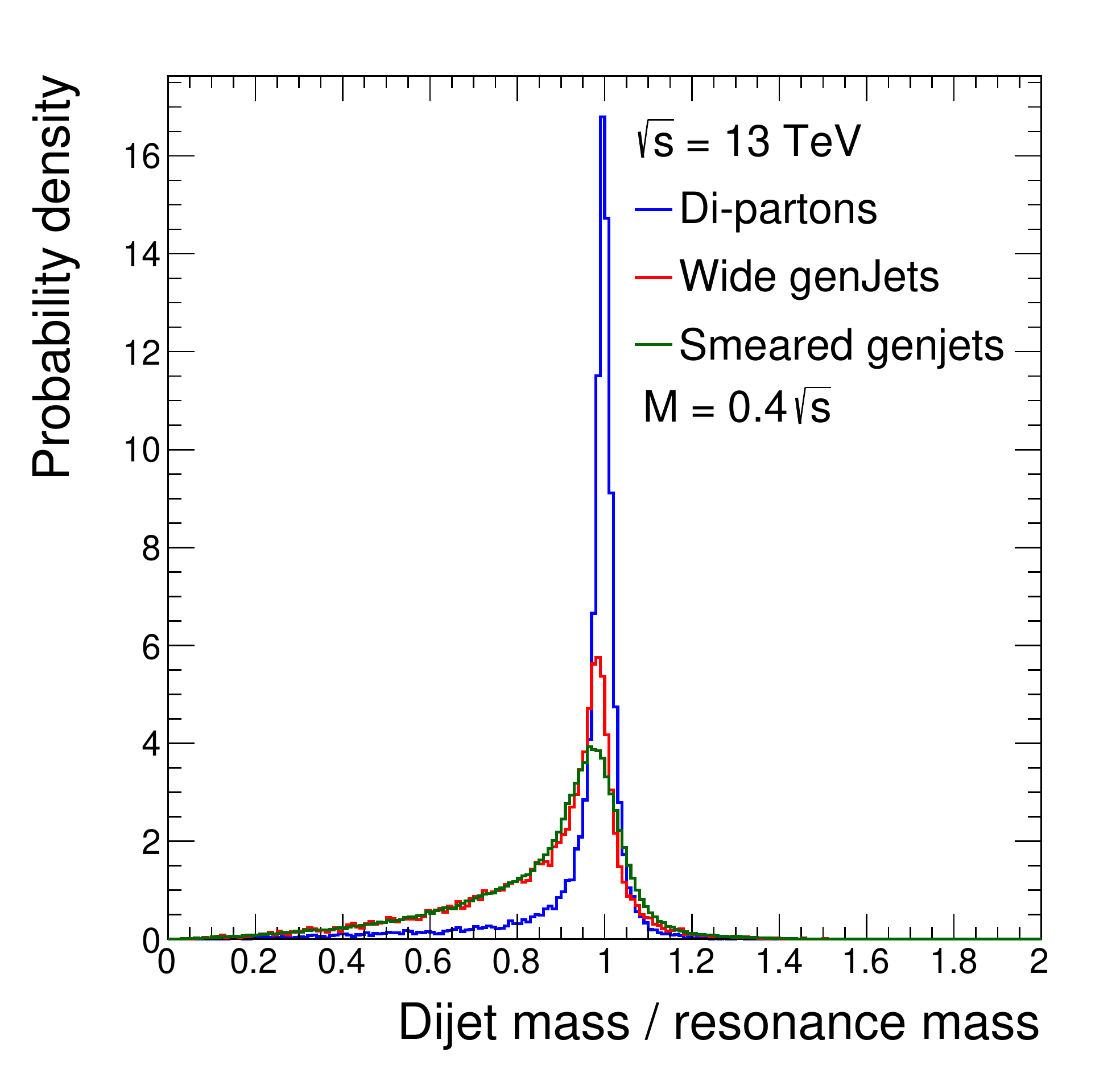}
\includegraphics[width=.32\textwidth]{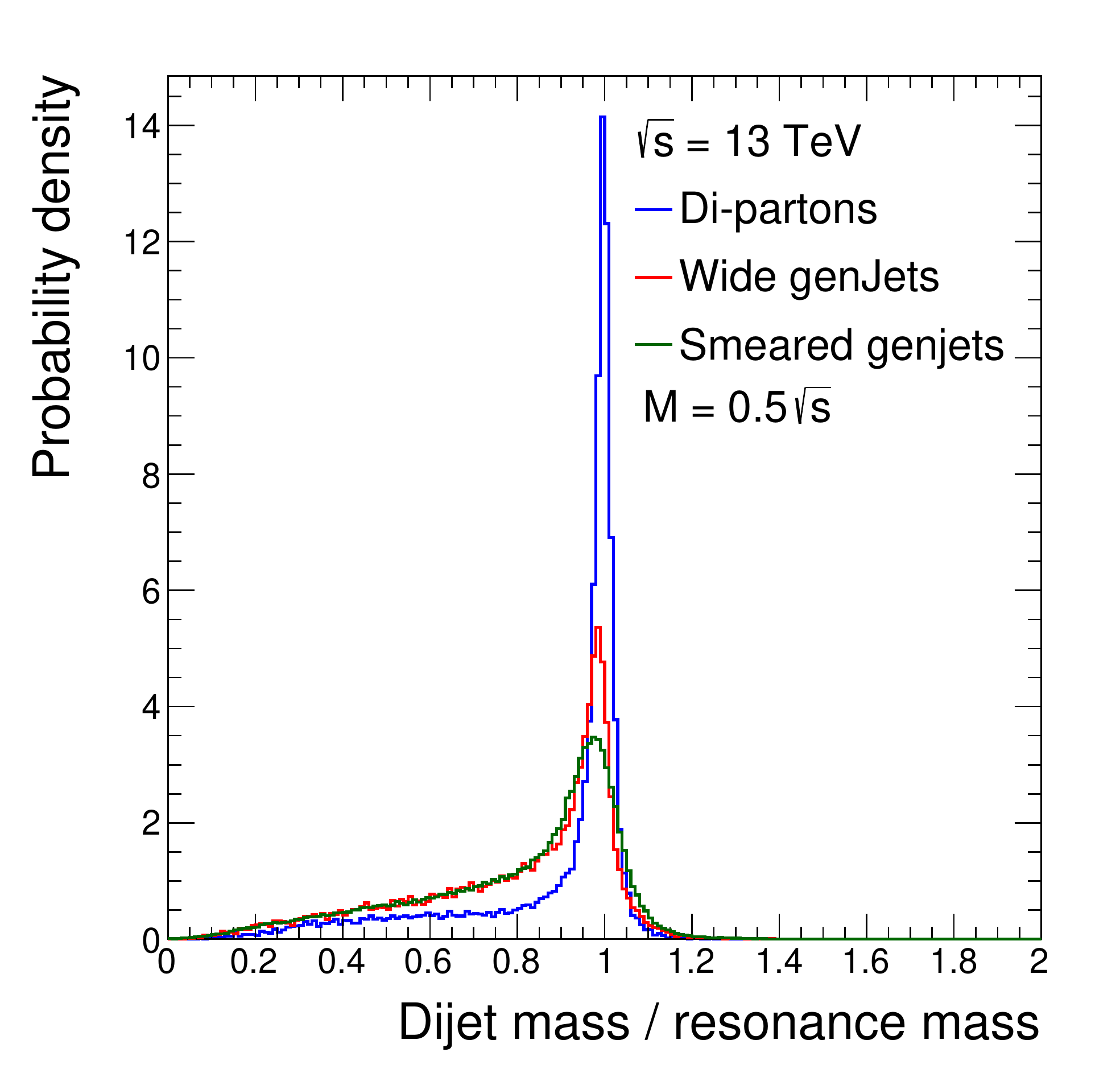}
\includegraphics[width=.32\textwidth]{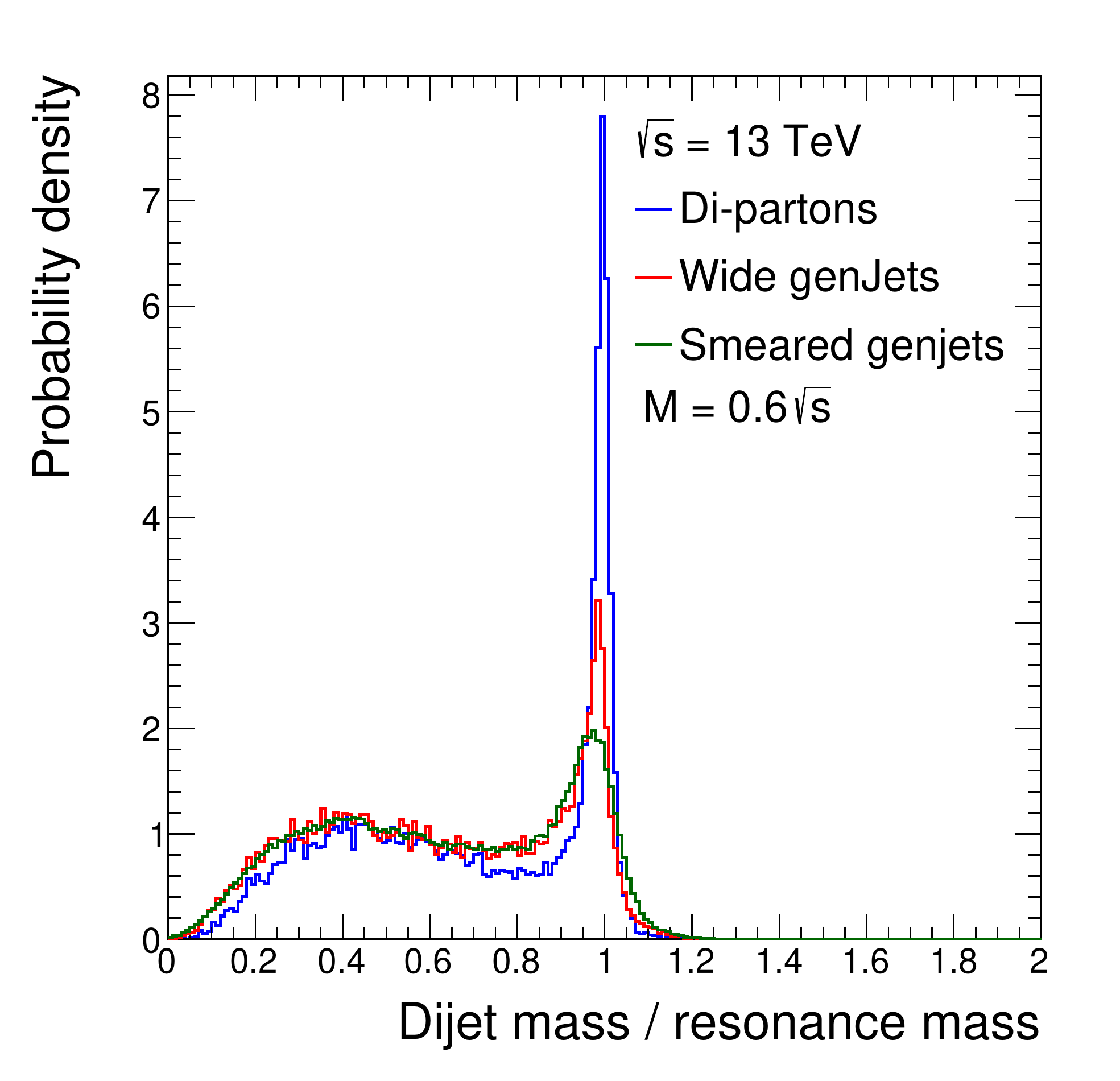}
\includegraphics[width=.32\textwidth]{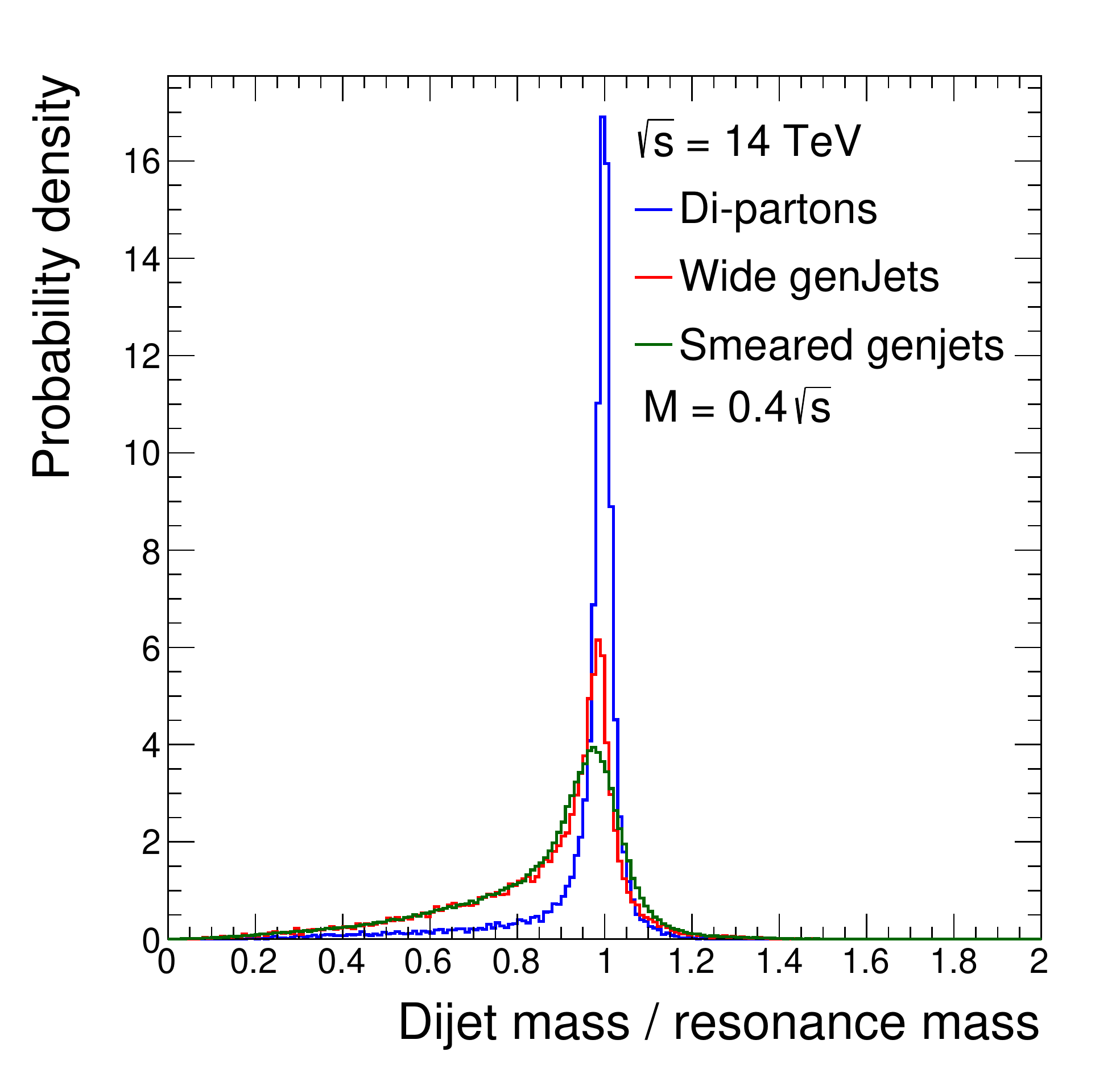}
\includegraphics[width=.32\textwidth]{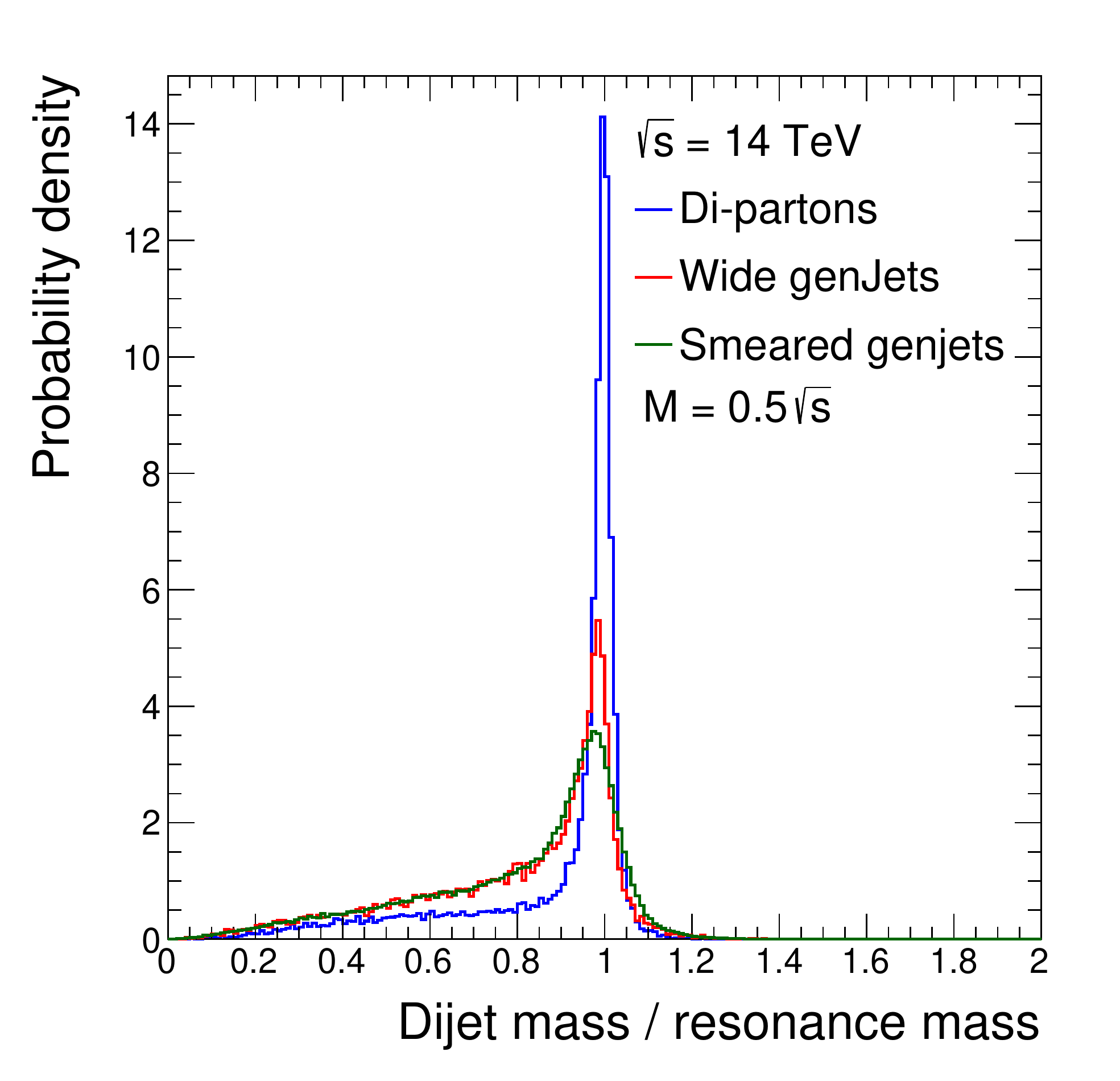}
\includegraphics[width=.32\textwidth]{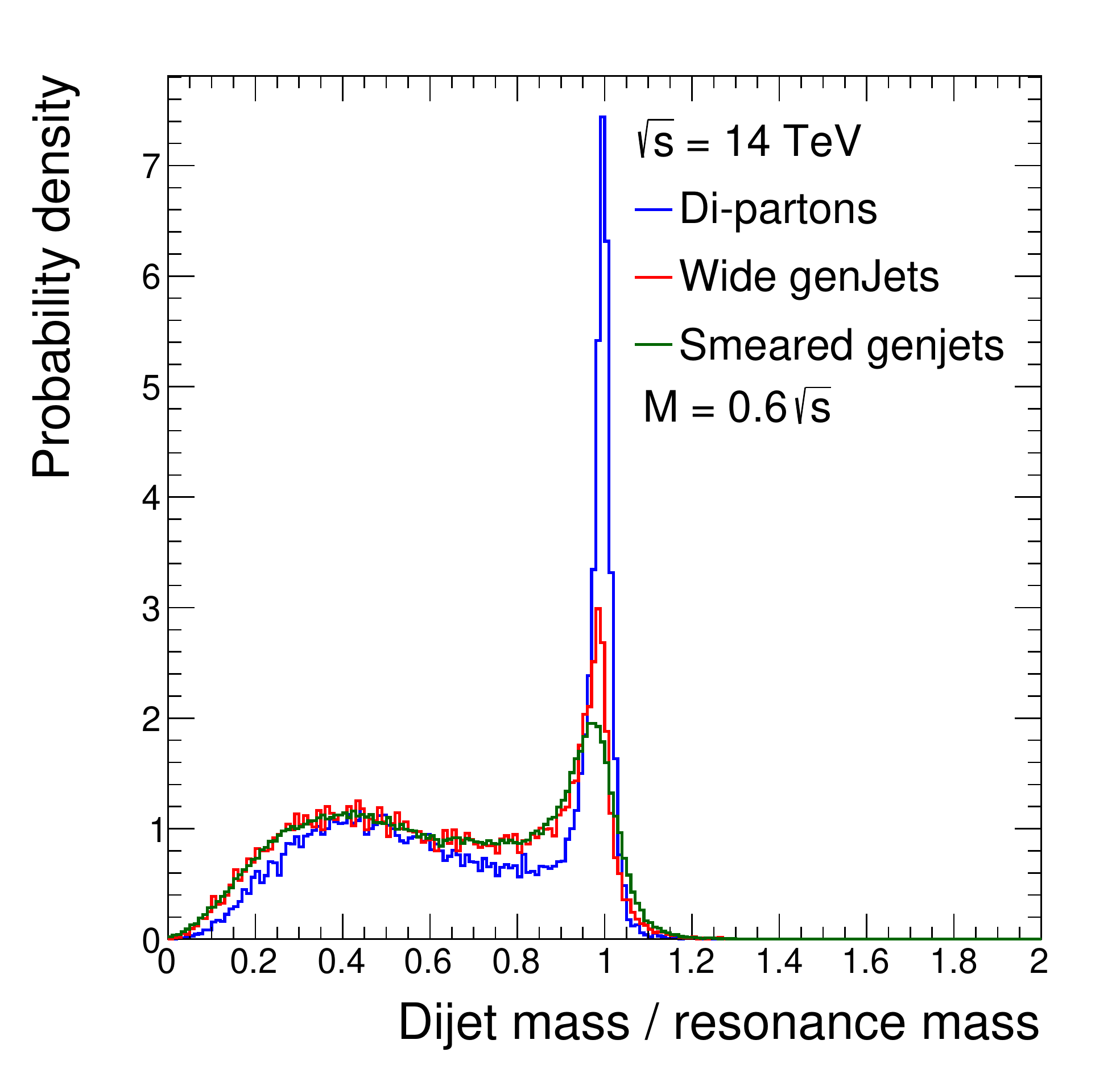}
\includegraphics[width=.32\textwidth]{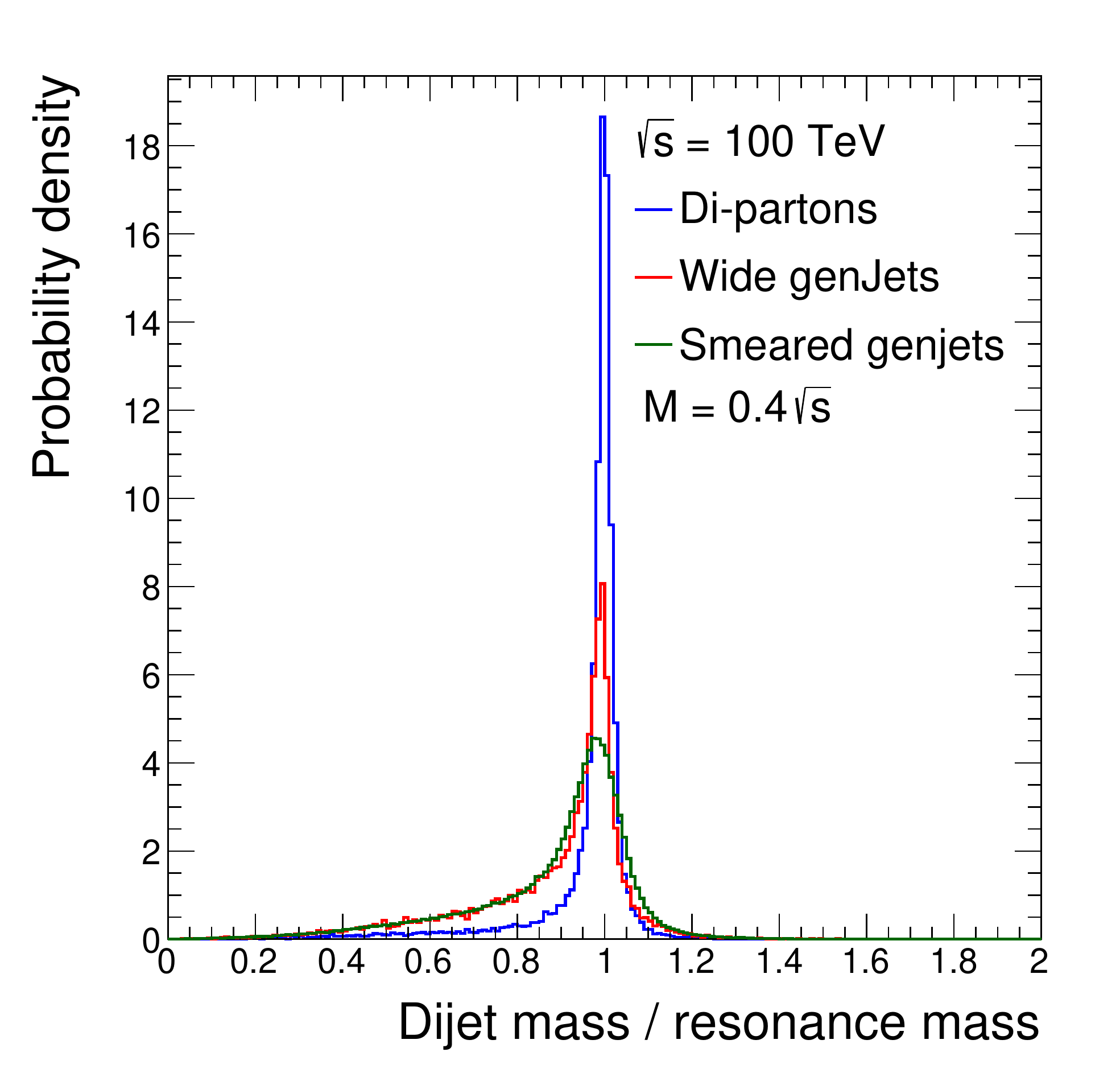}
\includegraphics[width=.32\textwidth]{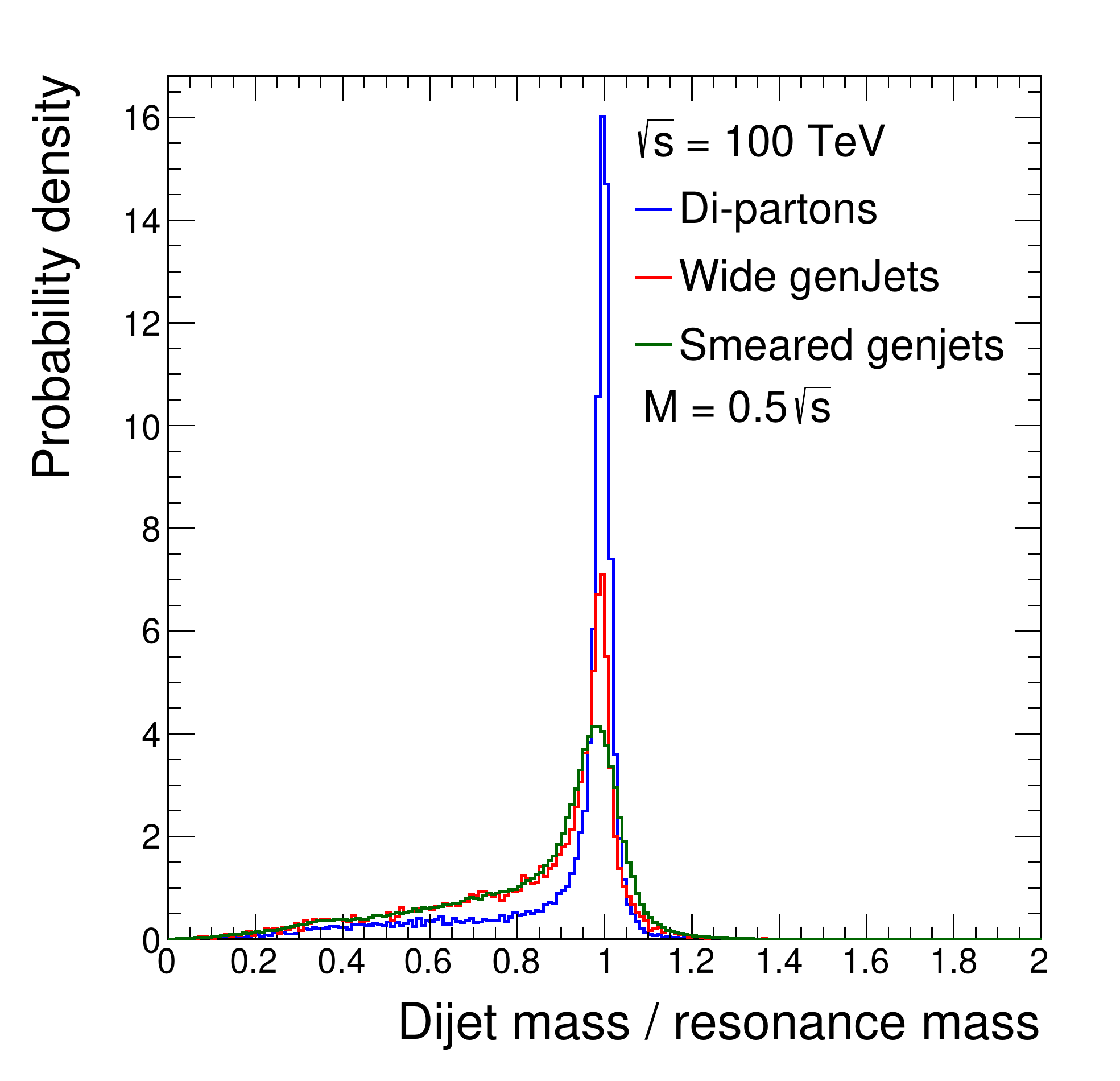}
\includegraphics[width=.32\textwidth]{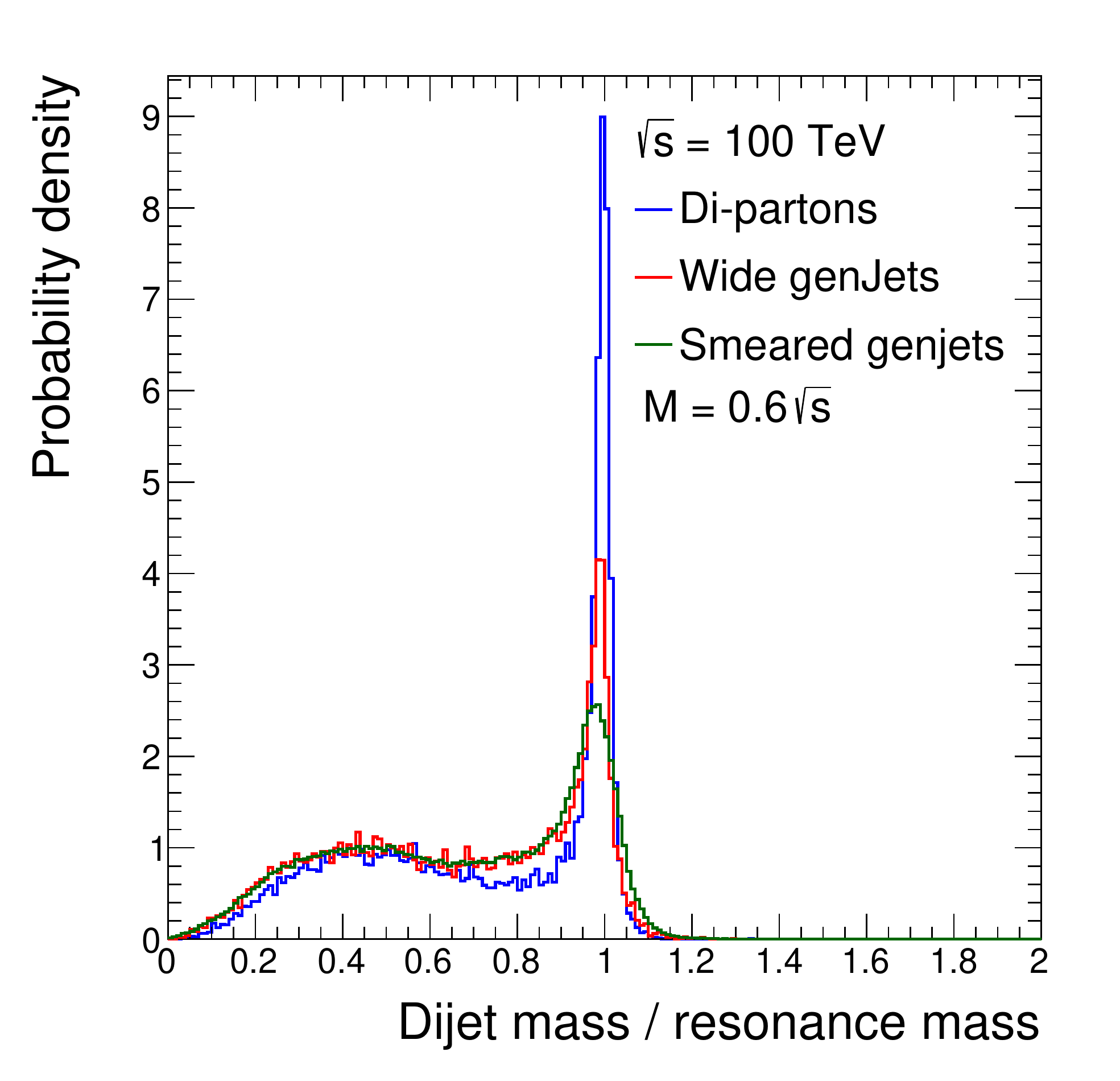}
\includegraphics[width=.32\textwidth]{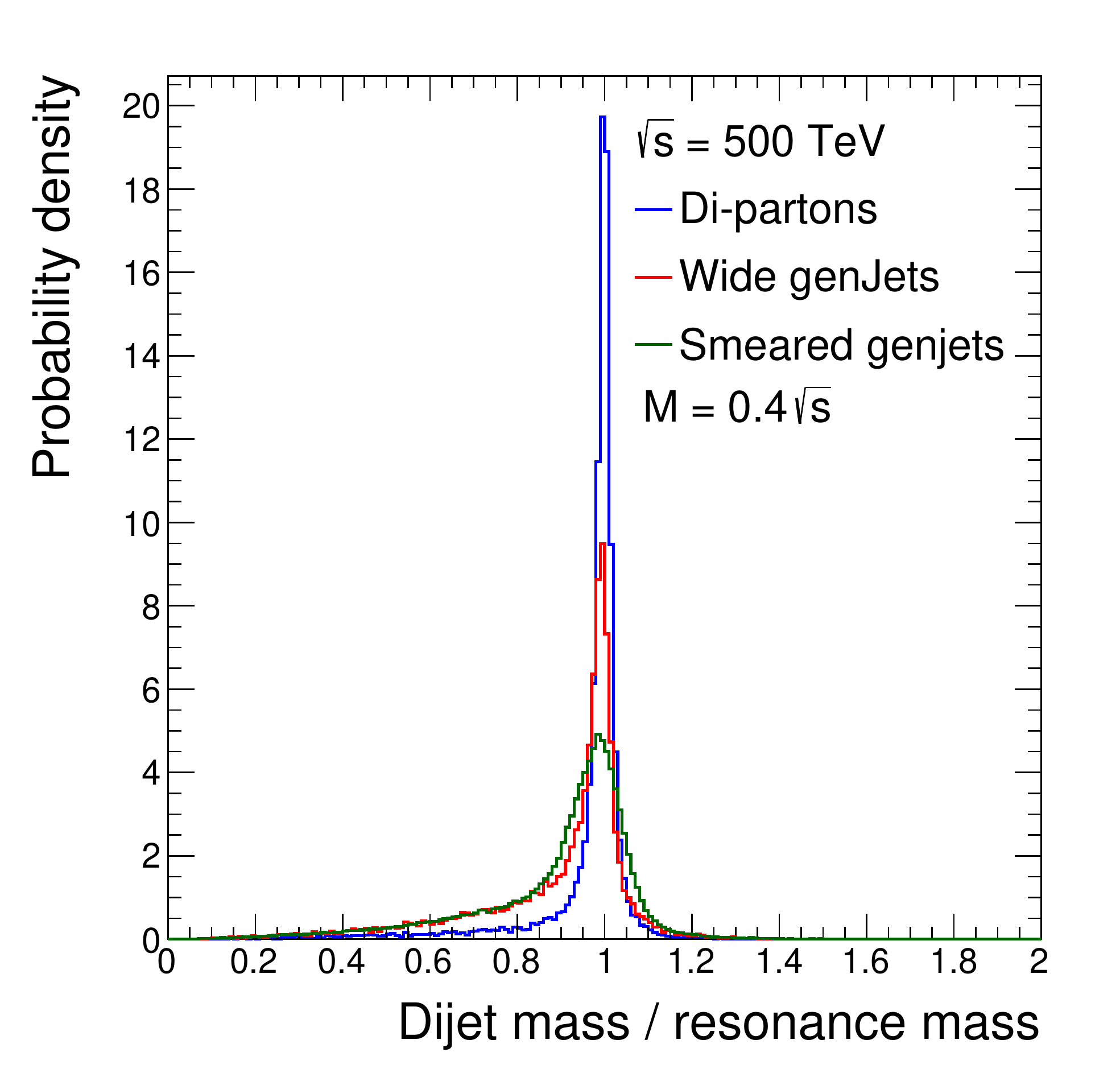}
\includegraphics[width=.32\textwidth]{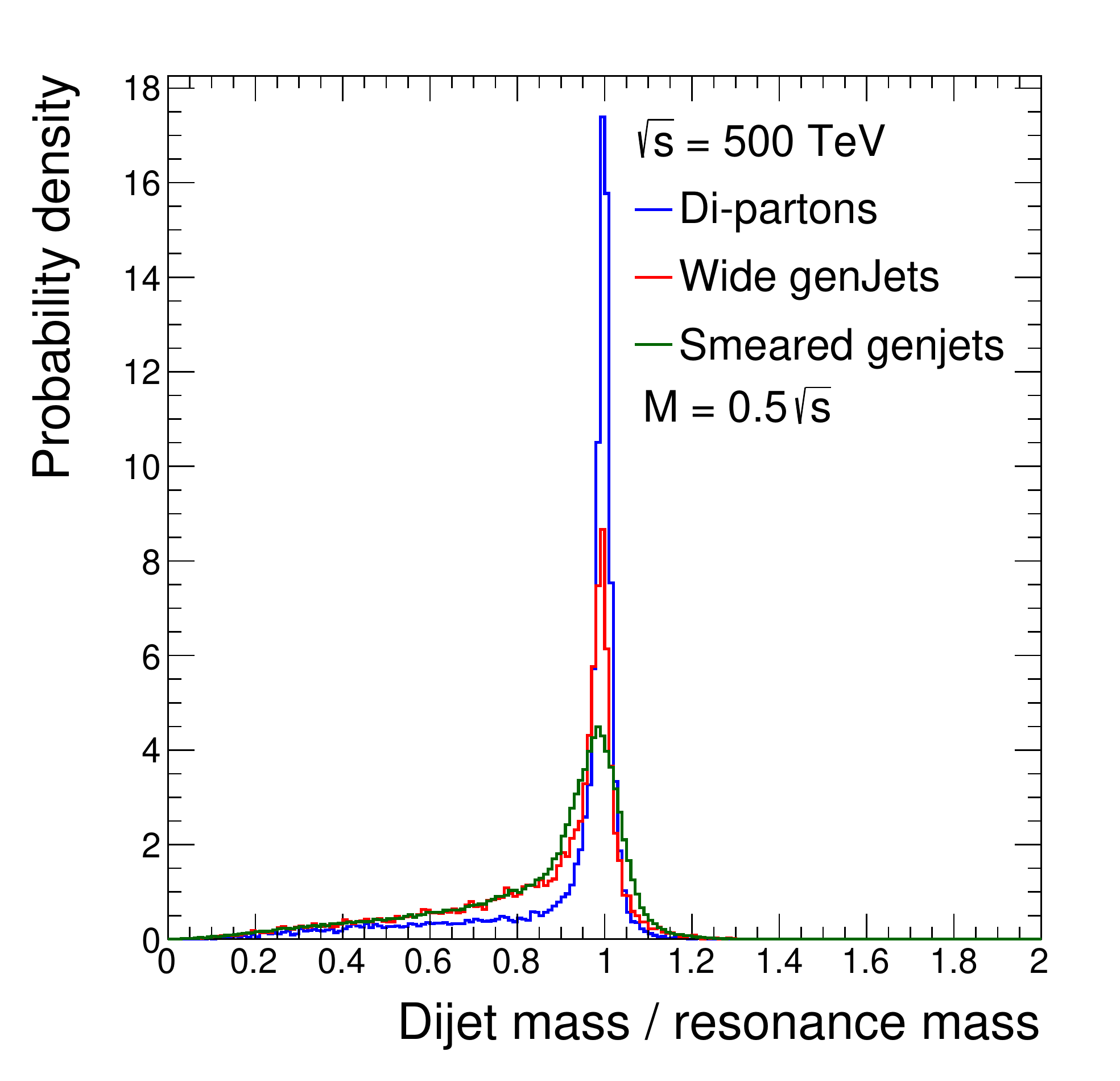}
\includegraphics[width=.32\textwidth]{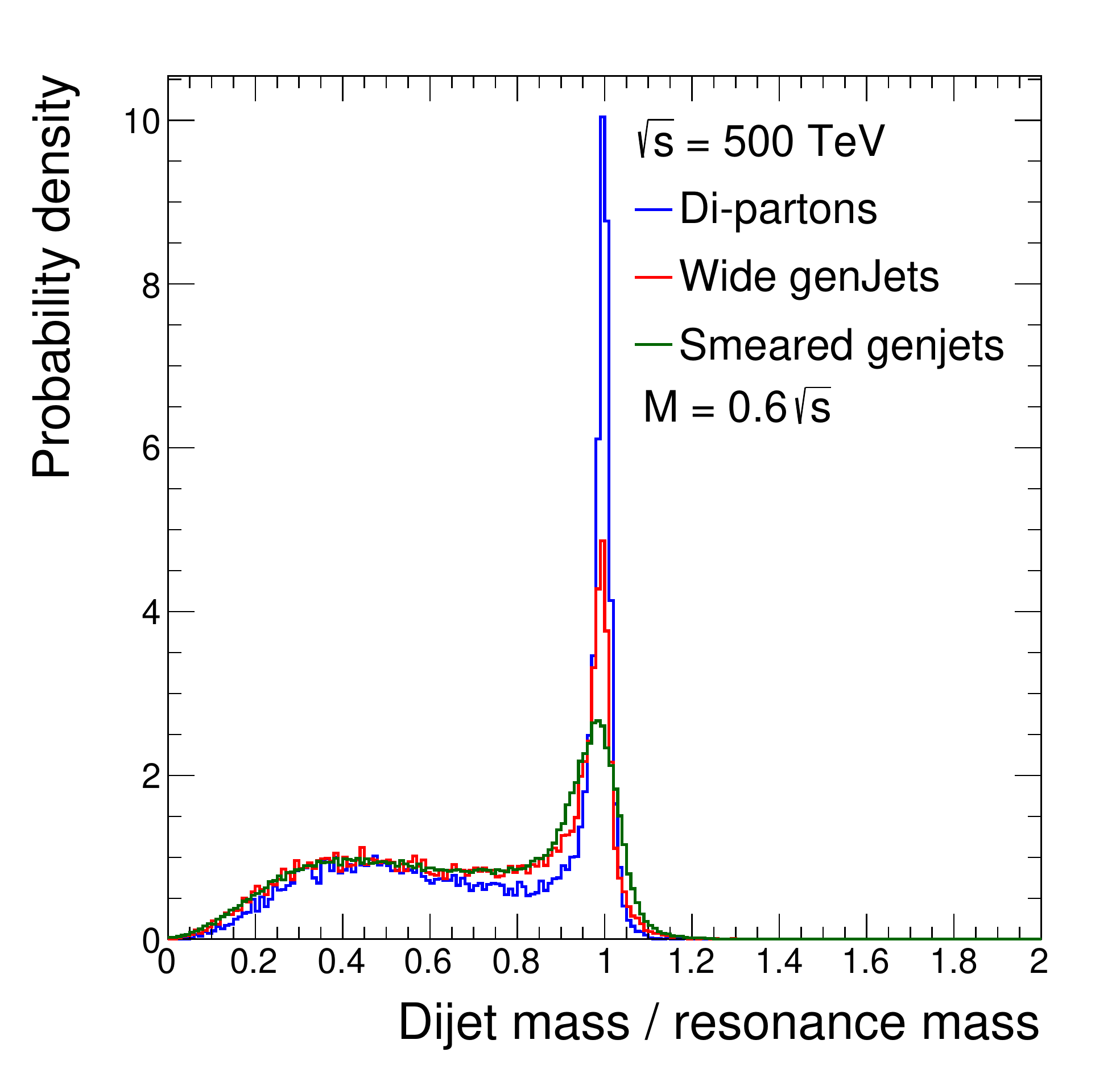}
\caption{\label{fig:AllSimLevels} Dijet mass distributions for excited quarks at parton level (blue), genjet level (red) and experimental level (green) from $pp$ collisions at $\sqrt{s}$ equal to 13 TeV (top row), 14 TeV (2nd row), 100 TeV (3rd row), and 500 TeV (bottom row), for resonance mass equal to 40\% (left column), 50\% (middle column), and 60\% of $\sqrt{s}$ (right column).}
\end{figure}

Fig.~\ref{fig:AllSimLevels} compares three levels of the excited quark dijet mass distributions. We make this comparison for resonance masses in the range $0.4 < M/\sqrt{s} < 0.6$, because that is near the critical mass $M/\sqrt{s}=0.5$ where $pp$ colliders can exclude or discover an excited quark~\cite{Harris:2022kls}, and is also the mass range where the generator-level distributions show a significant change in shape, as discussed in section~\ref{sec:massInvariance}.

First, we show parton level, the generator level closest to the collision sub-process,
the dijet mass distribution of di-partons shown in Fig.~\ref{fig:AllSimLevels} in blue. Naturally it is the narrowest of the distributions, and for $M/\sqrt{s}=0.4$ it is a truly narrow Breit-Wigner-like distribution, with a full width at half maximum $\Gamma\approx.03M$.  At larger masses the product of the steeply falling PDFs with the underlying Breit-Wigner process makes the distribution noticeably asymmetric, as discussed in section~\ref{sec:massInvariance}. For $M/\sqrt{s}=0.5$ the di-parton distribution has developed a long tail to low dijet mass, and for $M/\sqrt{s}=0.6$ that tail is the dominant feature of the distribution, containing the majority of the probability.

Second, we show genjet level, the generator-level after initial-state and final-state radiation, the parton shower, hadronization, and clustering of particles into jets, which results in the dijet mass distribution of wide genjets shown in Fig.~\ref{fig:AllSimLevels} in red. Wide genjets include most of the energy from excited quark decay products, but energy from wide-angle final-state radiation can be lost, when the particles from the hadronization of that radiation are outside the $\Delta R=1.1$ radius of the wide genjets.  As discussed in section~\ref{sec:energyInvariance}, this energy loss is the primary cause of the widening of the peak, a very small shift of the peak, and a significant tail to low mass, which are the majority of the differences between the di-parton and wide genjet distributions shown. As discussed in section~\ref{sec:massInvariance}, for $M/\sqrt{s}\leq 0.4$ this radiation is the dominant source of the tail at low mass, $M/\sqrt{s}=0.5$ is a transition point, where radiation and the product of PDFs with the underlying Breit-Wigner are roughly the same size contributions to the tail, and for $M/\sqrt{s}=0.6$ radiation is only a small component of the tail.

Third, we show the experiment-level estimates of fully reconstructed wide jets, the dijet mass distributions of smeared wide genjets shown in Fig.~\ref{fig:AllSimLevels} in green. We have used Gaussians with RMS given by Eq.~\ref{eq:ExpRes},  to smear the wide genjet dijet mass distributions. 
Fig.~\ref{fig:AllSimLevels} shows that this Gaussian experimental resolution widens the peak of the resonance shape but does not affect the tails.  This is because the experimental resolution is wider than the peaks of these narrow resonances, but significantly narrower than the long tail to low mass.  The experimental width of the peak is  determined by the experimental resolution, as expected for narrow resonances.  However, the long tail at low mass is completely independent of experimental resolution. As discussed in the paragraphs above and in sections~\ref{sec:energyInvariance} and \ref{sec:massInvariance}, these tails are determined completely at generator-level by the dijet reconstruction algorithm, the initial-state and final-state radiation, the underlying Breit-Wigner line shape, and the PDFs. 

Figure~\ref{fig:AllSimLevels} also demonstrates that the comparisons of the different levels of dijet mass shapes are essentially independent of $\sqrt{s}$. 

\section{Experiment-level Resonances}
\label{sec:shapes}

The remaining figures in this paper will all be made with smeared wide genjets, to approximate the expected measured shapes, cumulative probability functions, and efficiencies of actual resonances observed by an experiment at a pp collider. 

\begin{figure}[bthp]
\centering 
\includegraphics[width=.32\textwidth]{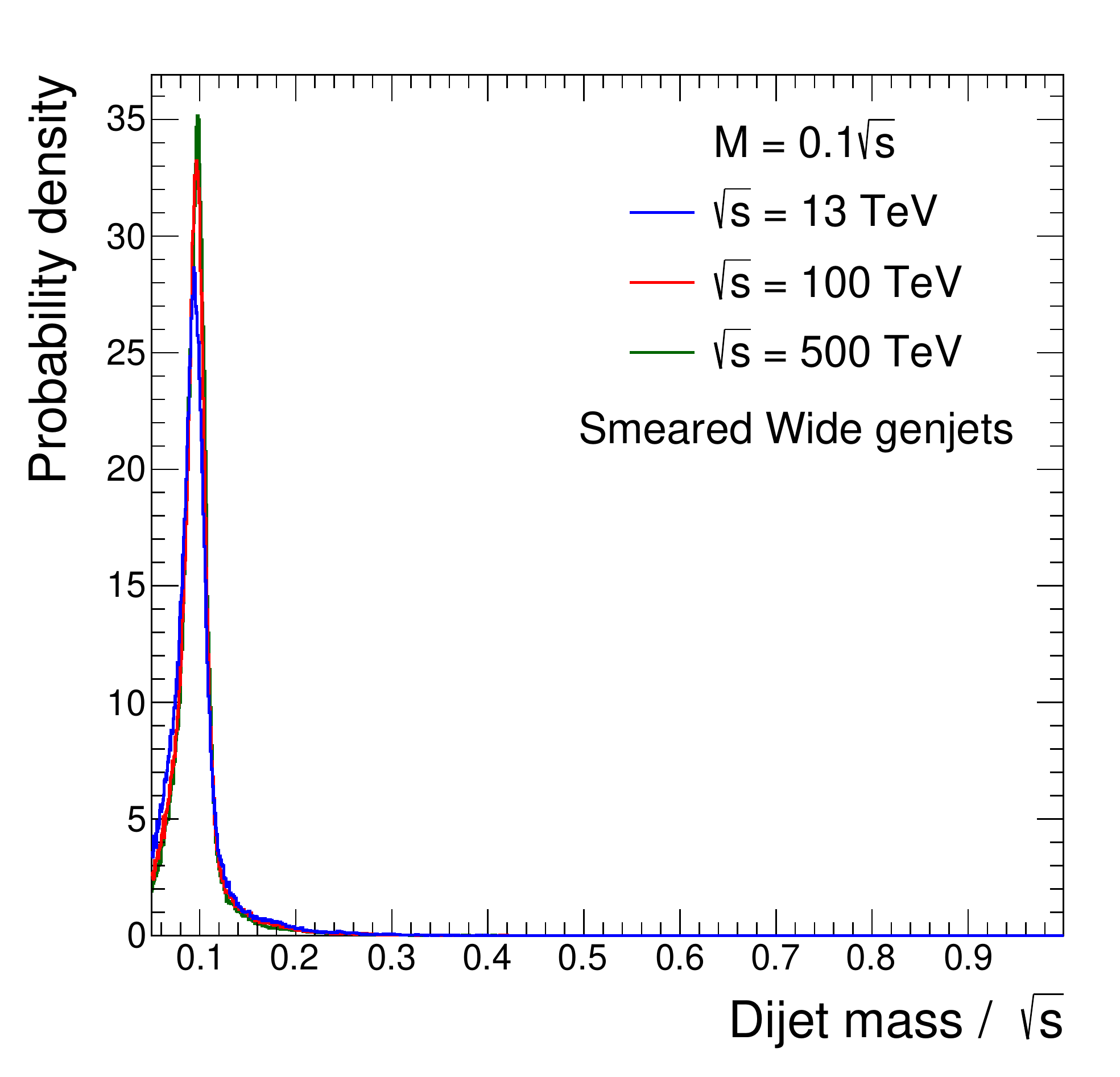}
\includegraphics[width=.32\textwidth]{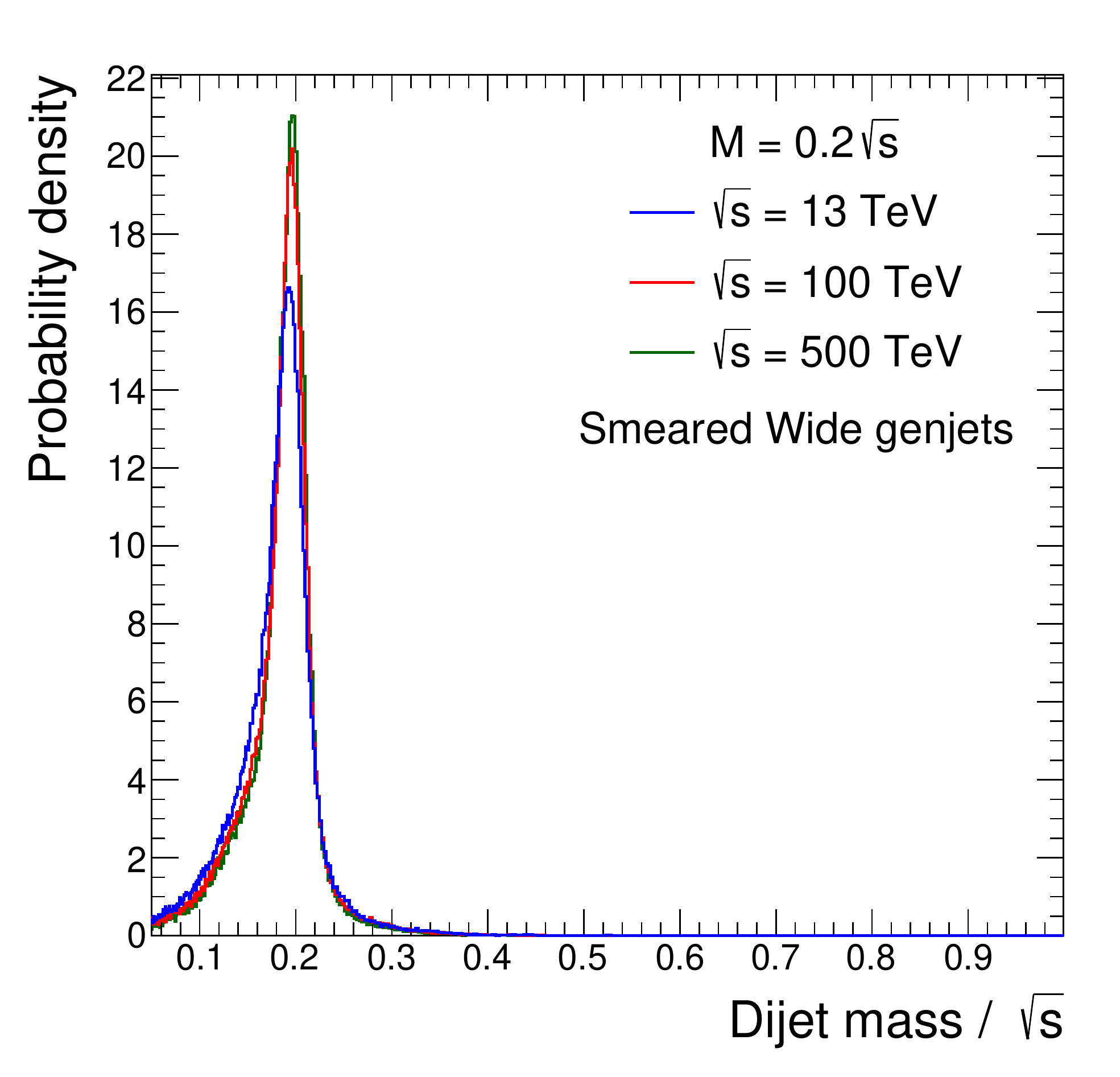}
\includegraphics[width=.32\textwidth]{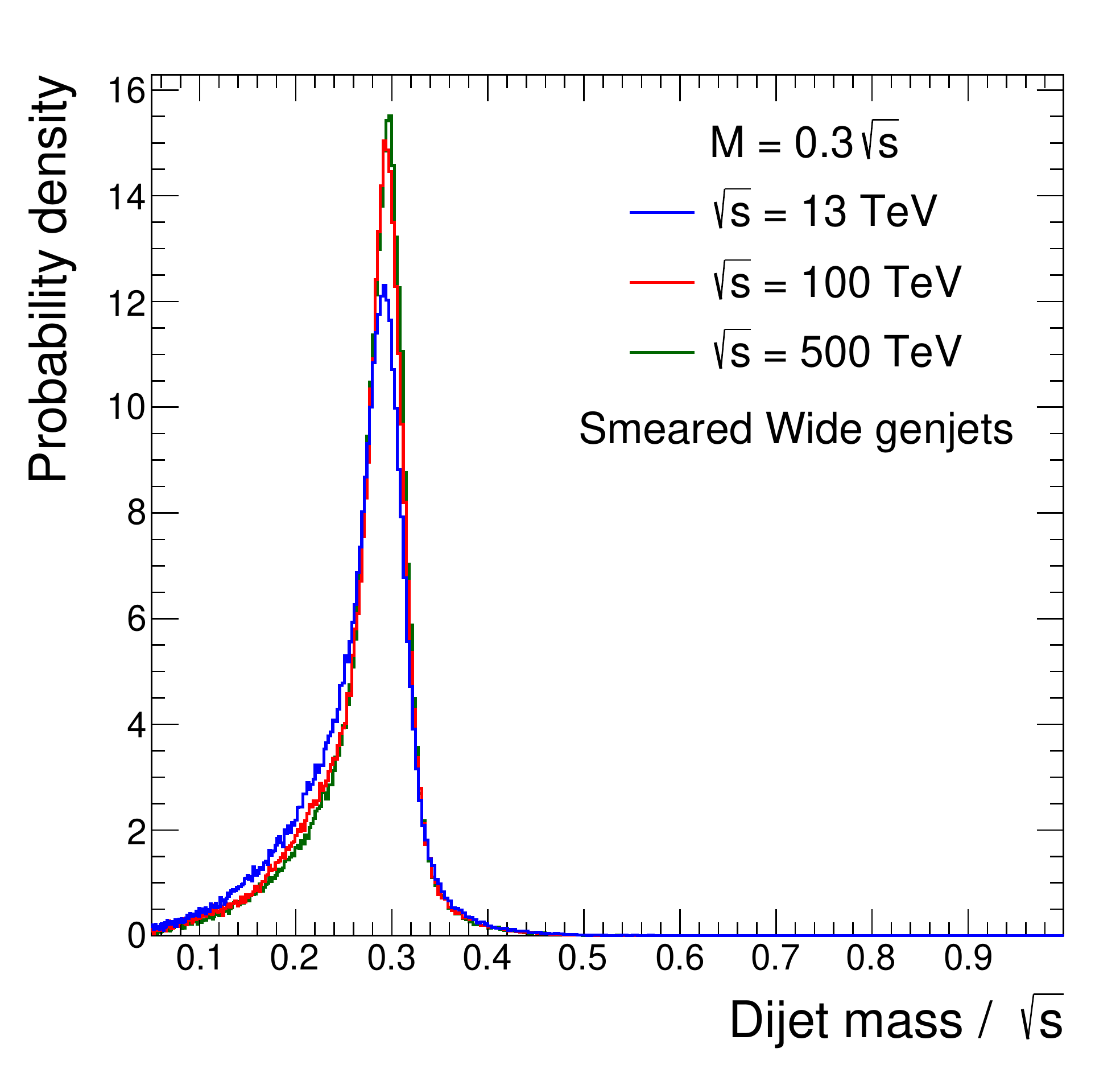}
\includegraphics[width=.32\textwidth]{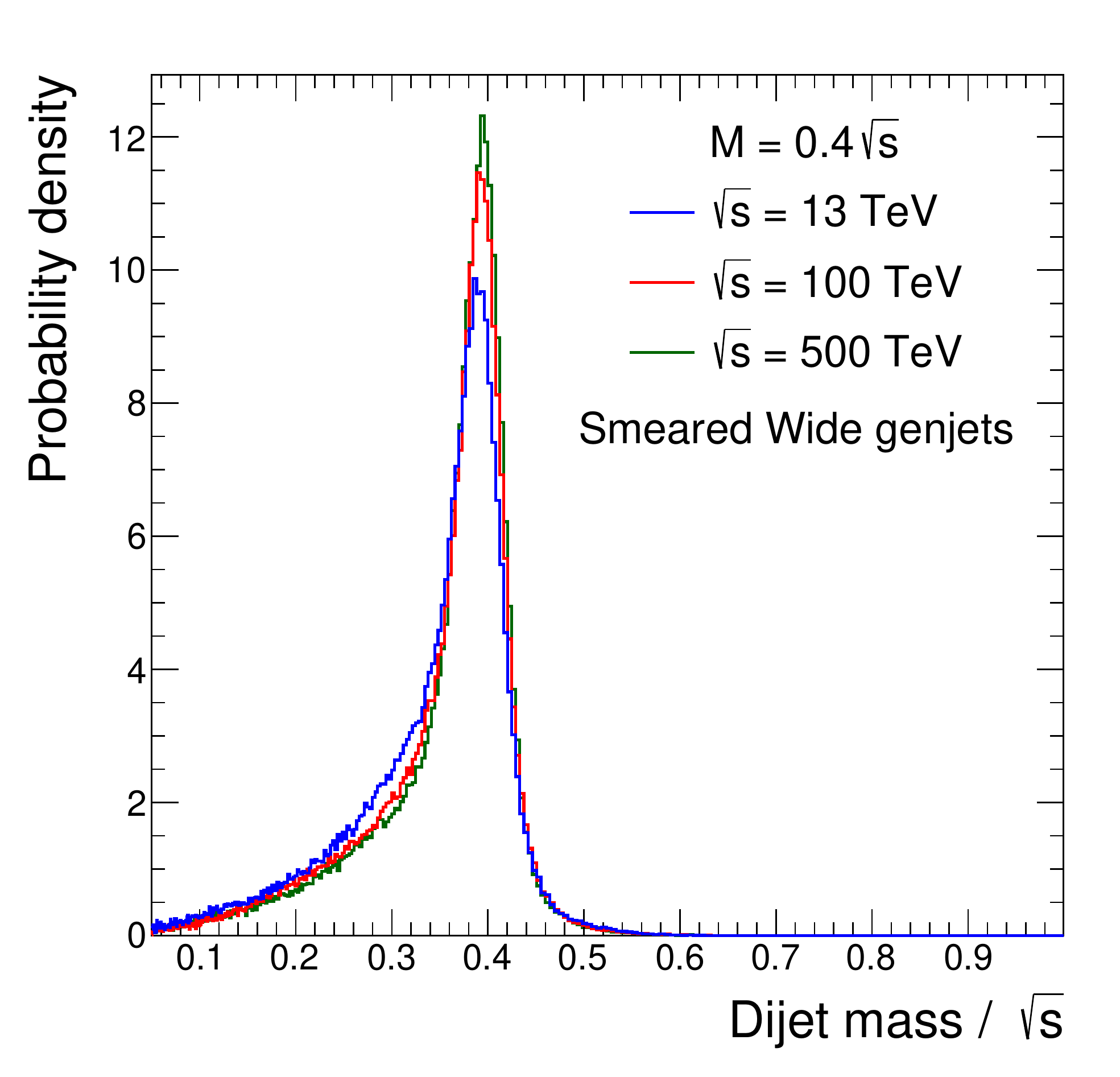}
\includegraphics[width=.32\textwidth]{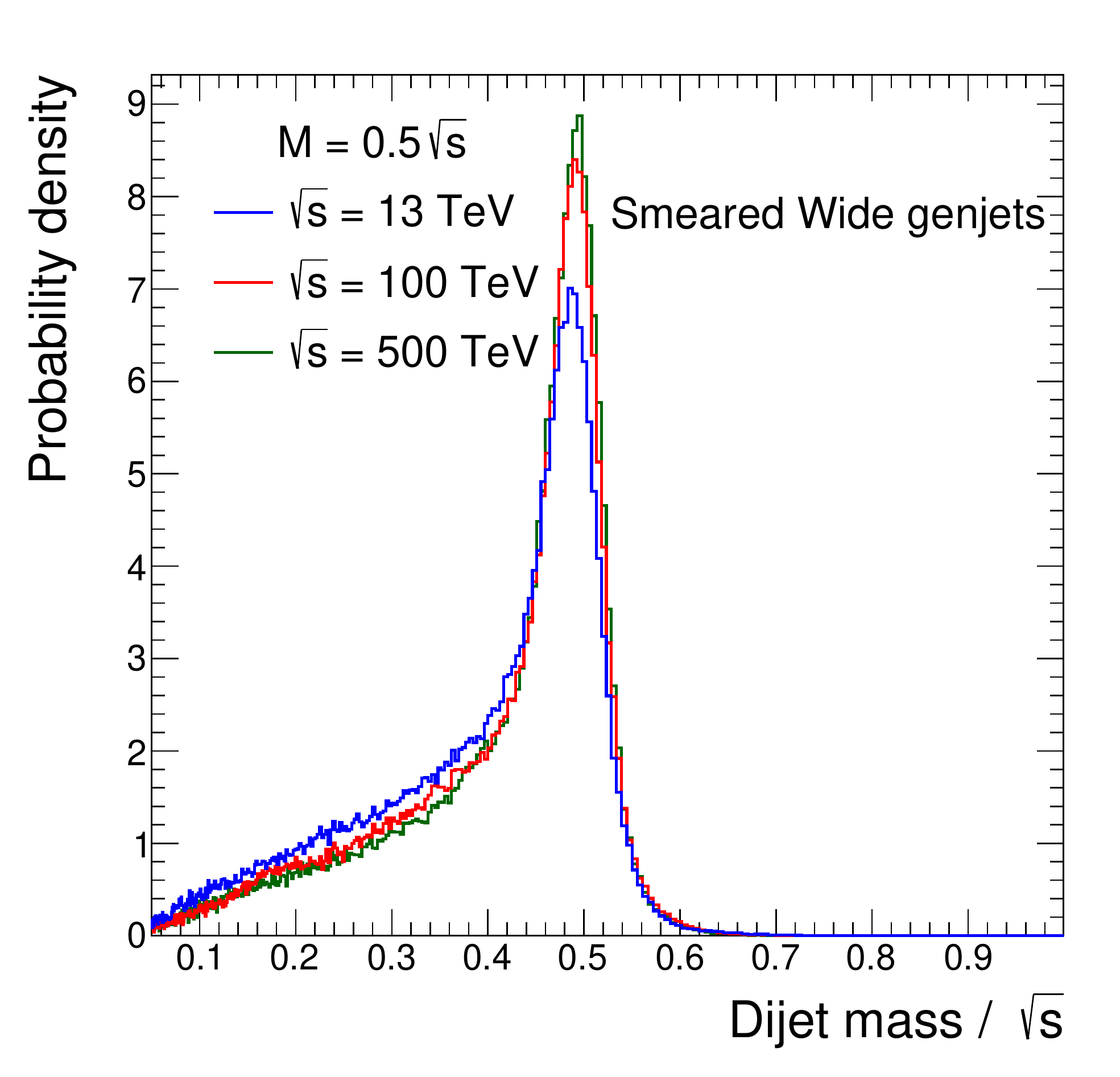}
\includegraphics[width=.32\textwidth]{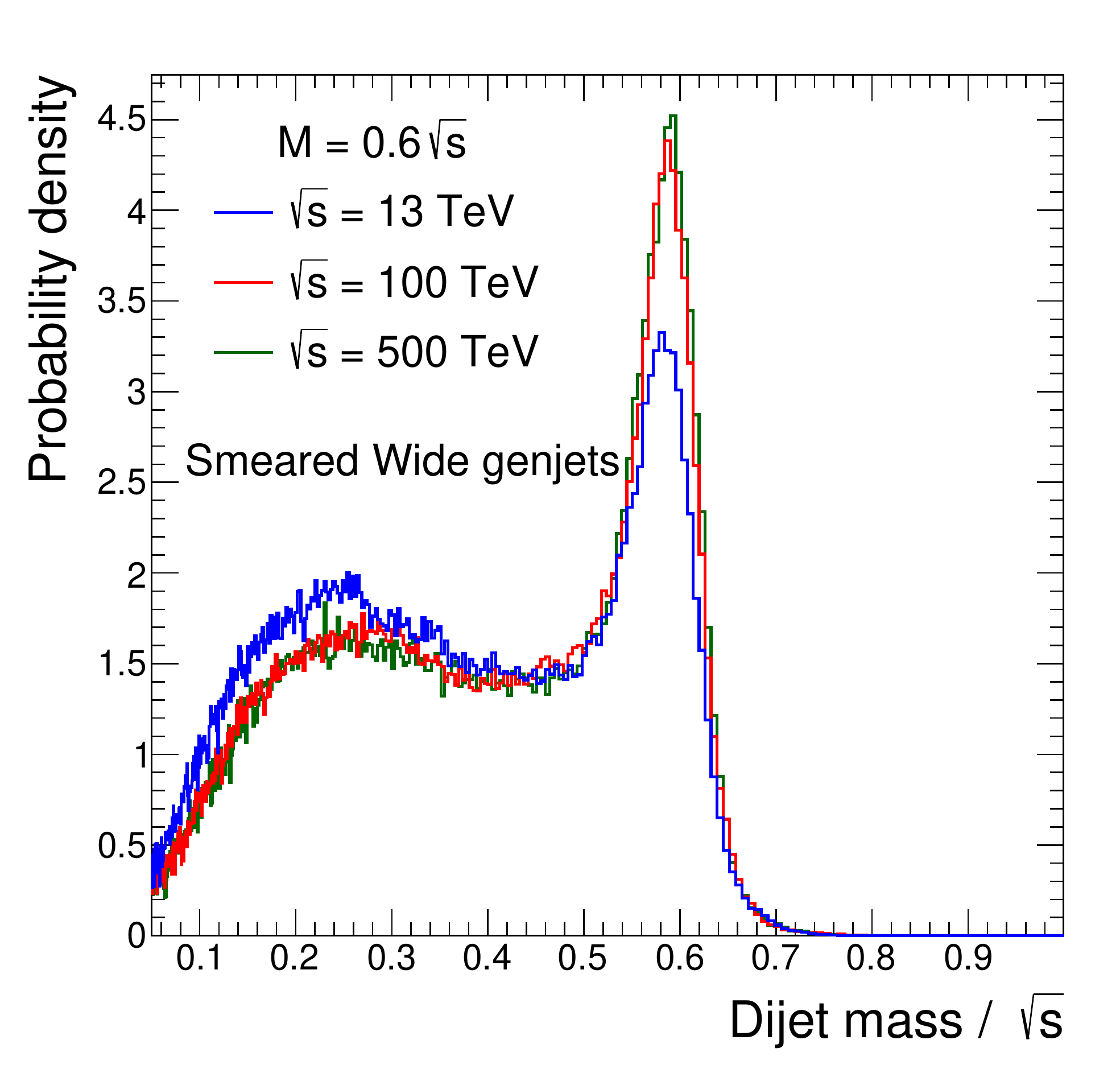}
\caption{\label{fig:SmearedMass} Comparisons of experiment level dijet mass distributions of excited quarks, with dijet mass plotted as a fraction of $\sqrt{s}$, from $pp$ collisions at $\sqrt{s}$ equal to 13 TeV (blue), 100 TeV (red), and 500 TeV (green) for resonance masses of 10\%, 20\% and 30\% of $\sqrt{s}$ (top row) and 40\%, 50\% and 60\% of $\sqrt{s}$ (bottom row).}
\end{figure}

\subsection{Collision energy and resonance mass invariances at experiment-level}

\begin{figure}[bhtp]
\centering 
\includegraphics[width=.32\textwidth]{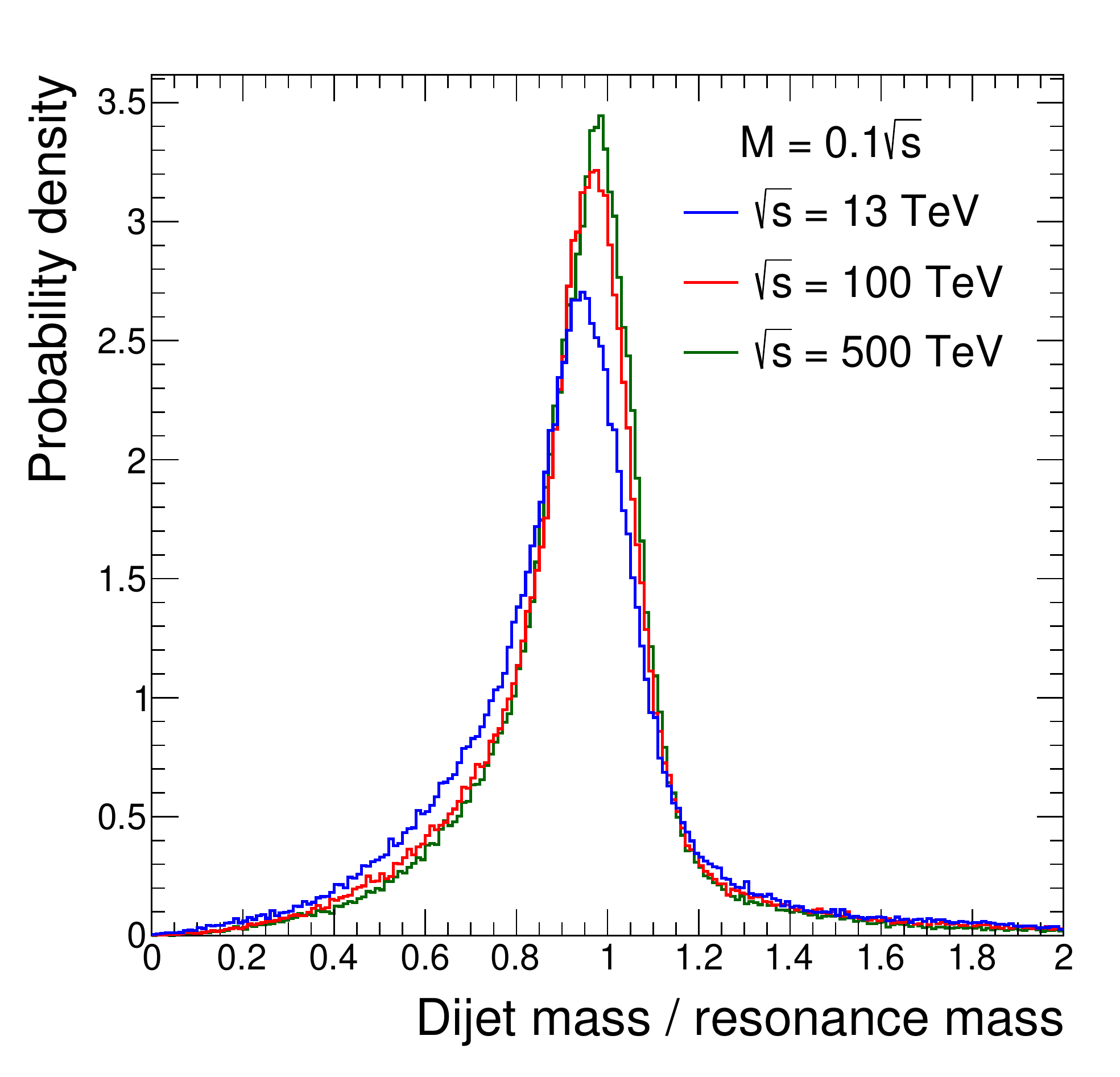}
\includegraphics[width=.32\textwidth]{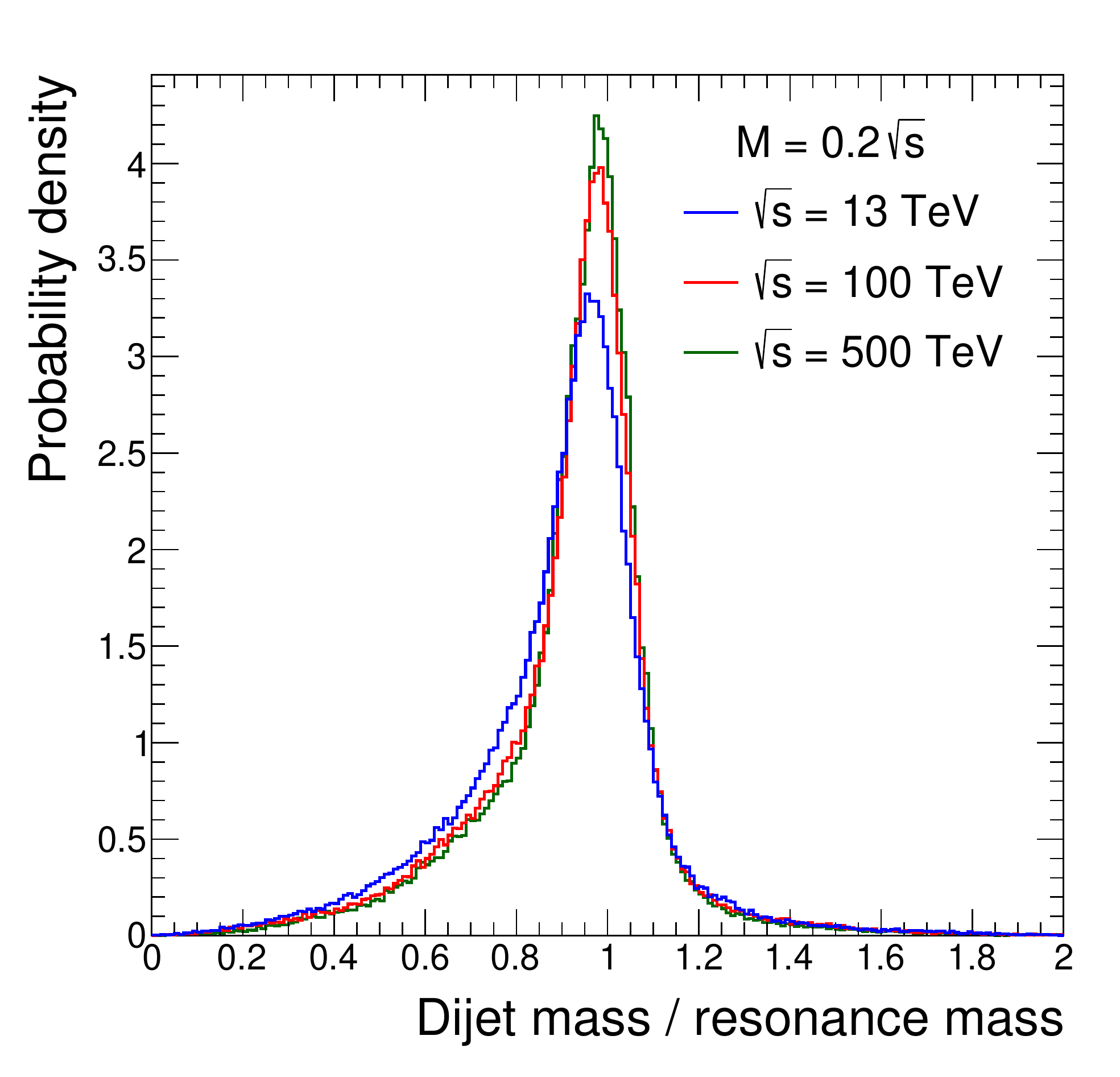}
\includegraphics[width=.32\textwidth]{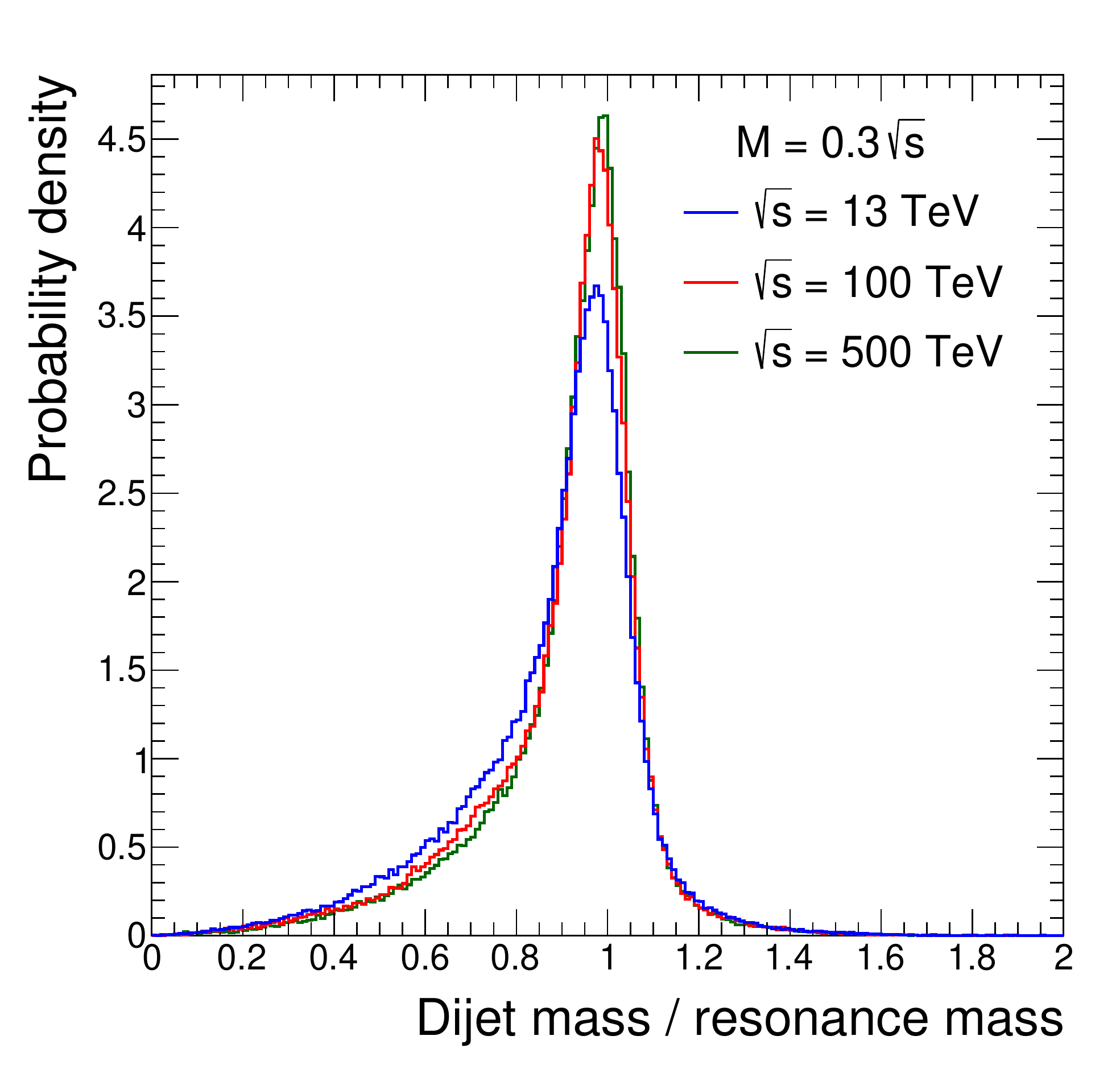}
\includegraphics[width=.32\textwidth]{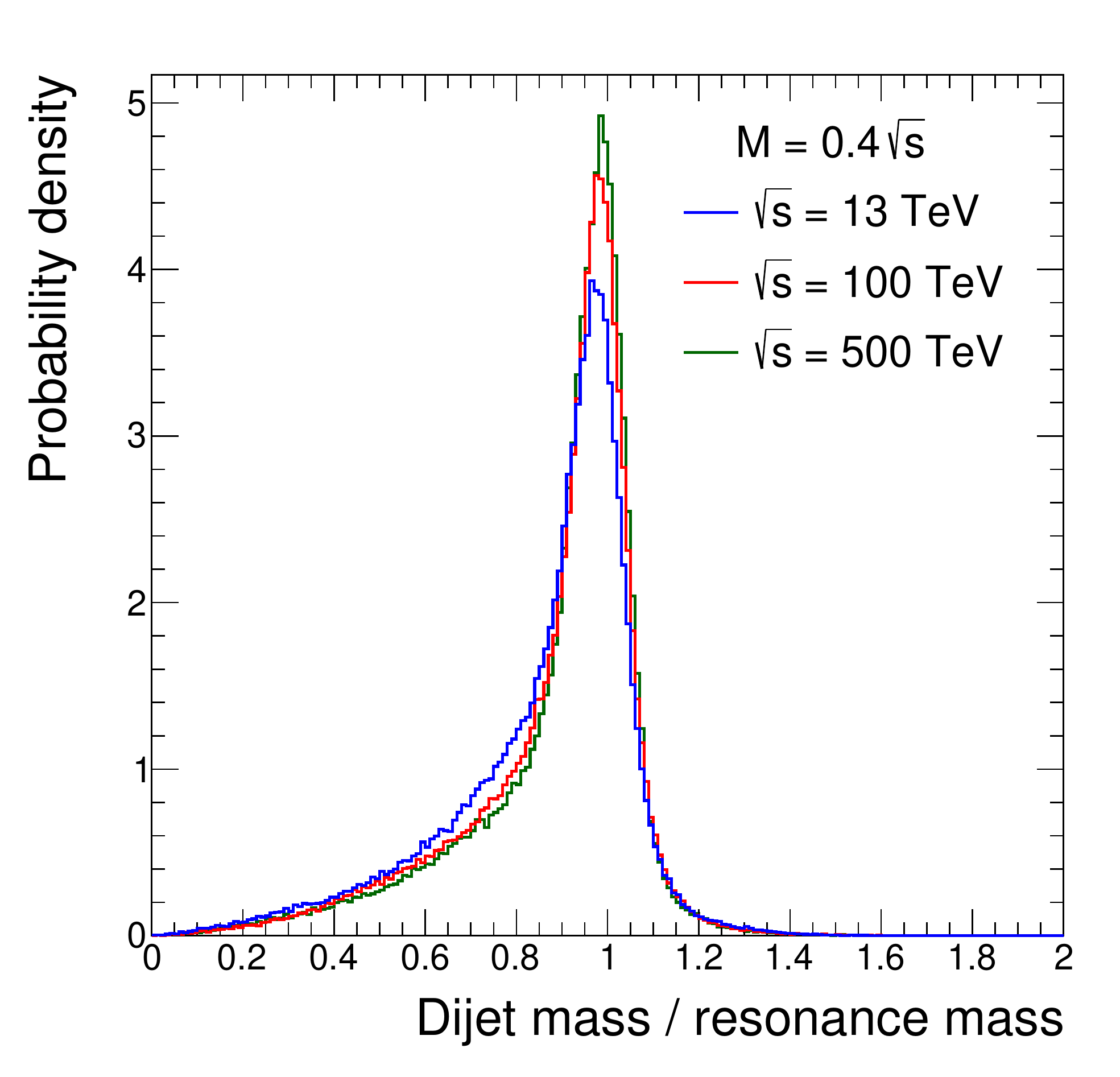}
\includegraphics[width=.32\textwidth]{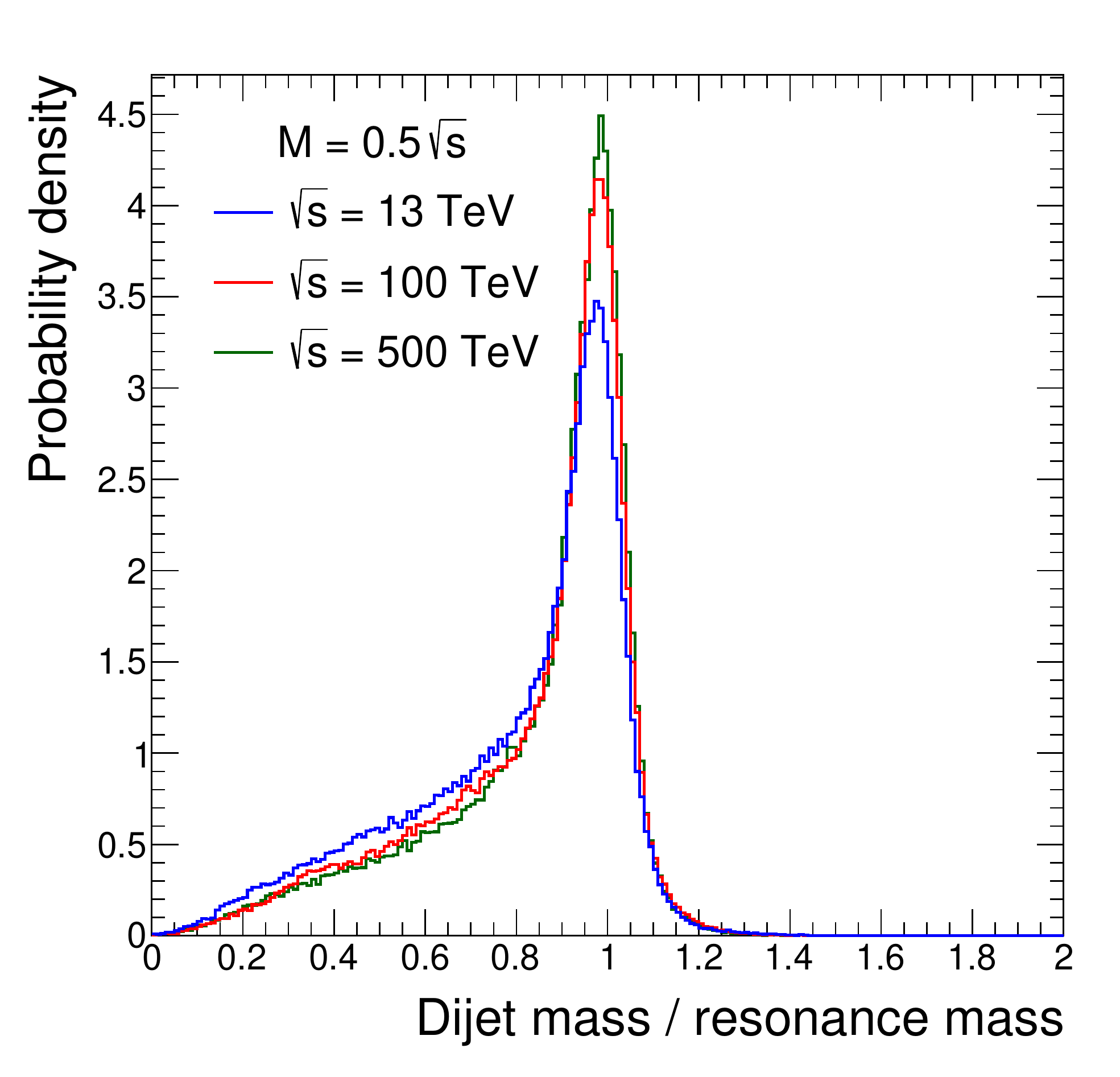}
\includegraphics[width=.32\textwidth]{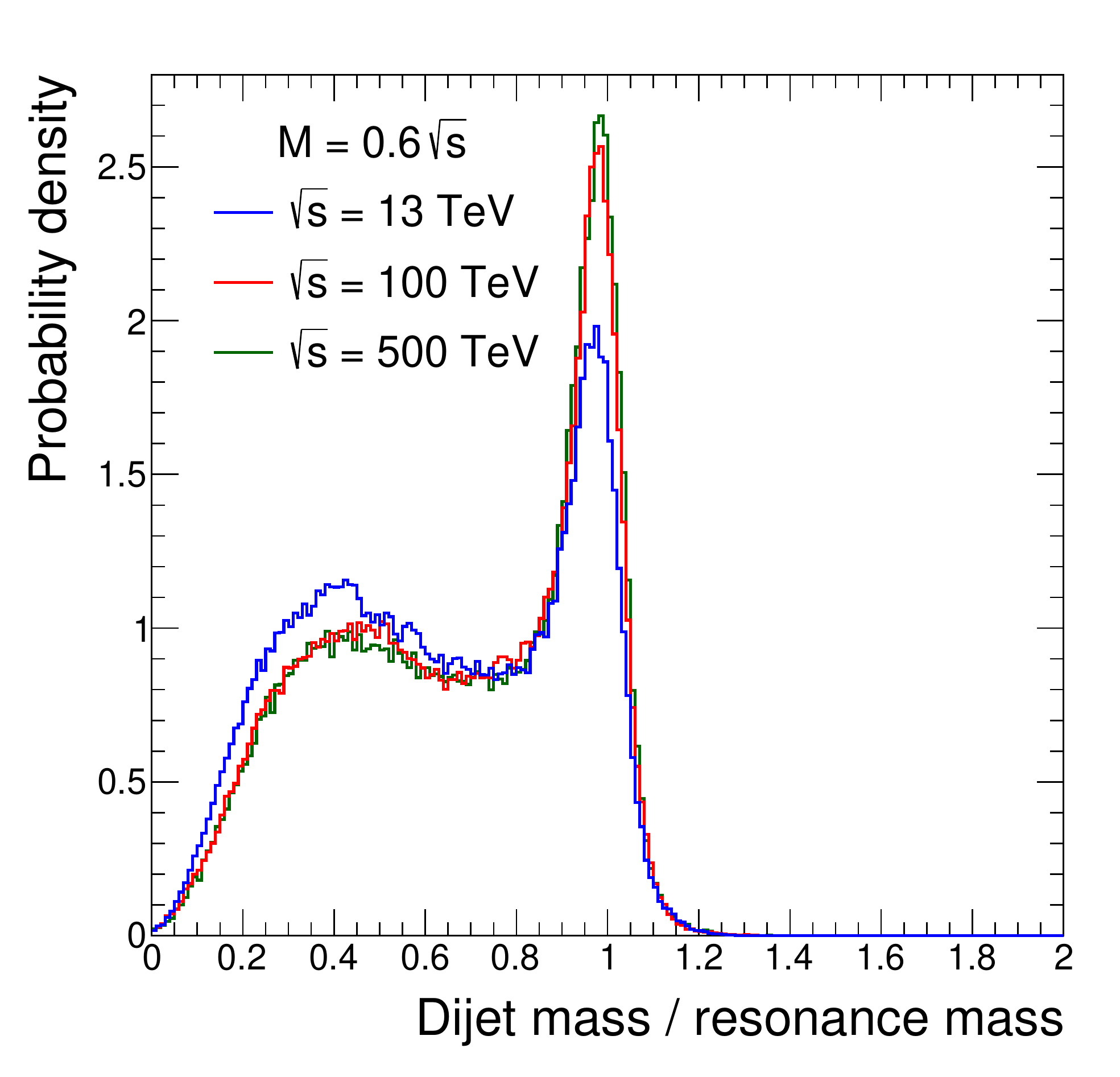}
\caption{\label{fig:SmearedMassFracFixedMass} Comparisons of experiment level dijet mass distributions of excited quarks, with dijet mass plotted as a fraction of resonance mass, from $pp$ collisions at $\sqrt{s}$ equal to 13 TeV (blue), 100 TeV (red), and 500 TeV (green) for resonance masses of 10\%, 20\% and 30\% of $\sqrt{s}$ (top row) and 40\%, 50\% and 60\% of $\sqrt{s}$ (bottom row).}
\end{figure}

The approximate invariance of the dijet mass distribution at the experiment-level to the choice of $\sqrt{s}$ is presented in Fig.~\ref{fig:SmearedMass} and Fig.~\ref{fig:SmearedMassFracFixedMass}. These two figures differ only in the choice of the dimensionless ratio plotted on the horizontal axis: $m/\sqrt{s}$ and $m/M$. They both demonstrate the approximate invariance of the experiment-level resonance shapes to variations in $\sqrt{s}$, which naturally follows from the invariance at the generator-level observed in Fig~\ref{fig:GenjetThreePlots} and discussed in section~\ref{sec:energyInvariance}, and the invariance of the experimental resolution with $\sqrt{s}$, which was explicitly assumed in section~\ref{sec:experimentalRes}.  The change in shape of the experiment-level resonances when $\sqrt{s}$ increases from 13 to 100 TeV, although small, is significantly more than the change in shape when $\sqrt{s}$ increases from 100 to 500 TeV.  This is because the decrease in width due to logarithmic evolution with $\alpha_s$ of the generator-level half width is being damped by convolution with a dominant and constant experimental resolution.

\begin{figure}[tbp]
\centering 
\includegraphics[width=.48\textwidth]{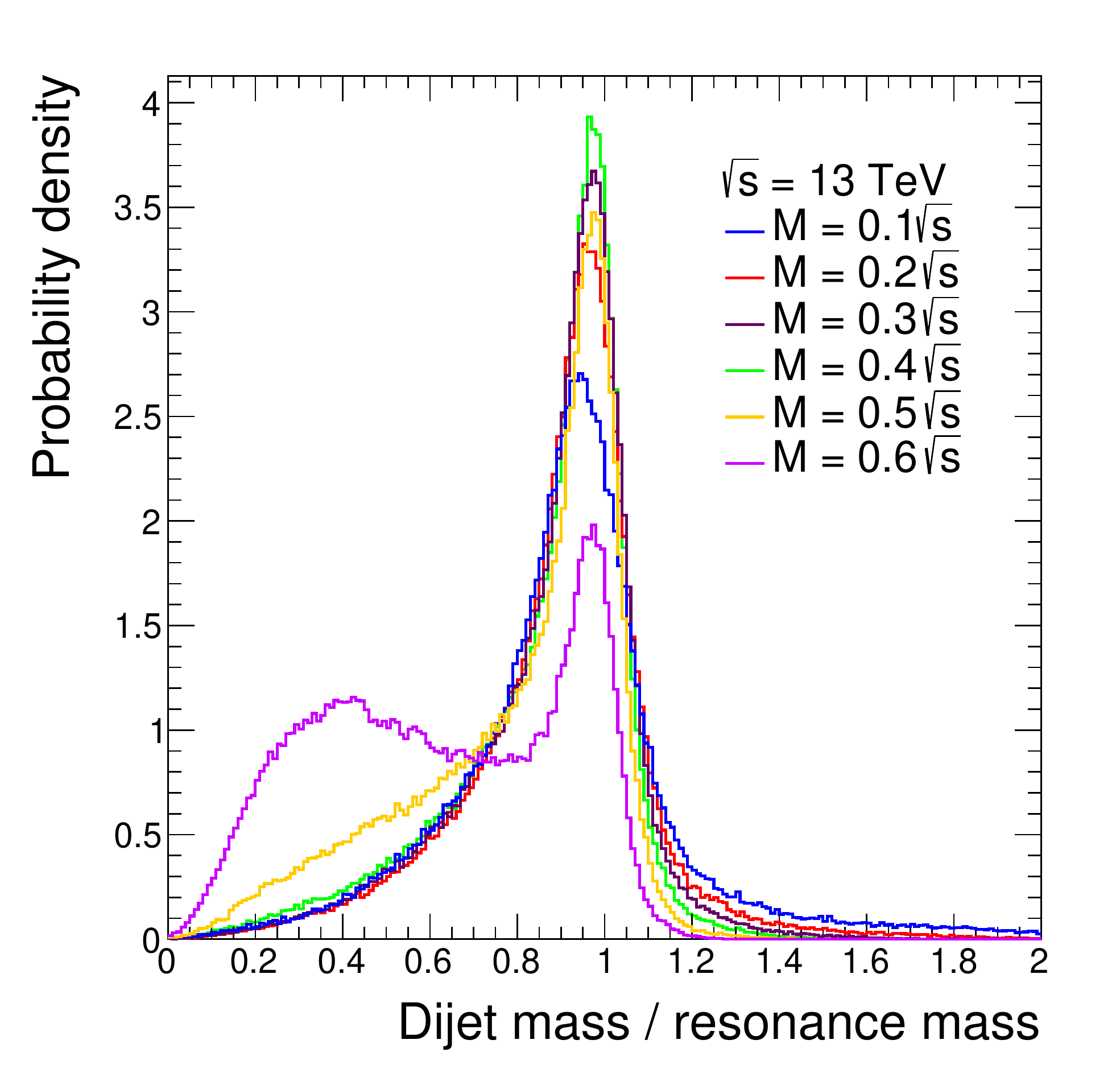}
\includegraphics[width=.48\textwidth]{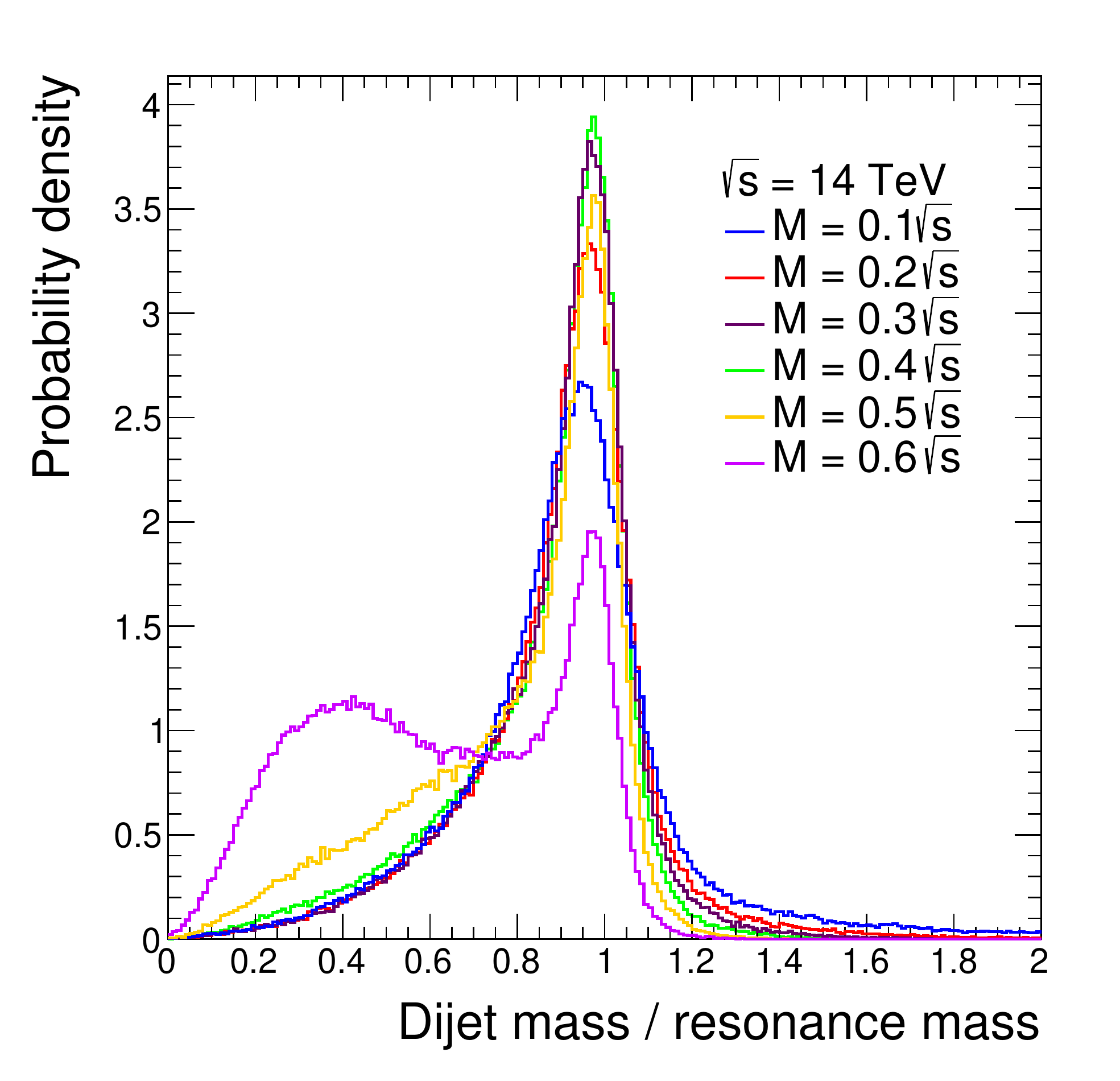}
\includegraphics[width=.48\textwidth]{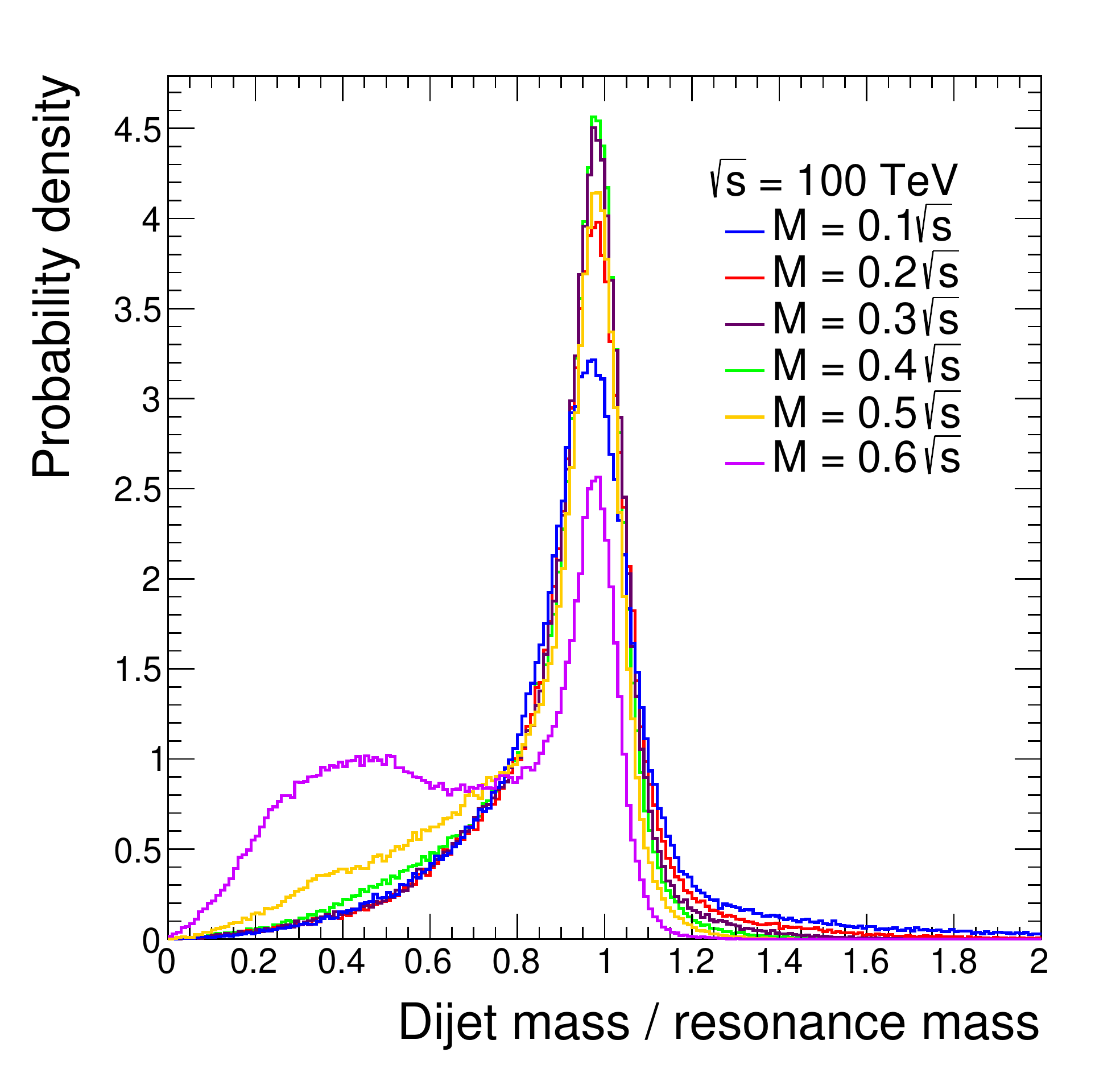}
\includegraphics[width=.48\textwidth]{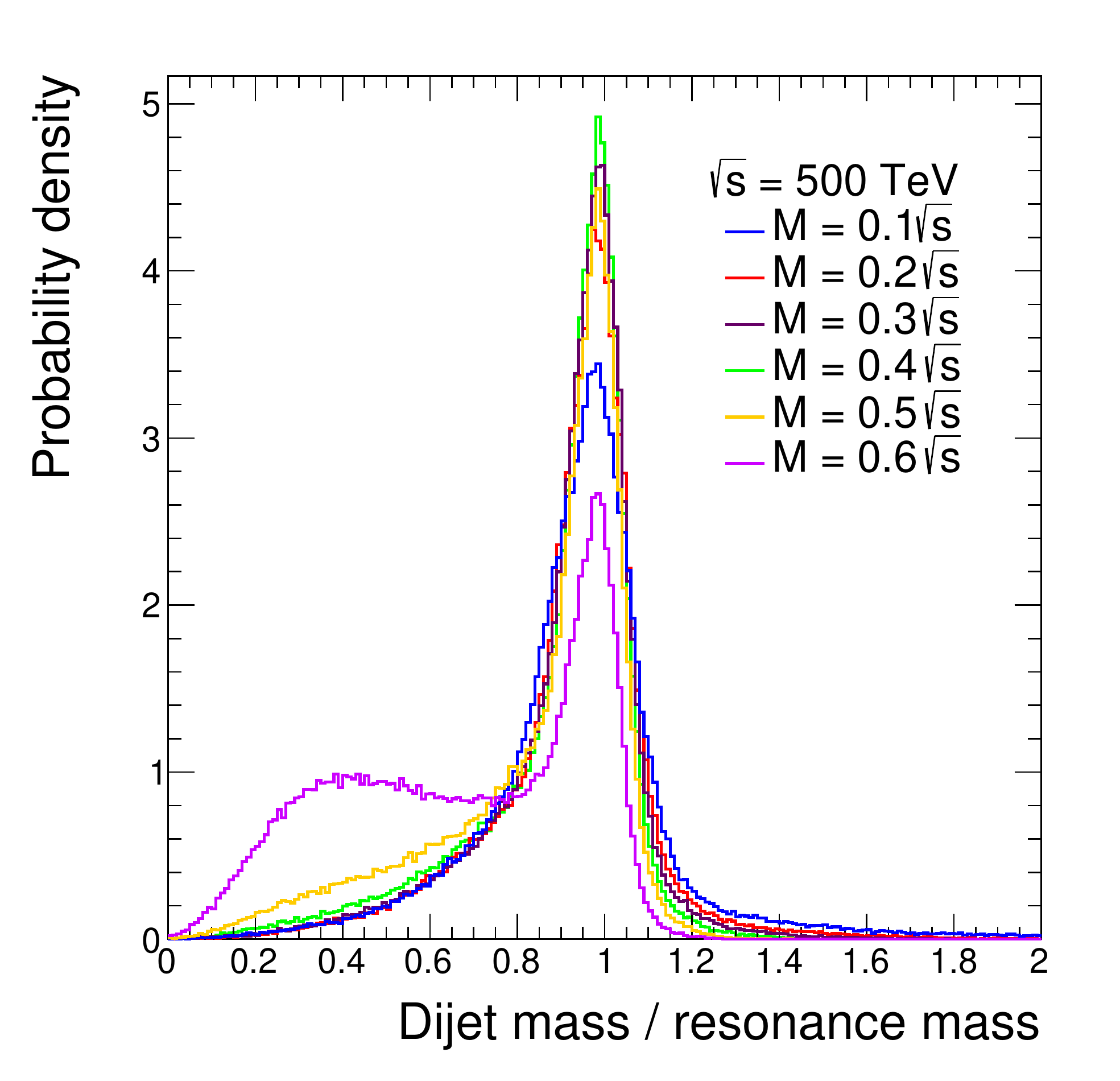}
\caption{\label{fig:SmearedMassFracFixedRootS} Comparisons of experiment level dijet mass distributions of excited quarks, with resonance mass equal to 10\%, 20\%, 30\%, 40\%, 50\% and 60\% of $\sqrt{s}$, from $pp$ collisions at $\sqrt{s}$ equal to 13 and 14 TeV (top row), and 100 and 500 TeV (bottom row).}
\end{figure}

The invariance of the dijet mass distribution at the experiment-level to the choice of resonance mass $M$, approximately valid for the mass range $0.1 < M/\sqrt{s} < 0.5$, is shown in 
Fig.~\ref{fig:SmearedMassFracFixedRootS}.  For all four values of $\sqrt{s}$ shown, only the resonance mass value $M/\sqrt{s}=0.6$ has a significantly different dijet mass distribution, because the significant probability within the long tail to low mass breaks the invariance. We discuss here the evolution of the shape with $M/\sqrt{s}$ for both the peak and the tails. 

On the peak of the dijet mass distribution, the probability density increases with $M/\sqrt{s}$
for the resonance masses in the range $0.1 < M/\sqrt{s} < 0.4$. This is because the experimental resolution in Eq.~\ref{eq:ExpRes} improves as $M/\sqrt{s}$ increases. The highest peak is always for $M/\sqrt{s}=0.4$, because when the mass reaches $M/\sqrt{s}=0.5$ noticeable numbers of events have migrated from the peak towards the low mass tail, so that the peak probability density for 
 $M/\sqrt{s}=0.5$ always lies between the peaks for $M/\sqrt{s}=0.2$ and $M/\sqrt{s}=0.3$. The migration is caused by the steepening PDFs, discussed in section~\ref{sec:massInvariance} and \ref{sec:allLevels}, that dominate the distribution for $M/\sqrt{s}=0.6$ which therefore has the lowest peak probability density. Despite this issue with the low mass tail, the width of the peak is steadily decreasing as $M/\sqrt{s}$ increases within the full range $0.1 < M/\sqrt{s} < 0.6$, because the peak is dominated by the experimental resolution.

On the tail at high dijet mass, the probability density decreases with increasing $M/\sqrt{s}$
for all the resonance masses in the range $0.1 < M/\sqrt{s} < 0.6$. The high mass tail has the natural shape of the underlying Breit-Wigner distribution, slightly enhanced by initial state radiation and experimental resolution, both of which decrease with increasing $M/\sqrt{s}$. 

On the tail at low dijet mass, the probability density is virtually indistinguishable for $0.1 < M/\sqrt{s} < 0.3$, because the process of final state radiation which dominates the tail in this region is independent of $M/\sqrt{s}$.  The first distribution to show a very small increase in the low mass tail due to the steepening PDFs is $M/\sqrt{s}=0.4$, which remains dominated by final-state radiation. Once again, as discussed in section~\ref{sec:massInvariance} and \ref{sec:allLevels}, the value $M/\sqrt{s}=0.5$ is the transition mass where the low mass tail is roughly equally composed of the effects of final-state radiation and the steepening PDFs, and for $M/\sqrt{s}=0.6$ the low mass tail is dominated by the effect of the steepening PDFs.  Nowhere within the low mass tail are there any indications of the effects of experimental resolution, which are only apparent in the height of the peak, the width of the peak, and possibly the high-mass tail.

\clearpage

\subsection{Cumulative distributions of experiment-level resonances}

\begin{figure}[htbp]
\centering 
\includegraphics[width=.32\textwidth]{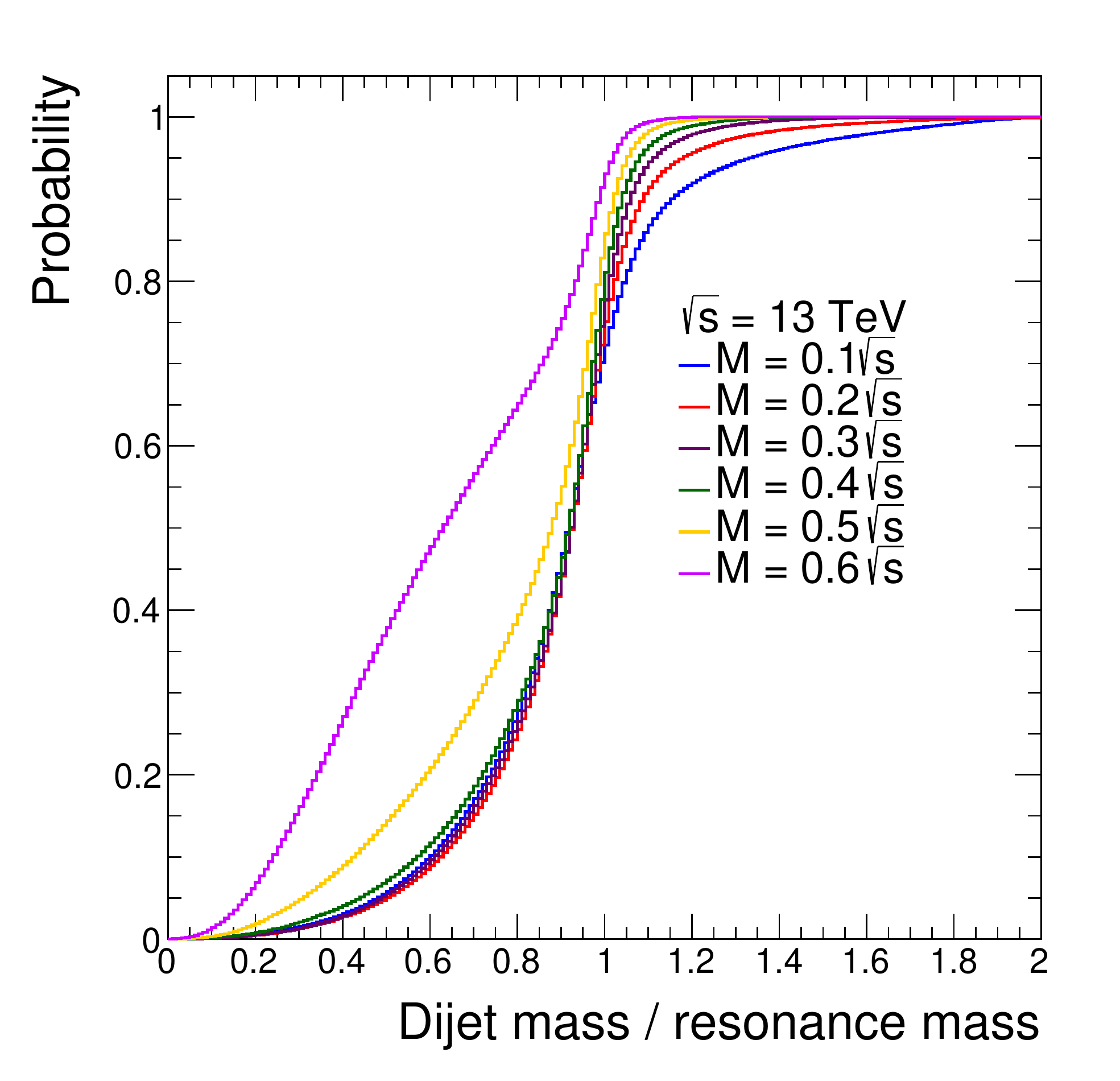}
\includegraphics[width=.32\textwidth]{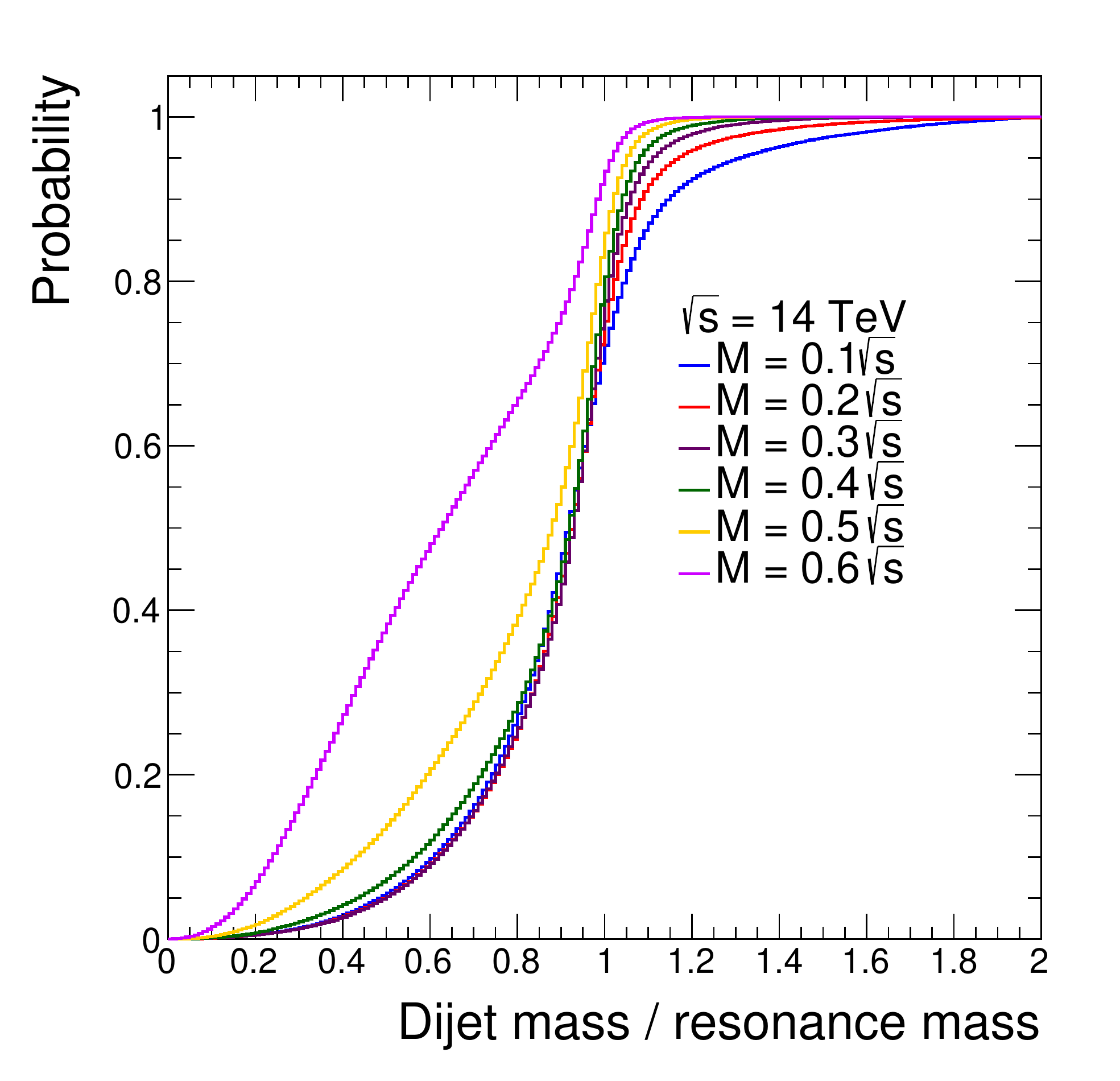}
\includegraphics[width=.32\textwidth]{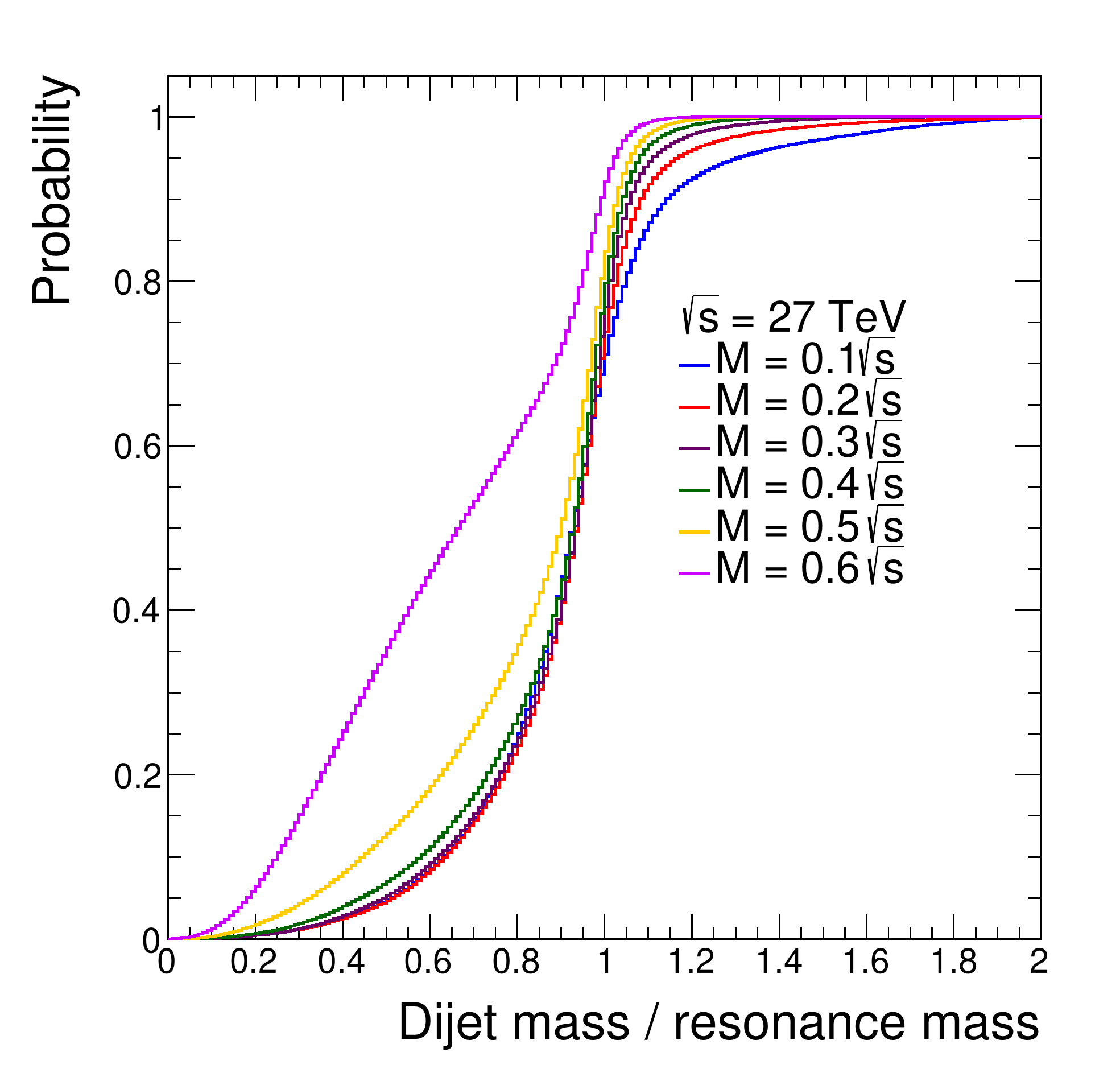}
\includegraphics[width=.32\textwidth]{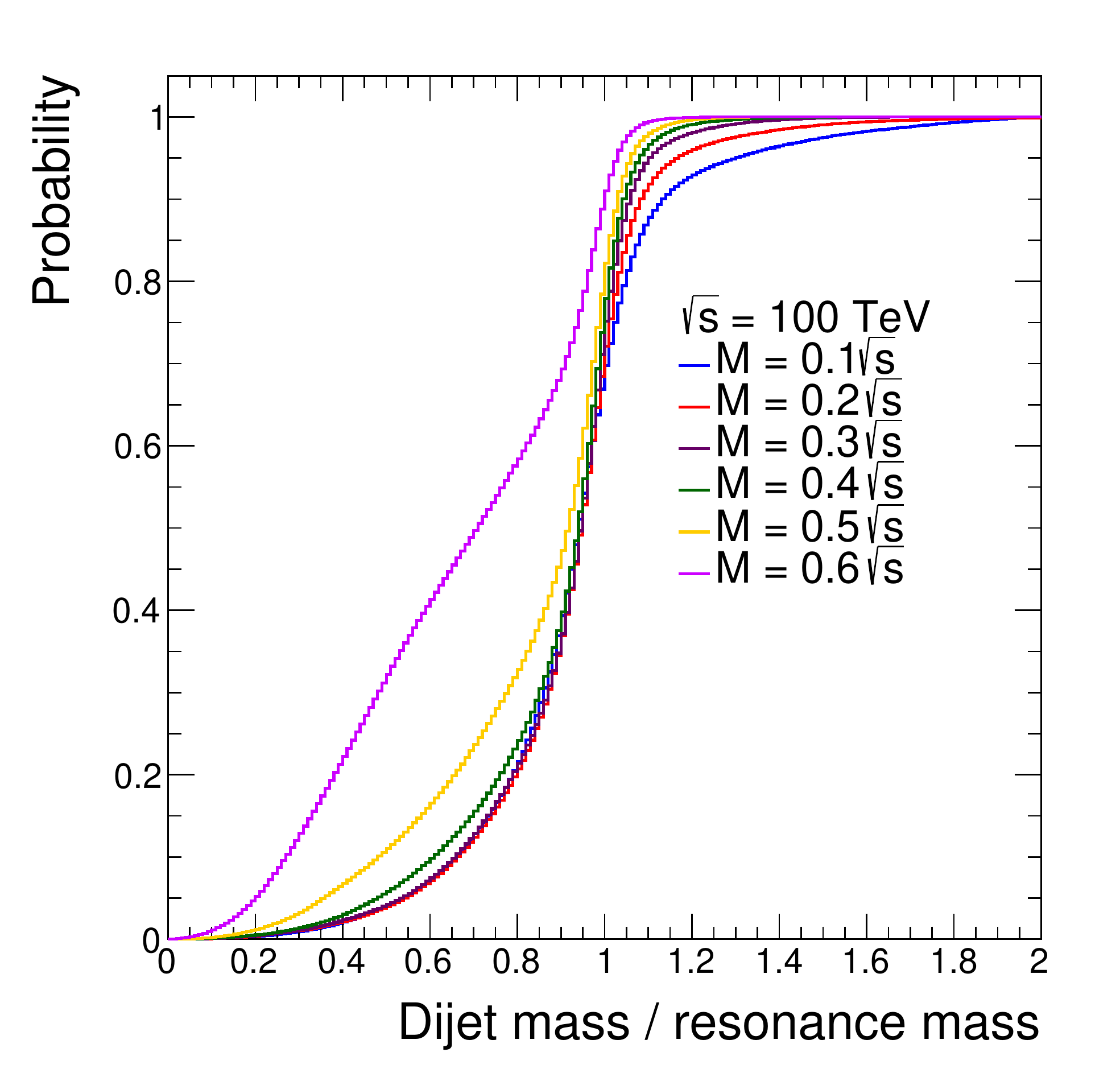}
\includegraphics[width=.32\textwidth]{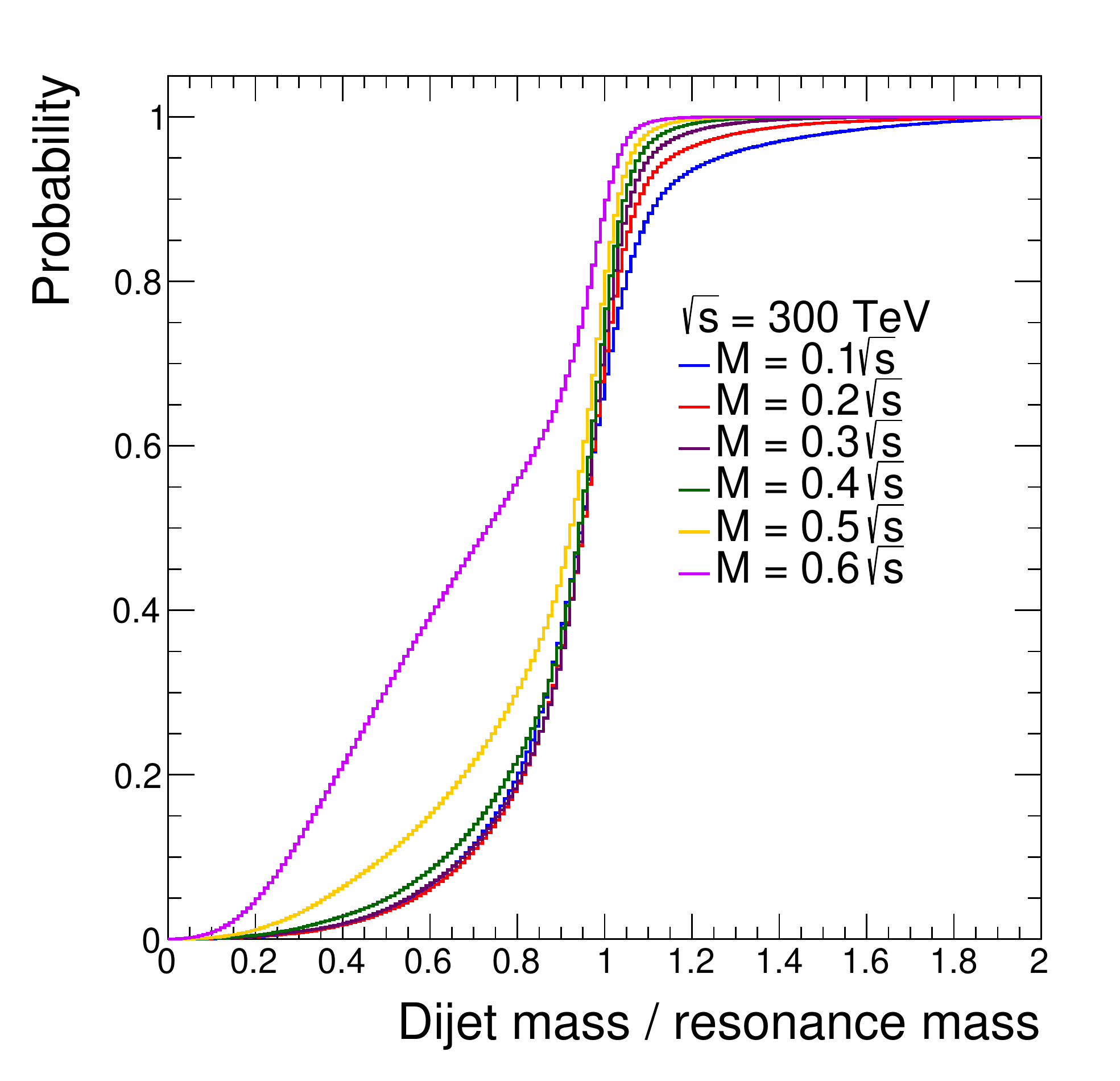}
\includegraphics[width=.32\textwidth]{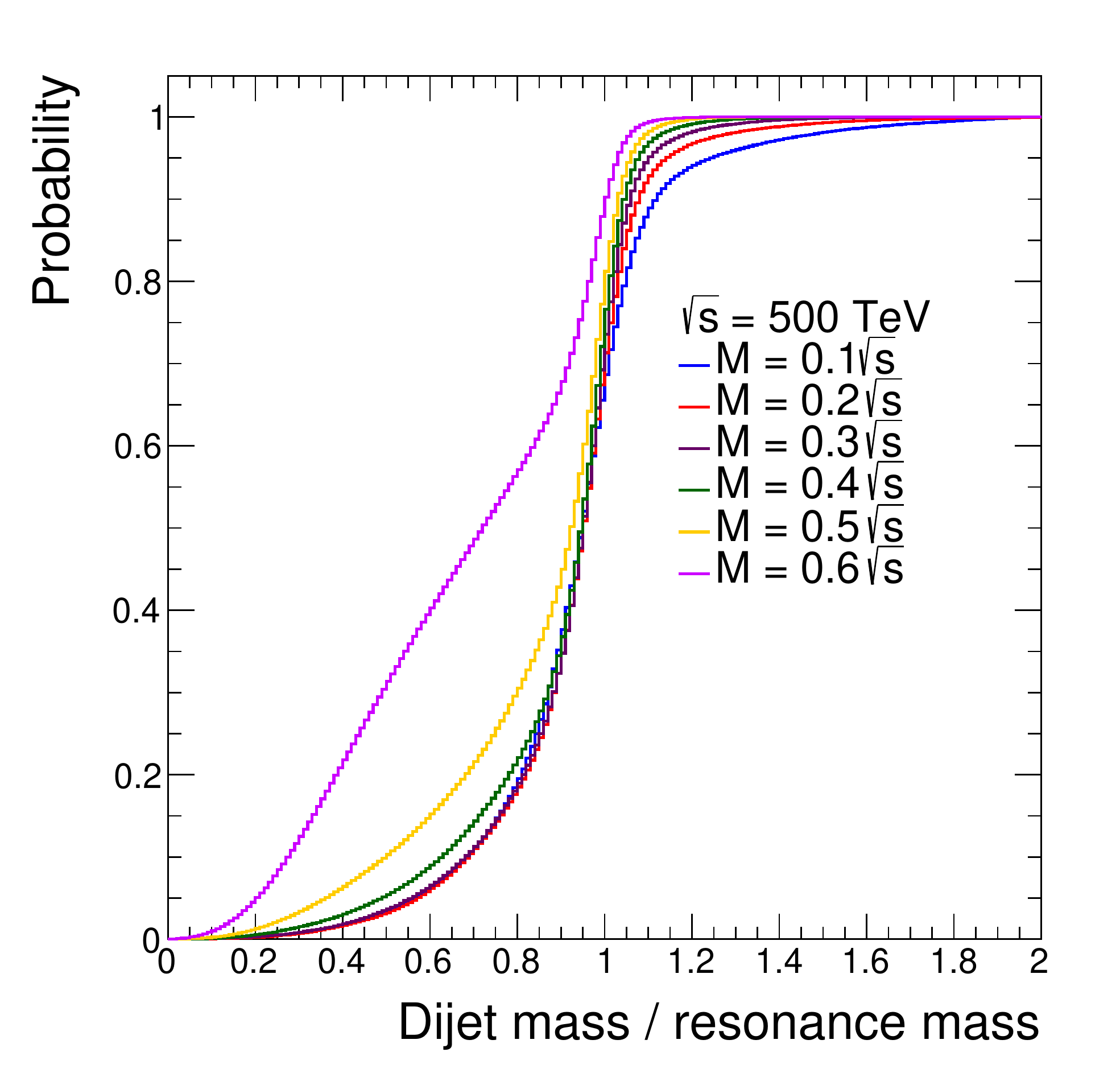}
\caption{\label{fig:CumulativeFixedRootS} Cumulative probability distributions from experiment level dijet mass distributions of excited quarks, for resonance mass $M$ equal to 10\%, 20\%, 30\%, 40\%, 50\% and 60\% of $\sqrt{s}$, from $pp$ collisions at $\sqrt{s}$ equal to 13, 14, and 27 TeV (top row), and 100, 300 and 500 TeV (bottom row).}
\end{figure}

\begin{figure}[tbp]
\centering 
\includegraphics[width=.32\textwidth]{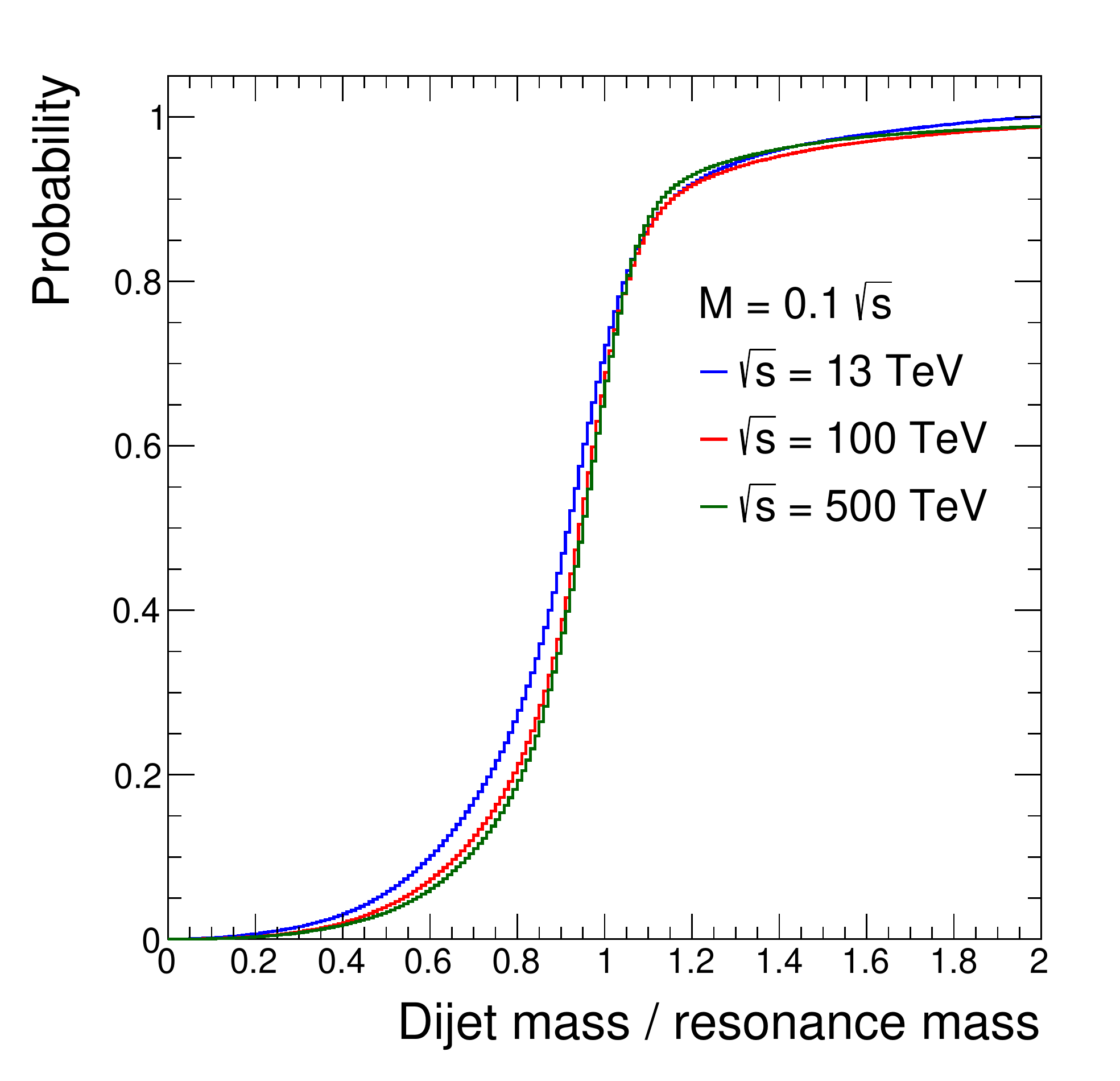}
\includegraphics[width=.32\textwidth]{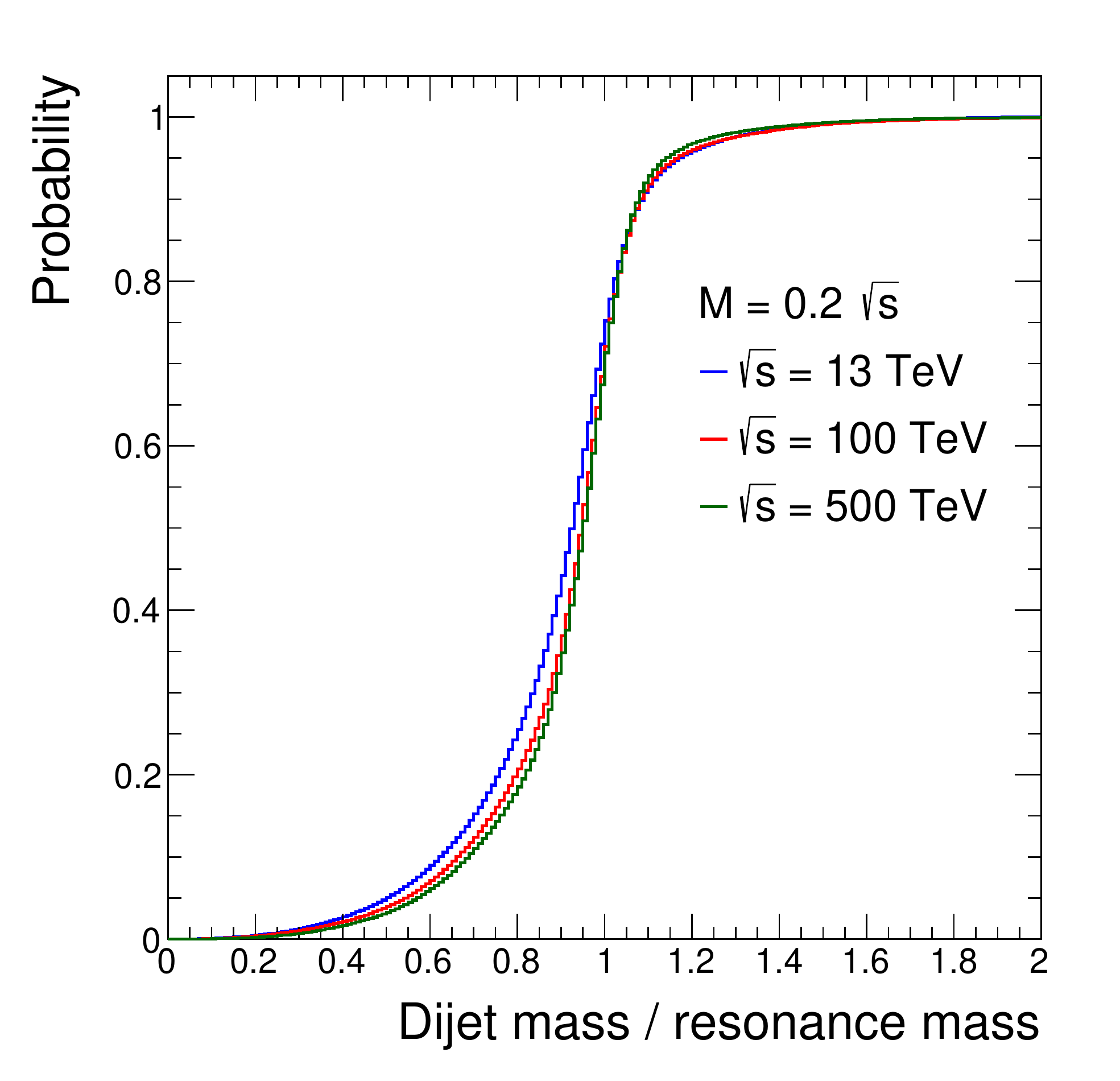}
\includegraphics[width=.32\textwidth]{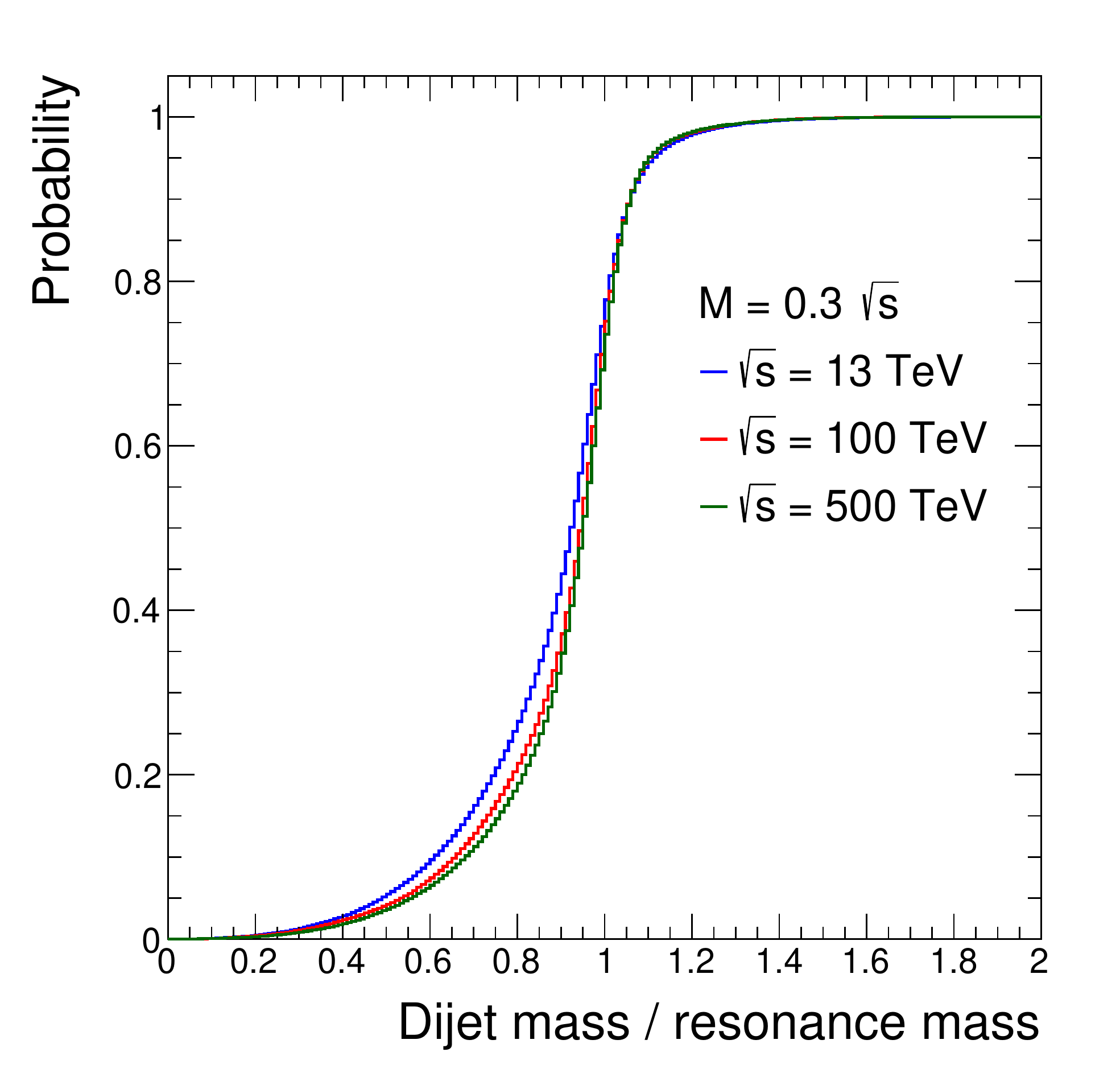}
\includegraphics[width=.32\textwidth]{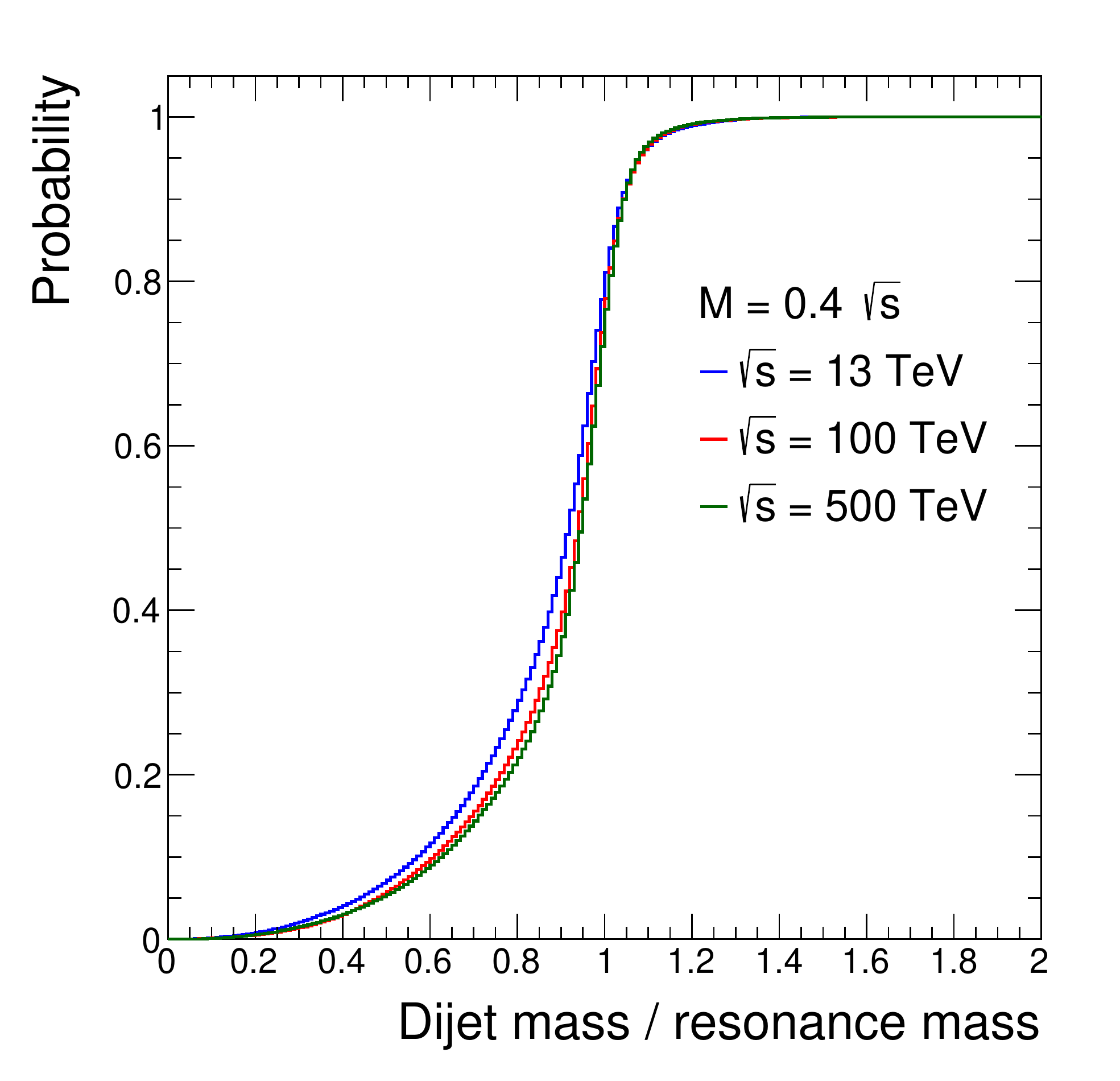}
\includegraphics[width=.32\textwidth]{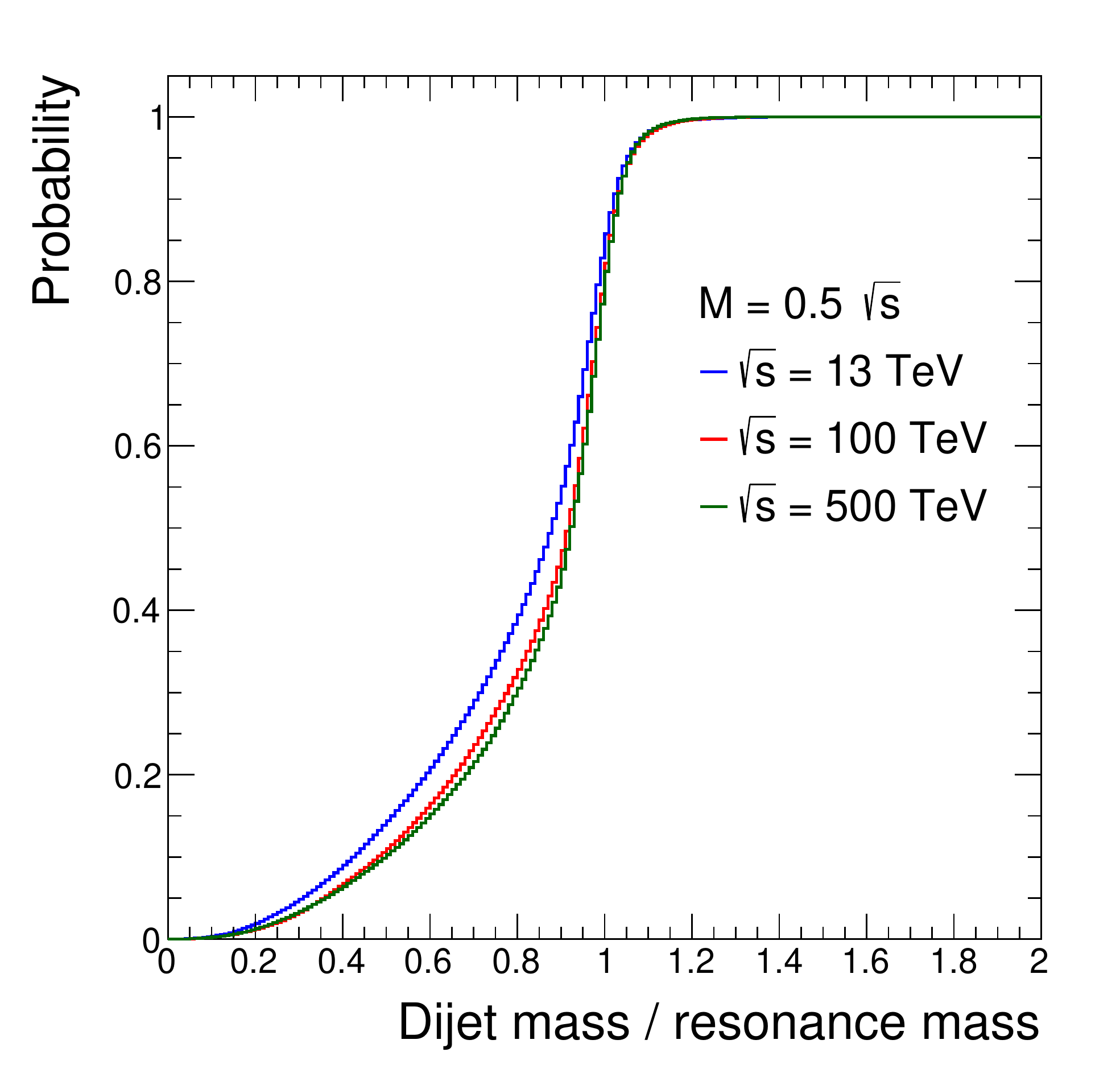}
\includegraphics[width=.32\textwidth]{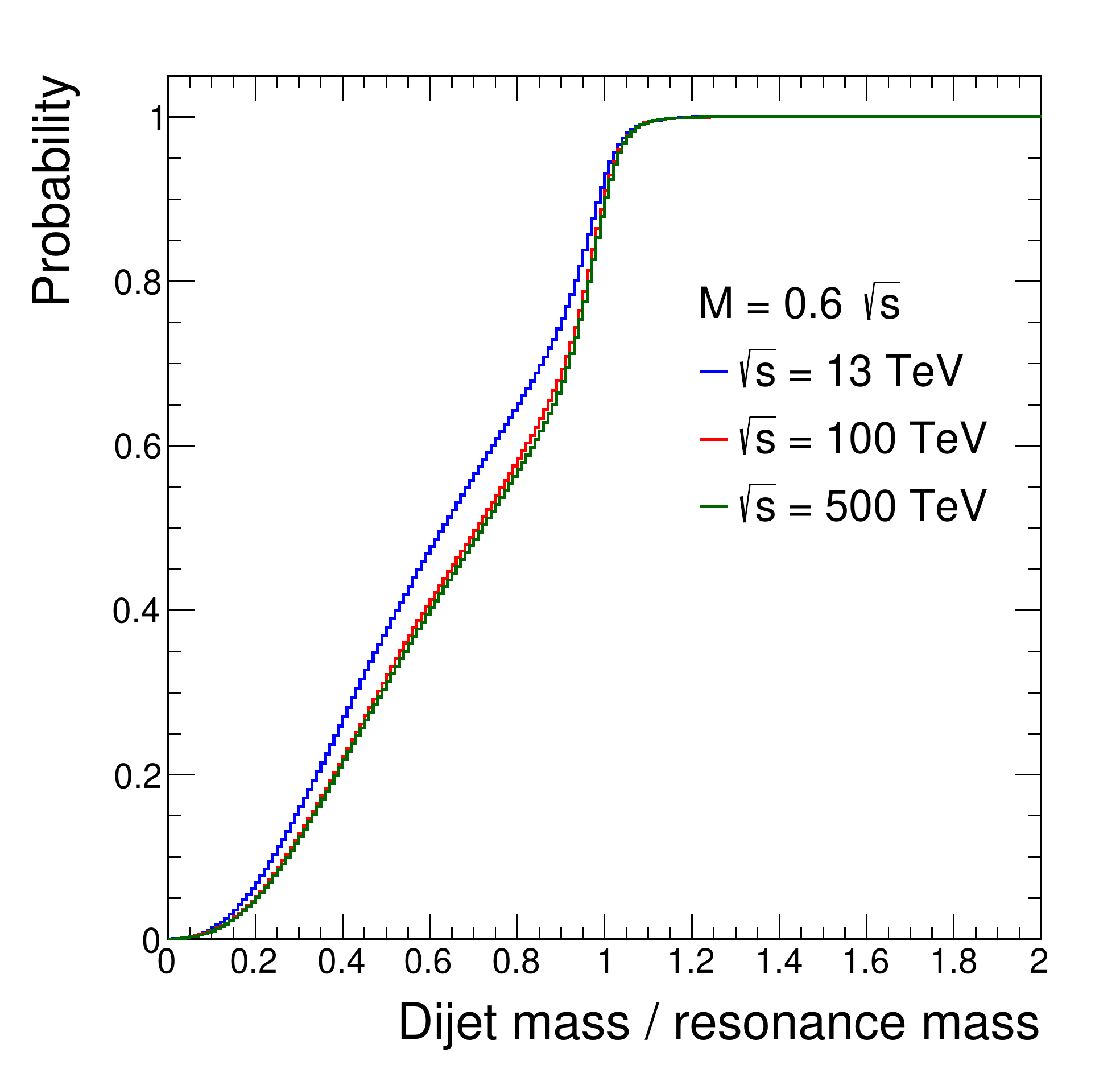}
\caption{\label{fig:CumulativeFixedMass} Cumulative probability distributions for experiment level dijet mass distributions of excited quarks, from $pp$ collisions at $\sqrt{s}$ equal to 13 TeV (blue), 100 TeV (red), and 500 TeV (green), for resonance masses of 10\%, 20\% and 30\% of $\sqrt{s}$ (top row) and 40\%, 50\% and 60\% of $\sqrt{s}$ (bottom row). }
\end{figure}

Cumulative probability distributions allow easier visualization and extraction of the acceptance of requirements made on dijet mass in a resonance search.  These distributions give the total probability, $p$, arising from integrating the probability density distribution, $dp/dm$, between a dijet mass of $0$ and $m$: $\int (dp/dm) dm$. They have the value $p=0$ for $m=0$ and the value $p=1$ for large enough $m$ to contain the entire distribution.  Cumulative probability distributions as a function of dijet mass, using the dimensionless ratio $m/M$, are shown in Fig.~\ref{fig:CumulativeFixedRootS} and Fig.~\ref{fig:CumulativeFixedMass}. 

Figure~\ref{fig:CumulativeFixedRootS} shows the cumulative probability from integrating Fig.~\ref{fig:SmearedMassFracFixedRootS} between 0 and a given value of the dimensionless ratio $m/M$. To within a probability of roughly 10\%, the cumulative probability is approximately invariant as a function of resonance mass for the mass range $0.1<M/\sqrt{s}<0.5$.

Similarly, Fig.~\ref{fig:CumulativeFixedMass} shows the cumulative probability from integrating Fig.~\ref{fig:SmearedMassFracFixedMass} between 0 and a given value of the dimensionless ratio $m/M$. Again, to within a probability of roughly 10\%, the cumulative probability is approximately invariant as a function of collision energy $\sqrt{s}$. 

\subsection{Acceptance of a dijet mass window}

A classic method to search for dijet resonance is to look for a bump in the dijet mass distribution within a window in dijet mass centered on the resonance peak.  This is because the majority of the sensitivity comes from a narrow interval in dijet mass surrounding the peak.  In our previous paper~\cite{Harris:2022kls} we used a dijet mass window, $0.836 < m/M < 1.164$, to estimate search sensitivity.   In that paper, the acceptance of that mass window was found to be $A=0.6$, for a dijet resonance from an excited quark at $M=6$ TeV and $\sqrt{s}=13$ TeV. We argued that acceptance was independent of resonance mass and collision energy, and used that acceptance to estimate the sensitivity to dijet resonances from excited quarks at $pp$ colliders.

\begin{figure}[tbp]
\centering 
\includegraphics[width=\textwidth]{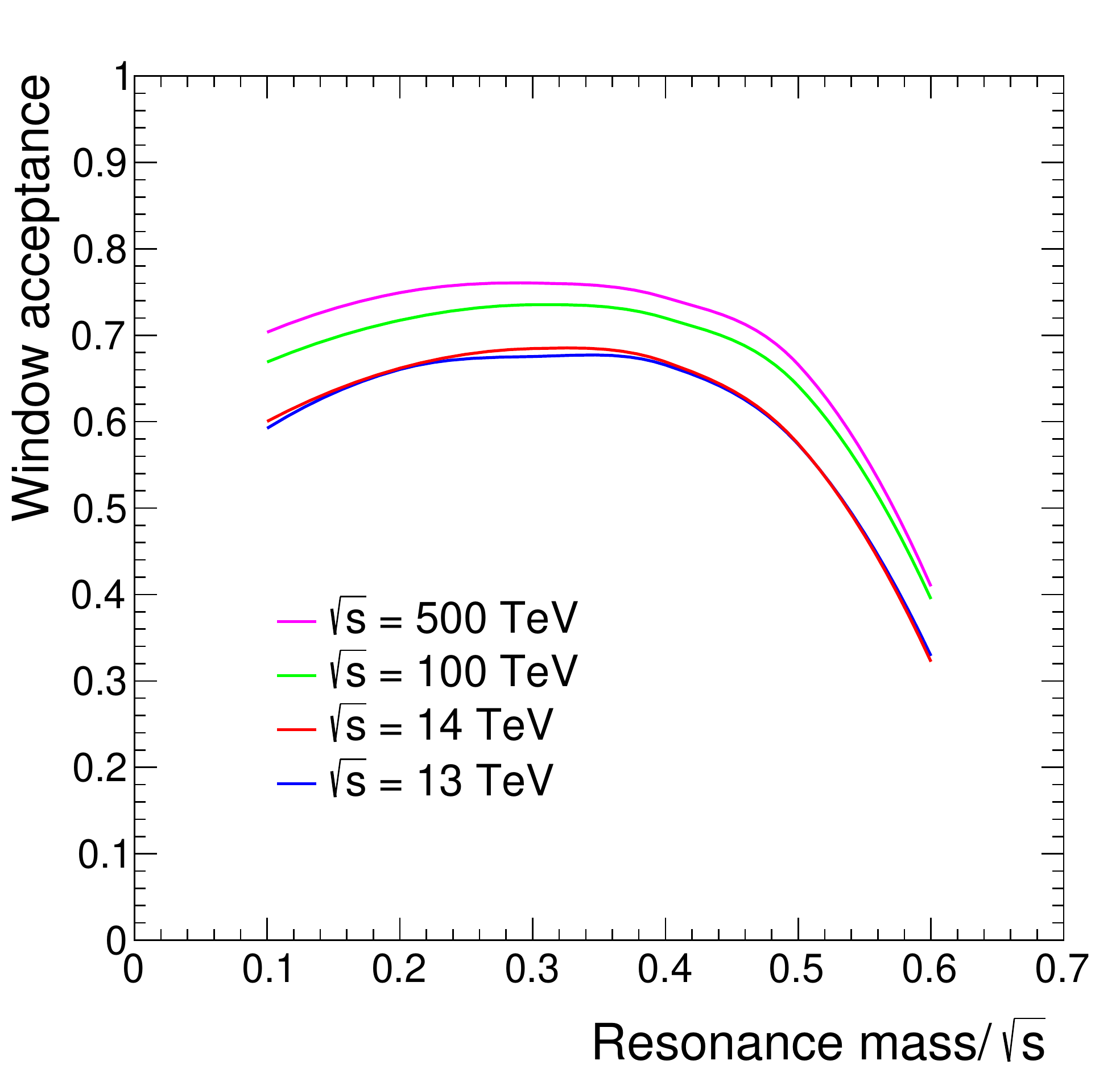}
\caption{\label{fig:acc} The experimental acceptance of the dijet mass window used in Ref.~\cite{Harris:2022kls}, as a function of resonance mass,  for excited quarks from $pp$ collisions at $\sqrt{s}$ equal to 13, 14, 100 and 500 TeV.}
\end{figure}

Figure~\ref{fig:acc} shows the $q^*$ acceptance of the dijet mass window, $0.836 < m/M < 1.164$, centered on the experiment level dijet resonance shapes in Fig.~\ref{fig:SmearedMassFracFixedRootS}.  The acceptance in Fig.~\ref{fig:acc} can also be easily extracted from Fig.~\ref{fig:CumulativeFixedRootS} by reading off the cumulative probabilities at the window boundaries in the ratio of dijet mass to resonance mass, $m/M = 1.164$ and $m/M=0.836$, and taking the difference.  Similarly, the acceptance for any choice of dijet mass window can be derived from Fig.~\ref{fig:CumulativeFixedRootS} by taking the difference between the cumulative probabilities  at the upper edge and lower edge of the mass window chosen.

Figure~\ref{fig:acc} shows the acceptance at four values of $\sqrt{s}$ as a function of resonance mass, which we will denote as $A_{\sqrt{s}}(M/\sqrt{s})$. There is less than about 10\% variation in window acceptance within the resonance mass range $0.1<M/\sqrt{s}<0.5$, approximately invariant as a function of resonance mass for the same mass interval where the resonance shapes and cumulative probabilities are approximately invariant as a function of resonance mass. The acceptance increases slowly with $\sqrt{s}$, $A_{13}<A_{14}<A_{100}<A_{500}$, but again doesn't change by more than about 10\%, approximately invariant as a function of $\sqrt{s}$. For the lowest $pp$ collision energy, LHC at $\sqrt{s}=13$ TeV, almost indistinguishable from HL-LHC at $\sqrt{s}=14$ TeV, the acceptance has a maximum value of $A_{13}(0.3)=0.68$ near $M/\sqrt{s}=0.3$ and is $A_{13}(0.5)=0.58$ for $M/\sqrt{s}=0.5$, which does not result in measurably different sensitivities than using $A=0.6$ as we did in our paper~\cite{Harris:2022kls}. For the highest $pp$ collision energy, $\sqrt{s}=500$ TeV, the acceptance has a maximum value of $A_{500}(0.3)=0.76$ at $M/\sqrt{s}=0.3$ and is $A_{500}(0.5)=0.67$ for $M/\sqrt{s}=0.5$, which again would not result in a measurable gain in sensitivity if we had used these values instead of the more conservative single choice $A=0.6$.  However, at $M\sqrt{s}=0.6$ the window acceptance for $\sqrt{s}=13$ and $14$ TeV has fallen to $A_{13}(0.6)=A_{14}(0.6)=0.32$, less than half its maximum value, and the window acceptance for $\sqrt{s}=100$ and $500$ TeV has fallen to $A_{100}(0.6)=A_{500}(0.6)=0.40$, just greater than half its maximum value.  As discussed in Ref.~\cite{Harris:2022kls}, this loss of acceptance means our quoted sensitivities for excited quarks with resonance masses near $M/\sqrt{s}=0.6$ need to be reduced by roughly 6\%, and our estimate of $q^*$ excluded mass at 95\% confidence level for HL-LHC with 3 ab$^{-1}$ in Refs.~\cite{Harris:2022kls,Bernardi:2022hny} needs to be reduced from $7.9$ TeV to roughly $7.6$ TeV.  

\section{Summary and conclusions}

We have presented and explained in detail the expected experimental dijet mass distributions from excited quarks at present and future $pp$ colliders. The natural width of the peak of the experimental distribution narrows only slightly with the collider energy $\sqrt{s}$ due to the running of the strong coupling constant. The expected width of the observed peak is consistent with the natural width convolved with a Gaussian experimental resolution. We have estimated and parameterized this resolution for wide jets from the CMS experiment, and scaled with $\sqrt{s}$ to estimate the resolution of future experiments. The expected experimental dijet mass distributions have a long tail at low dijet mass which is formed from two components, both of which come from physics at the generator-level. The first component of the low mass tail, which is visible and approximately invariant for all resonance masses, is the energy lost due to final-state radiation. That tail is significantly reduced by using the wide jet algorithm to reconstruct dijets, as opposed to using jets with a smaller cone size. This tail from final-state radiation interferes with accurate extraction of the Gaussian experimental resolution, making it approximate. The second component of the low mass tail, which is visible only at high resonance masses, is from the multiplication of the tail of the Breit-Wigner line shape by steeply falling parton momentum distributions of the proton, which are significantly steeper at high resonance mass. This second component contributes as much as final state radiation to the low dijet mass tail for resonance masses equal to half of $\sqrt{s}$, and dominates the dijet mass shape completely for resonance masses greater than 60\% of $\sqrt{s}$. Finally, the dijet mass distributions also have a short and small tail to high mass which becomes longer and larger at lower resonance masses, consistent with the effects of the natural line shape, initial state radiation and experimental resolution.

We have also presented cumulative probability distributions as a function of dijet mass, that can be used to extract the acceptance of requirements made on dijet mass within an excited quark search.  As an example of how to use these cumulative probability distributions, we have presented the excited quark signal acceptance of the dijet mass window which we used in our Snowmass 2021 study on the sensitivity to dijet resonances at $pp$ colliders~\cite{Harris:2022kls}.

We find an approximate universality of the differential dijet mass distributions, the cumulative probability distributions, and the window acceptance, which are all approximately invariant to changes in $\sqrt{s}$ and resonance mass for excited quark masses between 10\% and 50\% of $\sqrt{s}$. This universality, arising from the relatively constant natural width of the resonance and the approximate invariance of final-state radiation to changes in $\sqrt{s}$ and resonance mass, is violated by the steepening falloff of the parton momentum distributions of the proton when the resonance mass reaches and exceeds 60\% of $\sqrt{s}$.

\acknowledgments

This work was supported by the Turkish Energy, Nuclear and Mineral Research Agency. This work was also supported by the Fermi National Accelerator Laboratory, managed and operated by Fermi Research Alliance, LLC under Contract No. DE-AC02-07CH11359 with the U.S. Department of Energy. The U.S. Government retains and the publisher, by accepting the article for publication, acknowledges that the U.S. Government retains a non-exclusive, paid-up, irrevocable, world-wide license to publish or reproduce the published form of this manuscript, or allow others to do so, for U.S. Government purposes.  


\bibliographystyle{JHEP}
\bibliography{qstarShape}

\end{document}